\documentclass[floats,floatfix,showpacs,amssymb,prd,twocolumn,superscriptaddress,nofootinbib,longbibliography,reprint,aps]{revtex4-2}

\usepackage{amssymb,amsmath,verbatim,mathtools,needspace,enumitem,etoolbox,graphicx,physics,microtype,afterpage,xspace,tabularx,lmodern,multirow,bm}
\usepackage[normalem]{ulem}
\usepackage[dvipsnames]{xcolor}
\definecolor{linkcolor}{rgb}{0.0,0.3,0.5}
\usepackage[unicode, colorlinks=true, linkcolor=linkcolor, citecolor=linkcolor, filecolor=linkcolor, urlcolor=linkcolor, linktocpage, breaklinks]{hyperref}
\definecolor{romared}{RGB}{142,0,28}
\hypersetup{colorlinks=true,citecolor=romared,linkcolor=romared,urlcolor=romared}
\usepackage[all]{hypcap}
\usepackage[T1]{fontenc}
\usepackage[utf8]{inputenc}
\usepackage{aas_macros}
\usepackage{makecell}
\usepackage{soul}
\usepackage{booktabs}
\begin{document}

\author{Konstantinos Kritos}
\altaffiliation{\href{mailto:konstantinos.kritos@stonybrook.edu}{konstantinos.kritos@stonybrook.edu}, NASA Hubble Fellow.}
\affiliation{Department of Physics and Astronomy, Stony Brook University, Stony Brook, NY 11794-3800, US}
\affiliation{Department of Physics and Astronomy, Johns Hopkins University, 3400 North Charles Street, Baltimore, Maryland 21218, USA}

\author{Francesco Iacovelli}
\affiliation{Department of Physics and Astronomy, Johns Hopkins University, 3400 North Charles Street, Baltimore, Maryland 21218, USA}

\author{Rosalba Perna}
\affiliation{Department of Physics and Astronomy, Stony Brook University, Stony Brook, NY 11794-3800, US}

\author{Emanuele Berti}
\affiliation{Department of Physics and Astronomy, Johns Hopkins University, 3400 North Charles Street, Baltimore, Maryland 21218, USA}

\title{Compact Object Mergers and Micro-Tidal Disruption Events: \texorpdfstring{\\}{} A Multi-Messenger Probe of Dense Stellar Environments}

\begin{abstract}
    Dense stellar environments such as globular clusters and young massive clusters produce both compact-object mergers detectable in gravitational waves and tidal disruption events observable electromagnetically. We investigate compact-object mergers, tidal disruption events, and micro-tidal disruption events ($\mu$TDEs) in these environments using rapid Monte Carlo star-cluster simulations with the \href{https://github.com/Kkritos/Rapster}{\sc Rapster} code, extended with a unified treatment of neutron-star (NS) and black-hole (BH) spin evolution under disk accretion across the NS-to-BH transition, as well as new dynamical channels for disruptions by single compact objects and by BH-BH binaries. Simulating $10^5$ clusters across ten model variations and two formation histories, we find a BH-BH merger-rate density peaking at $z\approx2$--$3$ at several tens ${\rm Gpc}^{-3}\,{\rm yr}^{-1}$, consistent with the local LVK rate, while the cumulative number of $\mu$TDEs exceeds that of mergers by a factor of a few. Since we include only dynamically formed binaries, $\approx90\%$ of $\mu$TDEs proceed through single compact-object--star encounters, and our rates are conservative lower limits. Prograde accretion during $\mu$TDEs spins up compact objects, providing a purely dynamical, randomly oriented pathway for producing high-spin BHs, and can drive accretion-induced collapse of NSs to BHs that populate the lower mass gap.
\end{abstract}

\date{\today}
\maketitle

\tableofcontents

\section{Introduction}
\label{sec:Introduction}
 
Tidal disruption events (TDEs) provide a powerful probe of black holes (BHs) across a wide range of masses, from supermassive BHs in galactic nuclei to intermediate-mass and stellar-mass BHs in dense stellar systems. In a classical TDE, a star is torn apart when it approaches a BH within its tidal radius, producing a luminous electromagnetic (EM) flare and, in some cases, detectable gravitational-wave (GW) emission (see, e.g., \cite{Gezari:2021bmb}).
While most observed TDEs are associated with galactic nuclei, there is growing theoretical and observational interest in off-nuclear TDEs occurring in star clusters and other dense stellar environments.

At the same time, the LIGO-Virgo-KAGRA (LVK) collaboration~\cite{LIGOScientific:2014pky,Virgo:2014yos,Aso:2013eba} has published hundreds of GW events from compact-binary coalescences, including binary black holes (BBHs), neutron star--black hole (NSBH) binaries, and binary neutron stars (BNSs). 
The inferred local merger rate for BBHs is $27.5$--$49.4\,{\rm Gpc}^{-3}\,{\rm yr}^{-1}$ from the GWTC-5.0 catalog~\citep{LIGOScientific:2026ctl}, with correspondingly lower rates for NSBH and BNS systems; a broad range of formation channels remains consistent with these measurements~\cite{Mandel:2021smh}, and identifying the dominant astrophysical origin of these binaries remains an open problem.

Spin measurements offer one of the most promising diagnostics for distinguishing between formation channels. Hierarchical assembly in dense stellar clusters predicts a characteristic remnant spin $\chi_{\rm rem}\approx0.7$ (where $\chi$ is the dimensionless Kerr parameter of the BH) for near-equal-mass, non-spinning progenitors~\citep{Berti:2008af}, while sustained, coherent gas accretion in active galactic nucleus (AGN) disks can spin BHs up toward the Thorne limit~\cite{McKernan:2012rf,Yang:2019cbr,Tagawa:2020dxe,Kritos:2024kpn}. 
Previous works have analyzed LVK data and found evidence for spinning BH populations~\cite{Hotokezaka:2017esv,Piran:2018bbt,Galaudage:2021rkt,Chia:2021mxq}. Ref.~\cite{Kapil:2026hyn} considered the effect of tidal interactions in isolated binaries for tidally spinning up BHs.
More recently, Ref.~\cite{Bartos:2026xlt} recently reported evidence, using spin magnitudes alone, for a $\sim10\%$ high-spin subpopulation ($\chi\approx0.9$) among LVK binary BH mergers; this subpopulation is a better fit to an accretion-spin-up origin than to the hierarchical-merger prediction of $\chi\approx0.7$. 
This result sharpens the need for formation channels that can be distinguished observationally not only by spin magnitude but by spin orientation, a point we return to in Sec.~\ref{sec:comparison}.

Dense stellar systems such as globular clusters (GCs) and nuclear star clusters (NSCs; see Ref.~\cite{2020A&ARv..28....4N} for a review) are promising sites for TDEs and compact-object mergers involving stellar-mass black holes and intermediate-mass black holes (IMBHs).
Using Cluster Monte Carlo (CMC) simulations,~\citet{Kremer:2019zql,Kremer:2020cne,Kremer:2022xgm,Kremer:2023sof} have shown that dynamical interactions in star clusters can efficiently produce TDEs, including disruptions of main-sequence stars and compact objects. However, observational constraints remain challenging: recent work based on the absence of detected transients in wide-field surveys such as the Zwicky Transient Facility (ZTF)~\cite{2019PASP..131a8002B,Graham:2019qsw} has placed stringent upper limits on the TDE rate in star clusters of $\lesssim10^{-7}\,\rm yr^{-1}\, cluster^{-1}$~\cite{2024MNRAS.530.3043P}.
 
A particularly intriguing class of events involves the tidal disruption of white dwarfs (WDs) by IMBHs. These WD-IMBH TDEs are expected to produce both EM and GW signals, potentially detectable with next-generation observatories up to $\sim200\,\rm Mpc$~\cite{Sesana:2008zc}. The review by~\citet{Maguire:2020lad} summarizes the expected observational signatures, while numerical studies have explored the role of BH spin and the impact of nuclear reactions triggered during deep encounters~\cite{Rosswog:2009fh}. Typical WDs have masses~$\sim0.6\,M_\odot$ in the local Universe~\cite{2007MNRAS.375.1315K}, although heavier progenitors in dense clusters may produce more massive WDs. 
Analytical prescriptions for the EM emission from WD disruptions have also been developed~\cite{2020MNRAS.495.1061F}. 
\citet{Ye:2023fpb} estimate WD-IMBH tidal-capture rates of $\sim1\,{\rm Myr^{-1}}$ in nuclear star clusters, roughly two orders of magnitude higher than the $\sim0.01\,{\rm Myr^{-1}}$ they find for globular clusters, reflecting the higher central densities of NSCs. Since we consider only the GC channel for this class of events, our derived WD-IMBH rates should likewise be regarded as a lower limit relative to the NSC contribution.

At lower BH masses, so-called micro-TDEs ($\mu$TDEs), which are disruptions of stars by stellar-mass BHs or neutron stars (NSs), have emerged as another important class of transient events. First proposed by~\citet{Perets:2016pwr}, who estimated a rate of $10^{-7}$--$10^{-6}\,{\rm yr^{-1}\,galaxy^{-1}}$ from dynamically hardened binaries in dense clusters -- with a comparable contribution expected from the perturbation of wide binaries in the Galactic field -- these events are expected to occur in dense stellar environments and produce short-lived, high-energy transients \cite{Perets:2016pwr}. Hydrodynamic simulations of BH-star TDEs~\cite{Kremer:2019zql,
Wang:2021poh,Kremer:2022xgm,Ryu:2022,Kremer:2023sof,2019ApJ...877...56L,
Ryu:2023a,Ryu:2023b,Ryu:2024}, as well as more recent $N$-body simulations in young star clusters~\cite{2026A&A...707A.217R} have begun to characterize their rates and observable properties. In such environments, $\mu$TDEs may arise from tidally captured binaries whose merger timescales are shorter than subsequent dynamical encounter timescales~\cite{Liu:2025}. These events may also be associated with fast-evolving transients such as fast blue optical transients (FBOTs), which exhibit short durations, blue colors, and large offsets from their host galaxies~\cite{Ho:2023}. Because each $\mu$TDE deposits both mass and angular momentum onto the disrupting compact object, repeated $\mu$TDEs constitute a purely dynamical, cluster-based spin-up channel, complementary to, and, in principle, observationally distinguishable from, the AGN-disk channel discussed above.
 
Two recent population-synthesis studies bracket the current uncertainty in the $\mu$TDE rate. Using CMC simulations of the GC channel, \citet{Kremer:2019zql} find a local ($z=0$) BH-star $\mu$TDE rate of $\sim3\,{\rm Gpc^{-3}\,yr^{-1}}$, peaking at $\sim25\,{\rm Gpc^{-3}\,yr^{-1}}$ around $z\approx3$, with single-single encounters dominating over binary-mediated channels and the NS-star rate suppressed by two orders of magnitude relative to the BH-star rate. Using a comparable set of cluster simulations for the young massive cluster (YMC) channel but adopting a substantially higher primordial binary fraction, \citet{2026A&A...707A.217R} instead find that $\sim90\%$ of $\mu$TDEs originate from binary-mediated encounters, a total local rate of $\sim300\,{\rm Gpc^{-3}\,yr^{-1}}$ (with single-object encounters alone contributing $\sim10\,{\rm Gpc^{-3}\,yr^{-1}}$), and an NS-star fraction of $\sim20\%$, roughly three orders of magnitude higher than found by \citet{Kremer:2019zql}; most of their binary-mediated events involve primordial binaries undergoing common-envelope or mass-transfer episodes, with giants, WDs, and naked cores contributing only rarely, since most $\mu$TDEs are mediated by main-sequence stars. Because we do not include primordial binaries and consider only dynamically formed ones, our results are more directly comparable to the single-object rates of both studies and should be regarded as a conservative lower limit on the true $\mu$TDE rate.

In addition to single-passage $\mu$TDEs, dynamically formed BH-giant binaries in dense clusters provide another route to a compact-object accretor: physical collisions between a BH and a giant-branch star can produce a common-envelope-like event that leaves behind a BH-WD binary~\cite{Ivanova:2010ia}, which subsequent hardening or triple-induced Kozai-Lidov mass transfer can drive into an ultracompact, mass-transferring configuration. \citet{Yang:2026} recently confirmed, using CMC simulations, that BH+giant collisions are the dominant formation channel for such systems, identifying them as promising joint X-ray and LISA sources. We do not include this BH-giant collision channel in the present work.
 
Hydrodynamic simulations of tidal disruptions by BH-BH binaries themselves, rather than by single BHs, offer an additional motivation for this work. \citet{2019ApJ...877...56L} find that the outcome of a BBH-star encounter depends on the star's orbital energy relative to the binary's binding energy: high-energy encounters disrupt the star around a single BH member on a nearly parabolic orbit, while lower-energy encounters can instead form a circumbinary disk that is slowly accreted by both BHs, occasionally driving repeated, resonant disruptions of the same star. They find that such interactions can measurably alter both the magnitude and the direction of the disrupting BHs' spins -- for an initially non-spinning BH, the spin change is directed perpendicular to the incoming star's orbital plane -- and estimate a single-BH $\mu$TDE rate of $\sim10^{-4}\,{\rm yr^{-1}\,galaxy^{-1}}$, cautioning that more sophisticated $N$-body methods are needed to firm up this estimate. Motivated by this result and by the finding of Ref.~\cite{2026A&A...707A.217R} that a large primordial binary fraction can make binary-mediated disruptions competitive with single-BH ones, we introduce in this work a dedicated BBH-star TDE channel (Sec.~\ref{sec:bbh_star_tde}) within \href{https://github.com/Kkritos/Rapster}{\sc Rapster}, while finding, consistent with the single-BH-dominated picture of \citet{Kremer:2019zql}, that this channel remains subdominant in our dynamically formed (non-primordial) binary population (Sec.~\ref{sec:branching}).
 
TDEs are also promising multi-messenger sources. Recent work has provided updated predictions for GW emission from TDEs across a wide parameter space, building on the parabolic-encounter burst formalism of \citet{BerryGair2010} and highlighting their detectability with future detectors \citep{Toscani:2021bzr,Toscani:2025uar}. In addition, TDEs in AGN disks and other gas-rich environments may produce distinctive EM counterparts~\cite{McPike:2026ugj}. In some cases, repeated accretion or disruption events could even induce compact object transformations, such as accretion-induced collapse of neutron stars into BHs~\cite{Perna:2021fbq}. These processes may also contribute to the population of BHs in the low-mass gap~\cite{Gupta:2019nwj}.
 
Motivated by these theoretical developments, a growing number of candidate off-nuclear and IMBH-associated TDEs have been identified. These include sources such as eRASS J1421-29~\cite{Grotova:2025tdm},
NGC 6099 HLX-1~\cite{Chang:2025ucz}, EP20240222a~\cite{Jin:2025izu}, 3XMM J2150-05~\cite{Lin:2018dev}, and AT2024tvd~\cite{Yao:2025dbw}, as well as ESO243-49 HLX-1~\cite{Soria:2017ght}. More speculative candidates include EP250702a, which has been proposed as a possible WD-IMBH TDE based on its light curve alone~\cite{Li:2025mae}, and TDE 2025abcr, an event located at the outskirts of a massive galaxy~\cite{Stein:2026qmg}. The interpretation of such events remains uncertain, and alternative explanations, including ``TDE encores''~\cite{Ryu:2024utf}, may apply in some cases. Theoretical modeling of these systems often relies on few-body integrations, for which tools such as {\sc SpaceHub} provide fast and accurate solutions~\cite{2021MNRAS.505.1053W}.
 
In this work, we investigate compact-object mergers, TDEs, and $\mu$TDEs in dense stellar environments by carrying out rapid star-cluster simulations with \href{https://github.com/Kkritos/Rapster}{\sc Rapster}~\cite{Kritos:2022ggc}. We extend the code with a unified treatment of NS and BH spin evolution under disk accretion spanning the NS-to-BH transition, and with a new dynamical channel for tidal disruptions by single compact objects and binaries, and give particular emphasis to formation channels, rates, and observable signatures. We connect dynamical modeling with the growing population of off-nuclear transients and assess the implications for BH demographics across cosmic environments; in particular, we show that repeated $\mu$TDEs imprint a measurable, monotonically increasing trend of $\langle\chi_{\rm eff}\rangle$ with prior $\mu$TDE count, offering a dynamical alternative to AGN-disk accretion for the high-spin subpopulation reported by \citet{Bartos:2026xlt}; one that, unlike the AGN channel, predicts randomly oriented rather than preferentially aligned spins.

\section{Methods}
\label{sec:methods}

We investigate the formation channels, rates, and observable signatures of compact-object mergers, TDEs, and $\mu$TDEs in dense stellar environments by conducting rapid star-cluster simulations with the \href{https://github.com/Kkritos/Rapster}{\sc Rapster} (RAPid clu\textsc{ster} evolution) code~\cite{Kritos:2022ggc}. 
This section first introduces the code (Sec.~\ref{sec:code}), then describes the cluster initial conditions (Sec.~\ref{sec:cluster_ic}), the treatment of the NS (Sec.~\ref{sec:ns_population}) and WD (Sec.~\ref{sec:wd_population}) subpopulations, the TDE framework (Sec.~\ref{sec:tde_framework}), the unified spin-evolution model for compact objects accreting from a disk (Sec.~\ref{sec:spin_evolution}), and the calculation of the cosmological merger and $\mu$TDE rate density (Sec.~\ref{sec:merger_rate}) that connects our simulations to the population of off-nuclear transients and BH demographics discussed in Sec.~\ref{sec:Introduction}.

\subsection{The \textmd{\href{https://github.com/Kkritos/Rapster}{\sc Rapster}} code}
\label{sec:code}

\href{https://github.com/Kkritos/Rapster}{\sc Rapster} is a public, open-source Monte Carlo cluster-evolution code available on \textsc{GitHub}~\cite{RapsterGitHub}. Rather than following individual stellar orbits as direct $N$-body integrators do, the code advances a small set of cluster-averaged quantities (total mass, half-mass radius, central density, velocity dispersions, etc.) using the semi-analytic two-body relaxation theory of~\citet{Breen:2013vla}, supplemented with dedicated rate equations for three-body binary 
formation, binary hardening, exchange interactions, TDEs, GW-driven mergers in single-single and binary-single encounters, and triple formation and evolution, as described throughout this section. This approach trades the microphysical detail of direct $N$-body codes for computational speed, allowing a single realization to be evolved from cluster formation to a Hubble time in seconds to minutes, which makes large population studies over the cluster initial-condition space (Sec.~\ref{sec:cluster_ic}) computationally tractable. 
The~\citet{Breen:2013vla} treatment is calibrated against direct $N$-body simulations and is generally trusted for BH subsystems containing more than $\sim40$ BHs. To extrapolate to smaller BH populations, the code fixes the Coulomb logarithm of the BH subsystem, and the overall dynamics are evolved in the Newtonian regime, valid as long as the system compactness remains far from the strong-field limit.

Compact-remnant BH masses are computed from the zero-age-main-sequence mass and metallicity of each progenitor star. The default prescription
is the delayed core-collapse model of~\citet{Fryer:2011cx}, applied to stellar tracks evolved with the updated-BSE code~\cite{Banerjee:2019jjs}; three alternative prescriptions are selectable at run time, including the rapid~\citet{Fryer:2011cx} model and, for both the delayed and rapid core-collapse engines, remnant masses obtained by interpolating precomputed look-up tables generated with the population-synthesis code \textsc{SEVN}~\cite{Spera:2015vkd}, 
spanning zero-age main-sequence masses up to $340\,M_\odot$ over a grid of 12 absolute metallicities between $10^{-4}$ and $1.7\times10^{-2}$. The standard set of output files records, at each timestep, cluster- and subsystem-level properties, the hardening history of every dynamically formed binary, the parameters of every $\mu$TDE, and the full source parameters (masses, spins, formation/merger times and redshifts, assembly channel, and remnant properties) of every compact-binary merger. The code and its underlying semi-analytic framework are described and validated in Ref.~\cite{Kritos:2022ggc}, and have since been applied to a range of studies of dynamically assembled compact-binary populations and their electromagnetic counterparts.

\subsection{Star cluster model and initial conditions}
\label{sec:cluster_ic}

Star clusters form within $1\,\mathrm{Myr}$ and virialize within a dynamical time 
\begin{align}
    {\tau_{\rm dyn}}\approx0.02\,{\rm Myr}\,\left(\dfrac{r_{h,0}}{1\,\rm pc}\right)^{3/2}\left(\dfrac{M_{\rm cl,0}}{10^6\,M_\odot}\right)^{-1/2};
\end{align}
for the parameter space we focus on, this virialization time is $\lesssim5\,\mathrm{Myr}$. For each simulated cluster we draw uniformly and independently a formation redshift $z_{\rm cl,form}\sim\mathcal{U}(0,10)$, an absolute metallicity $\log_{10}Z\sim\mathcal{U}(-4,\log_{10}0.02)$, and an
initial cluster mass $\log_{10}(M_{\rm cl,0}/M_\odot)\sim\mathcal{U}(4,8)$.
The initial half-mass radius is likewise drawn log-uniformly between $100\,{\rm pc}$ and a mass-dependent lower bound, $r_{h,0} \sim \mathcal{U}\!\left[\log_{10}r_{h,\rm min},\,2\right]$, where
\begin{equation}
  r_{h,\rm min} = \max\!\left(0.1\,{\rm pc},\,
  {6.9\times10^{-3}}\,\frac{M_{\rm cl,0}/M_\odot}
  {(v_{\rm esc,max}/{\rm km\,s^{-1}})^2}\,{\rm pc}\right),
  \label{eq:rh-sampling}
\end{equation}
with $v_{\rm esc,max}=500\,\mathrm{km\,s^{-1}}$, chosen so that the central escape velocity, $v_{\rm esc,0} = 2\sqrt{0.4\,G\,M_{\rm cl,0}/r_{h,0}}$~\cite{1971ApJ...164..399S}, never exceeds $v_{\rm esc,max}$ at $r_{h,0}=r_{h,\rm min}$, which excludes clusters so compact that their central escape velocity would be too large; the same expression for $v_{\rm esc,0}$ is used throughout to test compact-object and merger-remnant retention. Each cluster is additionally assigned a galactocentric radius $R_{\rm gal,0}\sim\mathcal{U}(0,250)\,{\rm kpc}$, orbiting at a fixed galactic circular velocity of $220\,\mathrm{km\,s^{-1}}$, characteristic of a Milky-Way-like flat rotation curve. 

This broad, near-uniform coverage of initial conditions $(M_{\rm cl,0},\,r_{h,0},\,Z,\,z_{\rm cl,form},\,R_{\rm gal,0})$ is a computationally convenient sampling choice for the raw simulation ensemble, not itself the assumed astrophysical population: the physical cluster initial-mass function, $dN/dM_{\rm cl,0}\propto M_{\rm cl,0}^{-2}$ over the same mass range~\cite{2014CQGra..31x4006K,2019ARA&A..57..227K}, together with the YMC and GC formation-rate histories $\psi_{\rm cl}(z)$ (Sec.~\ref{sec:Results}), are applied afterward as importance-sampling weights (Sec.~\ref{sec:merger_rate}) to obtain the astrophysically weighted rates and distributions reported in Sec.~\ref{sec:Results}.

The initial binary fraction of low-mass (main-sequence) stars is set to $10\%$ by default, and the initial number of stars follows from the cluster mass and the mean stellar (system) mass of the assumed initial mass function (IMF) and binary fraction, $N_{\star,0}=M_{\rm cl,0}/\langle m\rangle$ with $\langle m\rangle\approx0.586\,M_\odot$ as derived from the standard initial mass function~\cite{Kroupa:2002ky}; the corresponding initial half-mass number density is $n_{h,0}=3N_{\star,0}/[4\pi(r_{h,0}/1.3)^3]$, where the factor $1.3$ converts the half-mass radius to the Plummer scale radius. Stellar evolution is followed with the SSE single-star model~\cite{Hurley:2000pk}, updated with the BSE prescription for stellar winds and with the latest pair-instability supernova (PISN) prescriptions~\cite{Banerjee:2019jjs}. Stellar masses are drawn from a Kroupa IMF in the range $0.08$--$200\,M_\odot$. While the upper limit of stellar masses is not well-known, we assume a universal IMF throughout.

Dynamically formed BH binaries are seeded through three-body binary formation: at each timestep we form binaries with a minimum hardness of $5$~\cite{Morscher:2014doa}, adopting a probability of $80\%$~\cite{1976A&A....53..259A} that a strong three-body interaction among single BHs results in a bound binary. First-generation (1g) BH natal spin magnitudes are assigned from a spin parameter $\chi_{\rm 1g}$, drawn uniformly in $[0,\chi_{\rm 1g}^{\rm max}]$; we explore $\chi_{\rm 1g}^{\rm max}=0.0$, $0.2$ (default), and $0.5$ in the population study of Sec.~\ref{sec:Results}.

Repeated BH mergers can build up an IMBH over time; alternatively, an IMBH can form promptly through runaway stellar collisions before the massive stars that participate in the collisions evolve into stellar-mass BH remnants. We optionally seed the cluster at $t=0$ with a non-spinning massive BH of initial mass $M_{\rm BH,0}=10^{-3}\,M_{\rm cl,0}$, following Refs.~\cite{PortegiesZwart:2002iks,Rasio:2003sz}. This choice is consistent with the hierarchical picture of YMC formation of Ref.~\cite{2014CQGra..31x4006K}, and with the finding of Ref.~\cite{Miller:2012ys} that clusters with velocity dispersion $\gtrsim40\,\mathrm{km\,s^{-1}}$ cannot support themselves through binary heating alone and rapidly assemble an IMBH via runaway growth. 
We take the onset of core-collapse-driven stellar evolution (first massive-star supernovae) at $t_{\rm SE}=3\,\mathrm{Myr}$ and the completion of BH formation from massive-star remnants at $t_{\rm BH,form}=10\,\mathrm{Myr}$ (see Ref.~\cite{Woosley:2002zz}, Table~I); the same formation timescale is adopted for NSs, which we note as a simplifying caveat of the model.

\subsection{Neutron star population}
\label{sec:ns_population}

\subsubsection{Formation and natal kicks}
\label{sec:ns_formation}

Neutron stars are included in \href{https://github.com/Kkritos/Rapster}{\sc Rapster} as compact objects and are added directly to the mass vector once their progenitor stars evolve off the main sequence with the rest of the BHs. The fraction of stars that form a NS remnant is~\cite{Fryer:2011cx}
\begin{equation}
  f_{\rm NS}^{\rm form} \approx \int_{8\,M_\odot}^{18\,M_\odot}
  \Phi(m_\star)\,dm_\star,
\end{equation}
where $\Phi(m_\star)\equiv N_\star^{-1}\,dN_\star/dm_\star$ is the IMF, normalized such that $\int\Phi(m_\star)\,dm_\star=1$ over the full mass domain $m_{\rm min}=0.08\,M_\odot$ to $m_{\rm max}=200\,M_\odot$. The number of NSs formed is then $N_{\rm NS}^{\rm form}=f_{\rm NS}^{\rm form}N_\star$.

Neutron stars receive a strong natal kick at the moment of supernova core collapse~\cite{Kapil:2022blf}. The kick is drawn from a Maxwellian distribution with one-dimensional (per-axis) velocity dispersion $\sigma_{\rm kick}=265\,\mathrm{km\,s^{-1}}$, following the fit of~\citet{Hobbs:2005yx} to Galactic pulsar proper motions; for BHs, by default, the kick is attenuated by fallback accretion during core collapse. A NS is retained in the cluster only if its natal kick velocity is smaller than the cluster's escape velocity, otherwise it is dynamically ejected and removed from the simulation. We treat NSs as first-generation remnants with zero spin at birth~\cite{Lorimer:2001vd} (Sec.~\ref{sec:ns_isd}).

\subsubsection{Initial mass and spin distribution}
\label{sec:ns_isd}

The initial NS mass distribution follows the bimodal Gaussian model of Ref.~\citep{Rocha:2023xwp}, which identifies two populations: a low-mass component centered at $\mu_1=1.351\,M_\odot$ with dispersion $\sigma_1=0.084\,M_\odot$, associated with recycled pulsars and double-NS systems, and a high-mass component centered at $\mu_2=1.816\,M_\odot$ with dispersion $\sigma_2=0.260\,M_\odot$ inferred from GW observations. The mixing fraction of the low-mass component is $0.539$. The initial spin of every NS is set to $\chi=0$ (non-rotating), consistent with the expectation that NSs are generally born slowly spinning (see Ref.~\cite{Gerosa:2013laa}, footnote~1, and Ref.~\cite{Perna:2008}) 
and that any pre-accretion spin is negligible compared to the spin subsequently imparted by disk accretion (Sec.~\ref{sec:spin_evolution}).

\subsubsection{Compact-binary electromagnetic counterparts}
\label{sec:ns_em}

For binaries involving at least one NS, we determine at the post-processing stage whether the merger is plausibly associated with an electromagnetic counterpart using the analytic remnant-mass and ejecta-mass fitting formulae of Refs.~\cite{Pannarale:2015jia} [their Eq.~(2) and Table~1] and~\cite{Foucart:2018rjc}.

\subsection{White dwarf population}
\label{sec:wd_population}

\subsubsection{Formation rate}
\label{sec:wd_formrate}

We model stellar lifetimes as $t_\star = t_{\star1}/m_\star^{p}$, with $p=2.5$ and $t_{\star1}=10^4\,\mathrm{Myr}$ the lifetime of a solar-mass star; or equivalently, $m_\star=(t_{\star1}/t_\star)^{1/p}$. The transition stellar mass separating NS and WD progenitors is metallicity-dependent, with a fiducial value $m_{\rm WN}\sim6$--$9\,M_\odot$. Assuming the cluster forms in an instantaneous burst at $t=0$, the first WDs appear once stars with mass below $m_{\rm WN}$ complete their evolution, at $t_{\rm WN}=t_{\star1}/m_{\rm WN}^{2.5}$. The cumulative number of WDs formed by time $t\ge t_{\rm WN}$ is
\begin{align}
  N_{\rm WD}^{(+)}(t) &= \int_{m_\star(t)}^{m_{\rm WN}} dN_\star
   = \int_{m_\star(t)}^{m_{\rm WN}} \Phi(m')\,N_\star\,dm' \nonumber\\
  &= -\int_{t_{\rm WN}}^{t} \Phi\!\left(m'(t')\right) N_\star
     \frac{dm'(t')}{dt'}\,dt' \nonumber\\
  &= \int_{t_{\rm WN}}^{t} \frac{N_\star}{p\,t'}\,
     \Phi\!\left[\left(\frac{t_{\star1}}{t'}\right)^{1/p}\right]
     \left(\frac{t_{\star1}}{t'}\right)^{1/p} dt',
\end{align}
and the corresponding WD formation-rate term is
\begin{equation}
\begin{aligned}
  \frac{dN_{\rm WD}^{(+)}(t)}{dt} &=
  \begin{cases}
  \dfrac{N_\star}{p\,t}\,
  \Phi\!\left[\left(\dfrac{t_{\star1}}{t}\right)^{1/p}\right]
  \left(\dfrac{t_{\star1}}{t}\right)^{1/p}, & t\ge t_{\rm WN}, \\[2mm]
  0, & t < t_{\rm WN},
  \end{cases}\\
  &\equiv \Gamma_{\rm WD}^{(+)}(t) 
\end{aligned}
  \label{eq:WD-formation-term}
\end{equation}
where we remind the reader that $\Phi$ is the IMF normalized to unity.

\subsubsection{Evolution equation: sources and sinks}
\label{sec:wd_evolution}

White dwarfs are removed from the single-object population either through TDEs onto BHs or NSs, or through two-body relaxation processes that repopulate the high-velocity tail of the WD Maxwellian velocity distribution on a timescale comparable to the half-mass relaxation time; WD-WD binaries can also form over time, sourcing a separate WD-WD population with its own source and sink terms. Since single-star and single-WD populations 
share comparable masses at late times, we expect them to have very similar density profiles: the stellar and WD populations thermalize with one another without triggering a mass-segregation (Spitzer) instability for the WD subpopulation, so both species can be treated as point particles of comparable mass from the point of view of two-body relaxation. In particular, we assume the WD evaporation rate per particle is identical to that of the stars, and impose energy equipartition, $m_{\rm WD}\langle v_{\rm WD}^2\rangle = \overline{m}_\star\langle v_\star^2\rangle$.

The single-WD population then evolves as
\begin{equation}
  \dot N_{\rm WD} = +\Gamma_{\rm WD}^{(+)}(t) - \Gamma_{\rm WD}^{\rm (TDE)}(t)
  - \Gamma_{\rm WD}^{\rm (ev)}(t),
\end{equation}
with formation terms given by Eq.~\eqref{eq:WD-formation-term}, evaporation
term
\begin{equation}
  \Gamma_{\rm WD}^{\rm (ev)}(t) = -\xi\,\frac{N_{\rm WD}(t)}{\tau_{\rm rh}},
\end{equation}
and TDE term $\Gamma_{\rm WD}^{\rm (TDE)}(t)$ given in Sec.~\ref{sec:wd_bh_tde}. At every timestep the number of WDs is updated as $N_{\rm WD}(t+dt) = N_{\rm WD}(t) + dt\,\dot N_{\rm WD}(t)$, where $dt$ is the adaptive timestep of \href{https://github.com/Kkritos/Rapster}{\sc Rapster}. We adopt a fiducial WD mass of $0.6\,M_\odot$ based on local observations~\cite{2007MNRAS.375.1315K}, corresponding to a carbon-oxygen degenerate star. 
We assume all WDs have the same fixed mass of $0.6\,M_\odot$ in our simulations.

\subsection{Tidal disruption event framework}
\label{sec:tde_framework}

\subsubsection{Two-body encounter formalism}
\label{sec:tde_twobody}

For a two-body hyperbolic encounter of total mass $M$ and relative velocity at infinity $v_\infty$, the impact parameter $b$ relates to the pericenter distance $r_p$ through gravitational focusing (e.g., Ref.~\citep{1969mech.book.....L}),
\begin{equation}
  b^2 = r_p^2 + \frac{2GM r_p}{v_\infty^2},
  \label{eq:impact-pericenter}
\end{equation}
where the first term is geometric and the second is the gravitational focusing term. If $b_c$ is the critical impact parameter for a given process to occur (corresponding to critical pericenter $r_{p,c}$), the cross section for that process is $\Sigma = \pi b_c^2$.

For tidal disruption of a star of radius $r_\star$ by a BH of mass $m_{\rm BH}$, the tidal radius is $r_T = r_\star(m_{\rm BH}/m_\star)^{1/3}$, and the corresponding critical impact parameter is, in the gravitational-focusing regime,
\begin{align}
  b_T^2 &= r_T^2 + \frac{2G(m_{\rm BH}+m_\star)\,r_T}{v_\infty^2} \nonumber\\
  &\approx \frac{2G(m_{\rm BH}+m_\star)\,r_\star}{v_\infty^2}
  \left(\frac{m_{\rm BH}}{m_\star}\right)^{1/3}.
  \label{eq:impact-param-TDE}
\end{align}

For each disruption event, we sample the pericenter uniformly in area, or equivalently -- since $b^2\propto r_p$ in the gravitational-focusing regime -- uniformly in $r_p^2$ over $[0,r_t^2]$, where $r_t$ is the relevant tidal radius. We define the penetration factor $\beta\equiv r_t/r_p>1$ required for disruption; interactions with very large penetration are exceedingly rare, so most BH-WD (and BH-star) TDEs occur close to, but within, the tidal radius.
In every single-object disruption, a fraction $f_A$ of the disrupted mass is retained in a bound accretion disk and subsequently deposited onto the compact object, with the remainder unbound; $f_A$ is a global model parameter (Sec.~\ref{sec:Results}) that sets the mass step $\Delta M=f_A\,m_\star$ (or $f_A\,m_{\rm WD}$) over which the spin-evolution equations of Sec.~\ref{sec:spin_evolution} are subsequently integrated for the disruptor.

\subsubsection{White dwarf--black hole TDE rate}
\label{sec:wd_bh_tde}

We compute the WD-BH TDE rate as an average of the cross section in Eq.~\eqref{eq:impact-param-TDE} over the Maxwellian distribution of the relative velocity $v_\infty$ between a typical WD and a typical BH in the cluster,
\begin{widetext}
\begin{align}
  \Gamma_{\rm WD}^{\rm (TDE)}(t) &= \pi\,n_{\rm BH}(t)\,N_{\rm WD}(t)\,
  \langle b_T^2 v_\infty\rangle \nonumber\\
  &\approx 2\pi G\!\left[\overline{m}_{\rm BH}(t)+m_{\rm WD}\right]
  n_{\rm BH}(t)\,N_{\rm WD}(t)\,r_{\rm WD}
  \left(\frac{\overline{m}_{\rm BH}(t)}{m_{\rm WD}}\right)^{1/3}
  \left\langle\frac1{v_\infty}\right\rangle,
  \label{eq:TDE-rate}
\end{align}
\end{widetext}
using Eq.~\eqref{eq:impact-param-TDE} in the gravitational-focusing limit, and where $v_\infty$ denotes the three-dimensional relative velocity of a BH and a WD at infinity. The required Maxwellian average is
\begin{align}
  \left\langle\frac1{v_\infty}\right\rangle
  &= \int_0^\infty dv_\infty\,\frac1{v_\infty}
   \sqrt{\frac2\pi}\frac{v_\infty}{\sigma_\infty}
   \exp\!\left(-\frac{v_\infty^2}{2\sigma_\infty^2}\right) \nonumber\\
  &= \sqrt{\frac2\pi}\,\frac1{\sigma_\infty}
   = \frac{\sqrt{2(3\pi-8)}}{\pi}\,\frac1{\langle v_\infty^2\rangle^{1/2}},
\end{align}
with $3\sigma_\infty^2=\langle v_\infty^2\rangle = \langle v_{\rm BH}^2\rangle+\langle v_{\rm WD}^2\rangle$ evaluated over the local Maxwellian velocity distributions, and $\sigma_\infty$ is the one-dimensional velocity dispersion parameter. At each timestep, we draw the number of BH-WD TDEs from a Poisson distribution with mean $\Gamma_{\rm WD}^{\rm (TDE)}(t)\,dt$. For each realized TDE, the disrupting BH mass is drawn consistently with Eq.~\eqref{eq:TDE-rate} and is increased by a fraction $f_{A}$ the disrupted WD mass, with default value $f_{A}=0.30$; the same BH is allowed to participate in more than one TDE within a single timestep.
We note that the TDE rate could be up to a factor of two larger if tidal captures (in addition to prompt disruptions) are accounted for.

An identical formalism, with $m_{\rm WD}\to m_\star$ and $r_{\rm WD}\to R_\star$ sampled from the time-evolving stellar mass function (Sec.~\ref{sec:exchange_tde}), gives the rate of direct compact-object--star disruptions. 
Because single BHs vastly outnumber BH-BH binaries and BH-WD pairs, and stars vastly outnumber WDs at early cluster ages, this direct compact object (CO)-star channel is the dominant source of single-object TDEs in the simulations.

\subsubsection{Restriction to single-object disruptors}
\label{sec:tde_single}

We adopt the working hypothesis, to be verified against dedicated binary-single scattering experiments (e.g. with codes such as \textsc{REBOUND}~\cite{Rein2012REBOUND},
\textsc{Tsunami}~\cite{TraniTsunami} or \textsc{SpaceHub}~\cite{2021MNRAS.505.1053W}), that a negligible number of TDEs originate from BH members of hard BH-BH binaries. For a binary-single interaction between a hard BH-BH binary and a single star/WD, the interaction cross section is $\sigma_{\rm flyby} = 2\pi G M_{\rm tot}\,r_p\,v_{\rm rel,\infty}^{-2}$, evaluated in the gravitational-focusing regime appropriate for a hard binary (whose binding energy greatly exceeds the typical kinetic energy of a cluster member). The ratio of the TDE cross section for either binary member to the flyby cross section is
\vspace{-0.02in}
\begin{equation}
  \frac{\sigma_{{\rm TDE},1/2}}{\sigma_{\rm flyby}}
  = \frac{R_\star}{a_{12}}\,\frac{m_{1/2}+m_\star}{m_{12}+m_\star}
  \left(\frac{m_{1/2}}{m_\star}\right)^{1/3},
\end{equation}
\vspace{-0.02in}%
which in the equal-mass, $m\equiv m_1=m_2\gg m_\star$ limit reduces to
\begin{equation}
  \frac{\sigma_{\rm TDE}}{\sigma_{\rm flyby}}
  = \frac{R_\star}{2a}\left(\frac{m}{m_\star}\right)^{1/3}.
\end{equation}%
For solar-type stars, TDEs in binary-star interactions can be neglected as long as $a\gg R_\odot\,m^{1/3}$. As an illustrative case, a circular $10\,M_\odot$--$10\,M_\odot$ BH-BH binary at separation $a=R_\odot\simeq2.25\times10^{-8}\,\mathrm{pc}$ has GW merger timescale \cite{Peters1963}
\begin{widetext}
\begin{equation}
  \tau_{\rm gw}(e=0) = \frac{5c^5a^4}{512G^3m^3}
  \simeq 0.076\,\mathrm{Myr}\left(\frac{a}{R_\odot}\right)^4
  \left(\frac{m}{10\,M_\odot}\right)^{-3}.
  \label{eq:tau-gw}
\end{equation}
\end{widetext}
This is a very short timescale compared to the characteristic time for a disruptive binary-star encounter, $\tau_{\rm tde}=(n_\star\langle\sigma_{\rm tde}v_{\rm rel}\rangle)^{-1} \approx (7GR_\odot n_\star m^{4/3})^{-1}v_\star$, with the ratio of the two being
\begin{widetext}
\begin{align}
  \frac{\tau_{\rm tde}}{\tau_{\rm gw}} &\approx
  \frac{512}{35}\,\frac{G^2}{c^5}\,\frac{m^{5/3}v_\star}{R_\odot^5\,n_\star}
  \simeq 9\times10^3\left(\frac{m}{10\,M_\odot}\right)^{5/3}
  \left(\frac{v_\star}{10\,\mathrm{km\,s^{-1}}}\right)
  \left(\frac{n_\star}{10^6\,\mathrm{pc^{-3}}}\right)^{-1}.
  \label{eq:tau-ratio}
\end{align}
\end{widetext}
Since this ratio is $\gg1$, TDEs from binary members do not play a significant role: a star that approaches a hard BH-BH binary is far more likely to extract binding energy and be accelerated than to be disrupted by one of its BH members, since disruption would require an incoming stellar velocity in the far Maxwellian tail. This conclusion is corroborated by the scattering-experiment results of Ref.~\citep{2026A&A...707A.217R}, which find that only $\sim7\%$ of $\mu$TDEs are associated with binaries, { while $90\%$, the majority, are associated with multiples}; we further note that most BHs in the cluster are single, even though binaries individually carry a larger disruption cross section. We therefore restrict TDEs in the baseline model to single objects, and treat the BH-BH-star channel separately in Sec.~\ref{sec:bbh_star_tde}.

\vspace{0.1in}
\subsubsection{BBH-star tidal disruptions}
\label{sec:bbh_star_tde}
We additionally implement BH-BH (BBH)--star tidal disruptions as a new dynamical channel, competing at each timestep with the existing BBH-BH and BBH-BBH interaction rates. The encounter rate follows the standard two-body cross section,
\begin{widetext}
\begin{equation}
  \Gamma_{\rm BBH-\star} = n_\star\,\Sigma_{\rm int}\,v_{\rm rel}, \qquad
  \Sigma_{\rm int} = \frac{2\pi G(m_1+m_2+\langle m_\star\rangle)\,
  (k_{p,\rm max}a)}{v_{\rm rel}^2},
  \label{eq:bbh-star-rate}
\end{equation}
\end{widetext}
with $v_{\rm rel}=\sqrt{v_\star^2+v_{\rm BH}^2}$ and maximum interaction pericenter $k_{p,\rm max}a$ ($k_{p,\rm max}=2$), consistent with the cross section already adopted for BBH-BH and BBH-BBH encounters. Once an encounter is selected, its pericenter is sampled \emph{uniformly in $r_p$ itself},
\begin{equation}
  r_p \sim \mathcal{U}(0,\,k_{p,\rm max}a),
\end{equation}
rather than uniformly in impact-parameter-squared, because in the gravitational-focusing regime ($v_{\rm rel}^2\ll GM/r_p$, applicable here) the effective cross section scales linearly with pericenter, $\sigma(r_p)\propto r_p$, rather than as $r_p^2$.

The sampled $r_p$ is then compared against two physically distinct length scales to determine the outcome. The whole-binary tidal radius, treating the BBH as an unresolved point mass $m_1+m_2$ (monopole approximation, valid for $r_p\gtrsim a$),
\begin{equation}
  r_{t,\rm bin} = R_\star\left(\frac{m_1+m_2}{m_\star}\right)^{1/3},
\end{equation}
sets the threshold below which disruption can occur at all: for $r_p>r_{t,\rm bin}$ the star is deflected but not disrupted. Two disruption outcomes are then distinguished:
\begin{itemize}
  \item \emph{Soft TDE}, 
  for $a<r_p\le r_{t,\rm bin}$: the star is disrupted by the combined tidal field of the binary, and the bound debris mass $\Delta m = f_A\,m_\star$ is split between the two BHs in proportion to their mass:
  \begin{equation}
    \Delta m_i = \frac{m_i}{m_1+m_2}\,\Delta m, \qquad i=1,2.
  \end{equation}
  \item \emph{Hard TDE}, 
  for $r_p\le a$: the star penetrates inside the binary orbit and is disrupted by a single BH component, using the standard single-BH tidal radius $r_t=R_\star(m_i/m_\star)^{1/3}$, with the disruptor selected with probability
  \begin{equation}
    p_i \propto (m_\star+m_i)\,m_i^{1/3},
  \end{equation}
  identical to the single-BH TDE prescription of Sec.~\ref{sec:tde_single}.
\end{itemize}
In both channels, the BH/NS spin evolution during the accretion episode follows the same prograde disk-accretion model used for single-BH TDEs (Sec.~\ref{sec:spin_evolution}).

For resonant encounters, identified by the standard criterion
\begin{equation}
  r_p < r_{p,c} \equiv \frac{\max(m_1,m_2)}{m_1+m_2}\,a,
\end{equation}
we account for the multiple close passages occurring during the chaotic phase of the three-body interaction by resampling the pericenter independently up to $N_{\rm IMS}=20$ times -- the same parameter governing the number of intermediate resonant states elsewhere in \href{https://github.com/Kkritos/Rapster}{\sc Rapster} -- checking for disruption at each passage and stopping at the first success. This boosts the cumulative disruption probability of a resonant encounter from a single-passage estimate $P_1$ to
\begin{equation}
  P_{\rm res} = 1-(1-P_1)^{N_{\rm IMS}}.
\end{equation}

If no disruption occurs, whether or not the encounter is resonant, the star is treated as a generic third body in the existing hardening/ionization machinery, using its own sampled mass $m_3=m_\star$ (rather than $\langle m_\star\rangle$) and a velocity drawn from a Maxwellian consistent with energy equipartition,
\begin{equation}
  \sigma_\star = \sqrt{\frac{\langle m_\star\rangle\,v_\star^2}{3\,m_\star}},
\end{equation}
mirroring the analogous expression $\sigma_{\rm BH}=\sqrt{\langle m_{\rm BH}\rangle v_{\rm BH}^2/(3m_{\rm BH})}$ already used for single-BH third bodies. The binary hardens according to the standard heuristic~\cite{Quinlan:1996vp,Antonini:2016gqe}
\begin{equation}
  a \;\to\; \frac{a}{1+H\,m_3/(m_1+m_2)},
\end{equation}
which, since $m_\star\ll m_{\rm BH}$, means the stellar contribution to hardening is intrinsically suppressed by a factor $\sim m_\star/m_{\rm BH}$ relative to BH-driven hardening. Stars are explicitly prevented from participating in the BH-specific outcomes available to genuine third bodies -- three-body GW-capture mergers, dynamical exchanges, and single-object ejection bookkeeping -- so that a star's possible fates are restricted to tidal disruption, ionization of the binary, or ordinary hardening.

\subsubsection{TDEs during binary-star exchange interactions}
\label{sec:exchange_tde}

TDEs are also allowed during binary exchange interactions, encompassing three channels distinguished by type: 
star-star $\to$ BH-star, 
BH-star $\to$ BH-BH via a binary-single scattering,
and BH-star $\to$ BH-BH via a four-body BH-star--BH-star interaction. 
In each case the disrupted-star mass is drawn from the time-evolving stellar mass function, taken to be a Kroupa IMF truncated at a maximum stellar mass set by the current cluster age (Sec.~\ref{sec:wd_formrate}), with a minimum stellar age of $\approx3\,\mathrm{Myr}$ imposed. Stellar radii at zero-age main sequence follow the fit $R_\star = 0.93\,m_\star^{0.88}$, in solar units~\cite{Ryu:2020gxf}. 
During each star-star $\to$ BH-star and BH-star $\to$ BH-BH exchange interaction, we compute the probability that a TDE occurs as the ratio of the TDE cross section to the exchange cross section, which scales approximately as the ratio of the tidal radius to the semimajor axis of the resulting binary. When a TDE occurs, the BH that disrupted the star has its mass and spin updated following Sec.~\ref{sec:spin_evolution}, and the bound state is subsequently ionized.

\subsubsection{Mass bias in the disrupted-star sampling}
\label{sec:mass_bias}

In every TDE channel described above, the mass of the star selected for disruption is not drawn directly from the time-evolving stellar mass function $\Phi(m_\star|t)$ (Sec.~\ref{sec:exchange_tde}), but from a mass-biased probability density
\begin{equation}
  p(m_\star|t) \propto \Phi(m_\star|t)\,m_\star^{m_b},
  \label{eq:mass-bias}
\end{equation}
where the power-law index $m_b$ weights the sampling toward more massive stars. This bias is designed to capture two effects that our spatially-averaged rate formalism (Secs.~\ref{sec:tde_twobody}--\ref{sec:wd_bh_tde}) does not otherwise resolve: more massive stars have larger radii ($R_\star\propto m_\star^{0.88}$, Sec.~\ref{sec:exchange_tde}) and are therefore intrinsically easier to disrupt, and two-body mass segregation concentrates the most massive stars toward the cluster core, where the local CO density -- and hence the TDE encounter rate -- is highest. Rather than modeling mass segregation explicitly, we absorb both effects into the tunable exponent $m_b$ applied to the disrupted-star mass function. We explore $m_b=0$, $1$, and $2$: $m_b=0$ recovers the unbiased mass function (``no mass bias''), while increasing $m_b$ progressively favors the disruption of heavier stars.

Since the mass -- and correspondingly the angular momentum -- accreted per TDE scales with the disrupted-star mass $m_\star$ through the global accretion fraction $f_A$ (Sec.~\ref{sec:tde_twobody}), larger values of $m_b$ systematically increase the mass and spin-up imparted to the population of accreting compact objects per disruption event. This mass bias acts purely on which star is selected once a TDE occurs; it does not modify the underlying two-body encounter rate of Sec.~\ref{sec:tde_twobody}, which depends on the stellar number density $n_\star$ rather than on the sampled stellar mass. Because the resulting mass and spin growth of the CO population can feed back on the cluster's subsequent dynamical evolution (e.g., through binary hardening efficiency and gravitational-wave capture rates), $m_b$ can nonetheless indirectly affect the overall normalization of the compact-binary merger rate, as discussed in Sec.~\ref{sec:Results}.

\subsection{Spin evolution of compact objects under disk accretion}
\label{sec:spin_evolution}

\subsubsection{Unified angular-momentum framework}
\label{sec:spin_unified}

When a compact object accretes matter from a geometrically thin, optically thick Keplerian disk, the angular momentum deposited at the innermost stable circular orbit (ISCO) drives a secular change in the spin of the accretor. The underlying physics differs qualitatively between NSs and BHs: a NS is an extended body characterized by an equation of state (EOS), a finite moment of inertia, and a quadrupole moment that departs significantly from the Kerr value, whereas a BH is fully described by its mass $M$ and dimensionless spin $\chi\equiv Jc/(GM^2)$, with no internal structure. We treat both cases within a single unified framework that evolves the angular momentum $J$ as the primary state variable, allowing a smooth transition from the NS to the BH regime during accretion; we assume $J$ is conserved during the collapse of a NS into a BH. Throughout, the sign $s=+1$ ($s=-1$) labels prograde (retrograde) equatorial accretion. Even when the incoming stellar orbit has a generic inclination relative to the spin of the compact accretor, we assume the inner disk aligns or anti-aligns with the spin via the Bardeen-Petterson effect on a timescale much shorter than the duration of the accretion episode, so that the accretion can be treated as strictly equatorial~\cite{Gerosa:2020xly}.

\subsubsection{Equations of state and maximum mass}
\label{sec:eos}

The internal structure of a NS is determined by its EOS, relating pressure to the energy density of dense nuclear matter. We adopt two representative tabulated EOSs: the APR EOS~\citep{Akmal:1998cf}, is soft and constructed from variational calculations of nucleon-nucleon interactions, and the AU
EOS~\cite{Wiringa:1984tg}, based on the Urbana nucleon-nucleon potential, and softer than APR. Both yield mass-radius relations $R(M)$ obtained by numerical interpolation of the tabulated sequences.

The maximum mass supportable by a non-rotating (Tolman-Oppenheimer-Volkoff, TOV) configuration, $M_{\rm TOV}$, sets the stability threshold for non-rotating NSs. Rotation stabilizes NSs against collapse up to a supramassive limit that, according to the numerical-relativity survey of Ref.~\citep{Breu:2016ufb}, is EOS-insensitive to $\sim2\%$ and given by\footnote{Differential rotation can further increase this limit to about $1.3 M_{\rm TOV}$ for NSs collapsing on the viscous timescale~\cite{Perna:2025}, but since this is very short, we use the lower limit for $M_{\rm max,rot}$.}
\begin{equation}
  M_{\rm max,rot} \approx 1.2\,M_{\rm TOV}.
  \label{eq:supramassive}
\end{equation}
A NS with $M_{\rm TOV}<M\le M_{\rm max,rot}$ is termed supramassive: it is stabilized solely by rotation, and any loss of angular momentum will eventually trigger collapse to a BH. In our model, once the accreted mass exceeds $M_{\rm max,rot}$, the compact object is treated as a BH for all subsequent evolution. We adopt $M_{\rm TOV}=2.20\,M_\odot$ ($M_{\rm TOV}=2.13\,M_\odot$) for the APR (AU) EOS as the fiducial NS-BH threshold throughout this work.

\subsubsection{Moment of inertia}
\label{sec:moi}

The NS moment of inertia $I$ enters the spin evolution through $J=I\Omega$, with $\Omega=2\pi f$ the angular velocity. We adopt the quasi-universal fit of Ref.~\citep{Lattimer:2004nj},
\begin{equation}
  \frac{I}{MR^2} = 0.237\left(1+0.674\,\mathcal{C}+4.48\,\mathcal{C}^4\right),
  \label{eq:LS_moi}
\end{equation}
where $\mathcal{C}\equiv GM/(Rc^2)$ is the stellar compactness; this expression reproduces $I$ to better than $10\%$ across realistic EOS models~\citep{Lattimer:2004nj}. This relation allows the conversion between the dimensionless spin $\chi$ and spin frequency $f$,
\begin{equation}
  \chi = \frac{cI\Omega}{GM^2} = \frac{2\pi c\,I\,f}{GM^2}.
  \label{eq:chi_from_f}
\end{equation}
The total derivative $dI/dM$ needed for the spin-evolution equation (Sec.~\ref{sec:ns_evolution}) is computed as
\begin{equation}
  \frac{dI}{dM} = \left(\frac{\partial I}{\partial M}\right)_R
  + \left(\frac{\partial I}{\partial R}\right)_M
  \left(\frac{dR}{dM}\right)_{\rm EOS},
\end{equation}
with $dR/dM$ evaluated numerically from the tabulated EOS mass-radius relation.

\subsubsection{ISCO radius}
\label{sec:isco}

\paragraph{Neutron stars.}
The NS ISCO differs fundamentally from that of a BH of the same mass and spin because the NS exterior spacetime departs from the Kerr metric: the stellar quadrupole moment $\bar Q\equiv -QM/J^2$ is substantially larger than the Kerr value $\bar Q_{\rm Kerr}=1$, typically $\bar Q\sim2$--$10$ for realistic NSs \citep{Laarakkers:1997hb,Pappas:2013naa}. This enlarged oblateness shifts the ISCO outward relative to the Kerr prediction. Through the quasi-universal $I$-Love-$Q$ relations \citep{Yagi:2013awa}, the ISCO location can be expressed as a function of mass and spin frequency alone, independent of the EOS to a few percent. We adopt the universal fit of Ref.~\citep{Luk:2018xmt},
\begin{equation}
  R_{\rm ISCO}^{\rm NS}\,f = a_1 x + a_2 x^2 + a_3 x^3 + a_4 x^4,
  \label{eq:luklin}
\end{equation}
where $x\equiv Mf$ with $a_1=8.809$, $a_2=-9.166\times10^{-3}$, $a_3=8.787\times10^{-8}$, $a_4=-6.019\times10^{-12}$ (mass in $M_\odot$, frequency in Hz, radius in km); this fit is accurate to $\sim2\%$ over the astrophysically relevant range of $Mf$. Because Ref.~\citep{Luk:2018xmt} derived Eq.~\eqref{eq:luklin} for prograde (corotating) orbits, we use the same expression as an approximation for retrograde orbits, noting that the true retrograde ISCO lies at larger radius and that no EOS-universal retrograde fit is available in the literature.

\paragraph{Black holes.}
For a Kerr BH the ISCO is an exact GR result \citep{1972ApJ...178..347B}. In units of the geometric mass $\hat M\equiv GM/c^2$, it is given by the standard Bardeen-Press-Teukolsky (BPT) expressions [Eqs.~(2.19)--(2.21) of Ref.~\citep{1972ApJ...178..347B}] and depends only on $\chi$: the prograde ISCO shrinks from $6\hat M$ at $\chi=0$ (Schwarzschild) to $\hat M$ as $\chi\to1$ (extremal Kerr), while the retrograde ISCO grows from $6\hat M$ to $9\hat M$. At the NS-to-BH transition the ISCO jumps discontinuously from the NS value [Eq.~\eqref{eq:luklin}] to the Kerr value; this discontinuity is physical, since the quadrupole moment drops from $\bar Q\gg1$ to $\bar Q=1$ on the ringdown timescale $\sim GM/c^3\sim$~milliseconds \citep{Buonanno:2007yg}, far shorter than any accretion timescale.

\subsubsection{Specific energy and angular momentum at the ISCO}
\label{sec:e_j_isco}

\paragraph{Neutron stars.}
We compute the specific orbital energy $\tilde E$ and specific angular momentum $\tilde l$ (per unit rest-mass energy $mc^2$ and per unit $mc$, respectively) at the NS ISCO following Ref.~\citep{2010A&A...520A..16B}, who approximate the exact metric functions with a Schwarzschild background plus a frame-dragging correction,
\begin{equation}
  v_{\rm appr} = \frac{(\Omega_{\rm Schw}\mp N^\phi)\,r}{N},
  \label{eq:bejger_v}
\end{equation}
where $\Omega_{\rm Schw}=\sqrt M/r^{3/2}$ is the Schwarzschild orbital angular velocity (geometric units $G=c=1$), $N^\phi=2GJ/(r^3c^3)$ the frame-dragging angular velocity, $N=\sqrt{1-2\hat M/r}$ the lapse function, and the upper (lower) sign applies to prograde (retrograde) orbits. The specific energy and angular momentum follow as
\begin{align}
  \tilde E &= \left(N+N^\phi v_{\rm appr}\,r\right)\gamma, \\
  \tilde l &= \pm\,v_{\rm appr}\,r\,\gamma,
  \label{eq:bejger_El}
\end{align}
with Lorentz factor $\gamma=(1-v_{\rm appr}^2)^{-1/2}$ and $\tilde l>0$ ($\tilde l<0$) for prograde (retrograde) orbits. This approximation reproduces exact numerical values to $\sim1\%$ accuracy for $M\gtrsim0.5\,M_\odot$ across the full range of spin up to the mass-shedding limit \citep{2010A&A...520A..16B}.

\paragraph{Black holes.}
For a Kerr BH, $\tilde E$ and $\tilde l$ at the ISCO are exact \citep{1972ApJ...178..347B}. Writing $\tilde r\equiv r_{\rm ISCO}/\hat M$ and $s=\pm1$ for prograde/retrograde orbits, we evaluate the numerically stable form
\begin{equation}
  \tilde E = \frac{1-2/\tilde r + s\chi/\tilde r^{3/2}}
  {\sqrt{1-3/\tilde r + 2s\chi/\tilde r^{3/2}}},
  \label{eq:E_kerr}
\end{equation}
which avoids the numerical cancellation near $\chi\to1$ present in the original factored expression. The radiative efficiency, $\eta=1-\tilde E(\chi)$, increases monotonically with prograde spin from $\eta\approx5.7\%$ at $\chi=0$ to $\eta\approx42\%$ as $\chi\to1$, and $\eta\approx30\%$ at $\chi=0.998$. The specific angular momentum $\tilde l$ follows Eq.~(2.13) of Ref.~\citep{1972ApJ...178..347B}, which we likewise evaluate in fully dimensionless form ($\tilde r$, $\chi$) to avoid a dimensional inconsistency present in the factored form for retrograde orbits.

\subsubsection{Mass-shedding and spin limits}
\label{sec:spin_limits}

\paragraph{Neutron stars: Kepler frequency.}
A NS cannot rotate faster than the Keplerian (mass-shedding) frequency, at which centrifugal support balances equatorial surface gravity. We adopt the empirical fit of Ref.~\citep{Lattimer2007},
\begin{equation}
  f_K \approx 1045\,{\rm Hz}\left(\frac{M}{M_\odot}\right)^{1/2}
  \left(\frac{10\,\mathrm{km}}{R}\right)^{3/2},
  \label{eq:fkepler}
\end{equation}
accurate to a few percent across a wide range of EOS models. The corresponding breakup spin $\chi_K(M,R)$ follows from inserting Eq.~\eqref{eq:fkepler} into Eq.~\eqref{eq:chi_from_f}, and is always $\chi_K\lesssim0.65$ for typical NS parameters, well below unity, reflecting the fact that realistic NSs are much less compact than extremal Kerr BHs.

\paragraph{Black holes: Thorne limit.} A BH accreting from a thin disk cannot reach $\chi=1$, because photons emitted by the inner disk are preferentially captured at high spin, exerting a counter-spinning torque. Ref.~\citep{1974ApJ...191..507T} showed that this limits the equilibrium spin to $\chi_{\rm max}\approx0.998$, the ``Thorne limit,'' which we impose as a hard cap on the BH spin throughout the evolution.

\subsubsection{Spin evolution equations}
\label{sec:spin_eq}

\paragraph{Neutron stars.}
\label{sec:ns_evolution}
For a NS accreting from a thin disk, angular momentum conservation gives
\begin{equation}
  \frac{d(I\Omega)}{dM} = \frac{\tilde l_{\rm ISCO}}{\tilde E_{\rm ISCO}},
  \label{eq:angular_momentum_balance}
\end{equation}
where $dM$ is the increment in gravitational mass and the right-hand side is the specific angular momentum flux delivered per unit accreted rest-mass energy \citep{1994ApJ...424..823C,2010A&A...520A..16B}. Expanding the left-hand side,
\begin{equation}
  \frac{df}{dM} = \frac1I\left[\frac{\tilde l_{\rm ISCO}}
  {2\pi\,\tilde E_{\rm ISCO}} - f\,\frac{dI}{dM}\right],
  \label{eq:dfdM}
\end{equation}
where $dI/dM$ (Sec.~\ref{sec:moi}) accounts for the change in moment of inertia as mass and radius evolve along the EOS, and $\tilde l_{\rm ISCO}$, $\tilde E_{\rm ISCO}$ are evaluated using the approximation of Ref.~\cite{2010A&A...520A..16B} (Sec.~\ref{sec:e_j_isco}) together with fit from Ref.~\cite{Luk:2018xmt} (Sec.~\ref{sec:isco}). Once the spin frequency reaches the Kepler limit $f_K$, further accretion conserves $f=f_K(M,R)$: mass continues to be deposited, but the excess angular momentum is expelled (e.g., via disk winds or GW emission). In the supramassive regime ($M_{\rm TOV}<M<M_{\rm max,rot}$) the stellar radius is frozen at its value at $M_{\rm TOV}$, a limiting approximation adopted in place of a full rotating-NS structure calculation.

\paragraph{Black holes.}
\label{sec:bh_evolution} 

For a BH accreting from a thin disk, the spin evolution follows Ref.~\citep{1970Natur.226...64B}. Since the BH is fully characterized by $M$ and $\chi$, with no moment of inertia, it is natural to express the evolution equation in terms of the geometric angular momentum $J=\chi GM^2/c$. Using $dM_{\rm grav}=\tilde E_{\rm ISCO}\,dM_{\rm acc}$ and $dJ=\tilde l_{\rm ISCO}\,dM_{\rm acc}$, with $dM_{\rm acc}$ the accreted rest mass, and eliminating $dM_{\rm acc}$, one obtains
\begin{equation}
  \frac{d\chi}{dM} = \frac1M\left(\frac{\tilde l_{\rm ISCO}}
  {\tilde E_{\rm ISCO}\,\hat M} - 2\chi\right),
  \label{eq:dchi_dM_BH}
\end{equation}
where $\hat M=GM/c^2$ and all quantities are evaluated at the Kerr ISCO (Sec.~\ref{sec:e_j_isco}). This equation is expressed directly in terms of the gravitational mass; no additional factor of $\tilde E_{\rm ISCO}$ should be applied to the mass step $dM$, since it is already encoded in the derivation \citep{1970Natur.226...64B}. Starting from a Schwarzschild BH ($\chi=0$), prograde accretion of $\Delta M\approx(\sqrt6-1)\,M_i$ spins the BH up to the Thorne limit $\chi=0.998$ \citep{1970Natur.226...64B,1974ApJ...191..507T}. Retrograde accretion is likewise permitted: a prograde-spinning BH is first decelerated toward $\chi=0$, after which the orbit direction naturally flips to prograde -- since there is no preferred retrograde state at zero spin -- and the BH is subsequently spun up in the prograde direction, consistent with the analytic solutions of Refs.~\citep{Volonteri:2005fj,Berti:2008af}.

\subsubsection{Numerical integration}
\label{sec:numerics}

\begin{figure}
  \centering
  \includegraphics[width=\linewidth]{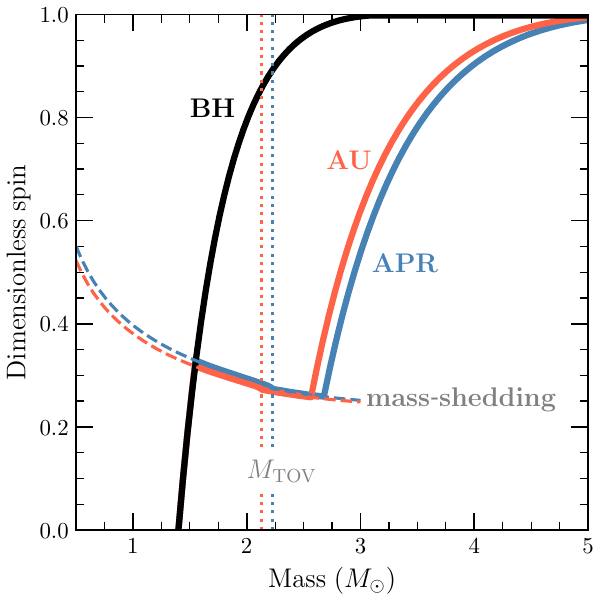}
  \caption{Evolution of the dimensionless spin magnitude of an initially $1.4\,M_\odot$, non-spinning compact object as it grows in mass through coherent accretion. The black track shows the evolution assuming the object is a BH throughout; for a NS, the spin saturates at the mass-shedding limit until the mass exceeds the TOV limit by a factor $\approx1.2$, after which the object collapses into a BH conserving angular momentum and the spin continues to grow along the Bardeen track until it reaches the Thorne limit $\simeq0.998$. Results depend on the adopted EOS; we show the APR (blue) and AU (orange) cases.} \label{fig:compact_object_accretion}
\end{figure}

Equations~\eqref{eq:dfdM} and~\eqref{eq:dchi_dM_BH} are integrated with a fourth-order Runge-Kutta (RK4) scheme using a fixed mass step $dM$. We use the geometric angular momentum $J=\chi(GM/c^2)^2$ (in units of km$^2$) as the single state variable carried through all phases of the evolution. This choice ensures continuity across the NS$\to$BH transition: at collapse, $J$ is inherited without discontinuity, and only the formulas used to evaluate $\tilde E$ and $\tilde l$ switch, from the approximation of Ref.~\cite{2010A&A...520A..16B} to the exact Kerr expressions. The NS and BH evolutions thus share a single integration loop, with a boolean flag selecting the appropriate physics at each mass step. For retrograde accretion, $\tilde l_{\rm ISCO}<0$ and $J$ decreases toward zero; when the angular momentum is exhausted ($J\to0$, $\chi\to0$) the orbit flips from retrograde to prograde, consistent with Ref.~\citep{1970Natur.226...64B}. At every step, the spin is capped at $\chi_K(M,R)$ for NSs (Sec.~\ref{sec:spin_limits}) and at $\chi_{\rm max}=0.998$ for BHs.
Figure~\ref{fig:compact_object_accretion} illustrates the resulting evolution of the dimensionless spin magnitude of an initially $1.4\,M_\odot$, non-spinning compact object as it grows in mass through coherent disk accretion. If treated as a BH throughout, the spin follows the Bardeen track of Sec.~\ref{sec:bh_evolution}; if treated as a NS, the spin instead saturates at the mass-shedding limit until the mass exceeds the TOV limit by a factor $\approx1.2$ [Eq.~\eqref{eq:supramassive}], at which point the star collapses to a BH conserving angular momentum, and the spin subsequently continues to grow along the Bardeen track until it reaches the Thorne limit of $\simeq0.998$. The details of the NS phase depend on the adopted EOS; we show results for the APR and AU EOS.

\subsection{Population-level rates, catalogs, and multi-messenger signals}
\label{sec:merger_rate}

\subsubsection{Astrophysical reweighting of the raw simulation ensemble}
\label{sec:reweighting}

The raw simulation ensemble of Sec.~\ref{sec:cluster_ic} spans $(M_{\rm cl,0},\,r_{h,0},\,Z,\,z_{\rm cl,form},\,R_{\rm gal,0})$ with a computationally convenient sampling density that is not itself the assumed astrophysical population. We recover physically meaningful rates and distributions by assigning each simulated cluster $i$, with initial conditions $\theta_i$, an importance weight
\begin{equation}
  w_i \propto \frac{\pi_{\rm astro}(\theta_i)}{\pi_{\rm sim}(\theta_i)}.
  \label{eq:weight-def}
\end{equation}
The simulation density $\pi_{\rm sim}(\theta)$ is uniform over the sampled domain of Sec.~\ref{sec:cluster_ic} and zero outside it, so it acts purely as a domain mask. The astrophysical density is
\begin{equation}
  \pi_{\rm astro}(\theta) \propto \psi_{\rm cl}(z_{\rm cl,form})\,
  M_{\rm cl,0}^{-2}\,r_{h,0}^{-1},
  \label{eq:astro-prior}
\end{equation}
combining the assumed cluster formation-rate history $\psi_{\rm cl}(z)$ (YMC or GC, Sec.~\ref{sec:Results}) with the $M_{\rm cl,0}^{-2}$ cluster initial-mass function of Sec.~\ref{sec:cluster_ic} and a $r_{h,0}^{-1}$ prior on the initial half-mass radius; metallicity and galactocentric radius are left unweighted beyond the bounds of the raw ensemble. Each weight is normalized so that $\sum_i w_i=1$ over the full cluster catalog, and every merger and $\mu$TDE event inherits the weight of its parent cluster. 

\subsubsection{Comoving and observer-frame rate densities}
\label{sec:rate_density}

We compute the comoving merger-rate density $R_m(z_m)\equiv dn_m/dt_m(z_m)$ by summing the contribution to the merger density at redshift $z_m$ from clusters formed at redshift $z_{\rm cl,form}$, integrated over the cluster formation history,
\begin{equation}
  R_m(z_m) = \int \frac{d^2n_m}{dt_m\,dn_{\rm cl}}(z_m,z_{\rm cl,form})\,
  dn_{\rm cl}(z_{\rm cl,form}).
\end{equation}
Changing variables to redshift throughout,
\begin{widetext}
\begin{equation}
  R_m(z_m) = \int \frac{d^2n_m}{dz_m\,dn_{\rm cl}}(z_m,z_{\rm cl,form})\,
  \frac{(1+z_m)H(z_m)}{(1+z_{\rm cl,form})H(z_{\rm cl,form})}\,
  \psi_{\rm cl}(z_{\rm cl,form})\,dz_{\rm cl,form},
  \label{eq:merger-rate-redshift}
\end{equation}
\end{widetext}
where $H(z)$ is the Hubble parameter (we adopt a flat $\Lambda$CDM cosmology with the Planck 2018 
parameters~\citep{Planck2020}), and the Jacobian factor $(1+z_m)H(z_m)/[(1+z_{\rm cl,form})H(z_{\rm cl,form})]$ follows directly from $dt/dz=-[(1+z)H(z)]^{-1}$. No explicit comoving-volume factor is required because clusters are simulated in their own comoving rest frames and the resulting mergers are binned in that same frame: the local simulation volume is not tied to the comoving volume at a particular comoving distance, so the simulated number density is already the physical comoving density.

We evaluate Eq.~\eqref{eq:merger-rate-redshift} as a weighted histogram over $N_z=50$ redshift bins of width $\Delta z$ spanning $z=0$--$10$: for every merger sample $j$ falling in bin $[z,z+\Delta z)$, formed in a cluster at $z_{\rm cl,form}$ with normalized event weight $w_j$ (Sec.~\ref{sec:reweighting}), we accumulate
\begin{widetext}
\begin{equation}
  R_m(z) = \mathcal{N}_{\rm cl}\!\!\sum_{j\,\in\,{\rm bin}(z)}\!\!
  \frac{(1+z_{m,j})H(z_{m,j})}
  {(1+z_{{\rm cl,form},j})H(z_{{\rm cl,form},j})\,\Delta z}\,w_j,
  \label{eq:rate-binned}
\end{equation}
\end{widetext}
where $\mathcal{N}_{\rm cl}=\sum_k\psi_{\rm cl}(z_{{\rm cl,form},k})\, \Delta z_{{\rm cl,form},k}$ is a Riemann-sum estimate of the total formation-rate density integrated over the formation-history grid, which converts the dimensionless weighted event counts into a properly normalized rate density. The micro-TDE rate density $R_{\mu\rm TDE}(z)$ follows from the identical procedure applied to the micro-TDE catalog in place of the merger catalog. The observer-frame cumulative rate~\cite{2018MNRAS.481.3278R},
\begin{equation}
  \mathcal{R}_o(<z) = \int_0^z \frac{dV_c}{dz'}\,
  \frac{\mathcal{R}(z')}{1+z'}\,dz',
  \label{eq:cumulative-rate}
\end{equation}
follows by multiplying $R_m(z)$ (or $R_{\mu\rm TDE}(z)$) by the differential comoving volume element and the $(1+z)^{-1}$ time-dilation factor and integrating -- cumulatively summing, in the binned implementation -- out to the redshift of interest.

\subsubsection{Monte Carlo universe realizations}
\label{sec:universe_realizations}

For each cluster-formation history (YMC or GC), we generate $N_{\rm real}=100$ independent Monte Carlo realizations of a single observed universe out to $z=10$. In each realization we (i) evaluate Eq.~\eqref{eq:cumulative-rate} at $z=10$ for both the merger and the micro-TDE channels to obtain the expected total number of observable events, $N_{\rm mer}$ and $N_{\rm TDE}$; (ii) draw that many events without replacement from the full simulated merger and micro-TDE tables, respectively, with each event selected with probability proportional to its normalized weight $w_j$ (Sec.~\ref{sec:reweighting}); and (iii) record the resulting catalog as one realization of the observable Universe. Where the expected count exceeds the number of nonzero-weight events available, we draw with replacement instead, since this can only occur when the astrophysical weighting has effectively excluded most of the raw ensemble for that population. We retain the full merger and micro-TDE catalogs for the first $N_{\rm keep}=10$ realizations for catalog-level analyses (Sec.~\ref{sec:Results}), while computing the mean and standard deviation of the event counts in each channel (BH-BH, NS-BH, NS-NS, BH-$\mu$TDE, NS-$\mu$TDE, and accretion-induced collapse, AIC) across all $100$ realizations. We adopt $M_{\rm TOV}=2.2\,M_\odot$ (APR EOS, Sec.~\ref{sec:eos}) throughout to classify merger and micro-TDE participants as NSs or BHs.

\subsection{Electromagnetic detectability of \texorpdfstring{$\mu$}{u}TDEs}
\label{sec:em_detectability_methods}

\subsubsection{Accretion luminosity}
\label{sec:disk_luminosity}

Following a TDE (Sec.~\ref{sec:tde_framework}) that deposits a bound mass $\Delta m=f_A\,m_\star$ (Sec.~\ref{sec:tde_twobody}) around a compact object,
the bound debris initially follows highly eccentric ballistic orbits with a broad distribution of specific orbital energies. As the most bound material returns to pericenter, relativistic apsidal precession and hydrodynamic interactions cause intersecting debris streams that dissipate orbital energy through shocks while largely conserving angular momentum. This process, known as debris circularization, gradually transforms the eccentric fallback stream into a rotationally supported accretion disk. For stellar-mass BHs, where relativistic precession is typically strong owing to the small pericenter distances, circularization can be efficient, although its timescale depends on the penetration factor, BH spin, and the thermodynamic properties of the debris. Once a disk is formed, accretion onto the BH powers the electromagnetic emission, with the subsequent evolution governed by the viscous timescale rather than the ballistic fallback time
$t_{\rm fb}\sim\pi r_p^3/\sqrt{2G(m_{\rm BH}+\Delta m)R_\star^3}$.
We thus model the subsequent accretion as viscous, on a timescale $t_{\rm visc} \sim t_{\rm fb}/\alpha$ with dimensionless viscosity parameter $\alpha=0.1$, giving a characteristic accretion luminosity
\begin{equation}
  L_{\rm disk} = \eta(\chi)\,\Delta m\,c^2/t_{\rm visc},
  \label{eq:Ldisk}
\end{equation}
where $\eta(\chi)=1-\tilde E(\chi)$ is the radiative efficiency at the ISCO of the accreting compact object (Sec.~\ref{sec:e_j_isco}), evaluated for the appropriate (prograde or retrograde) orbital orientation (Sec.~\ref{sec:spin_unified}). A fraction of this power can additionally emerge as a relativistic jet, $L_{\rm jet}=0.3\,\chi^2\,L_{\rm disk}$, scaling with the square of the accretor's spin as expected for a Blandford-Znajek-powered outflow  (see Ref.~\cite{1977MNRAS.179..433B} for the choice of 30\% efficiency in the jet luminosity).

{
\setlength{\tabcolsep}{14.8pt}
\begin{table*}
\centering
\caption{Model populations used in this study. All models share
a uniform natal spin magnitude distribution, one-dimensional natal velocity kick parameter ${\rm WK}=265\,{\rm km\,s^{-1}}$ for NSs, ${\rm KP}=0$ (natal kick model, 0 for fallback prescription), the NS mass function prescription from Ref.~\cite{Rocha:2023xwp}, an initial binary fraction $f_b=0.1$ for stars with mass $<10\,M_\odot$, maximum ZAMS star mass $m_{\rm max}=200\,M_\odot$, and the APR EOS.
The ``default'' model is highlighted in bold. The RP denotes the remnant-mass prescription and is 0 for the SEVN remnant mass prescription, and 1 for the SSE/updated-BSE prescription (default). Finally, the IMBH seed is taken as a fraction of $10^{-3}$ of the initial star cluster mass.}
\label{tab:populations}
\begin{tabular}{lccccccc}
\toprule\midrule
Label & $\chi_{1g}$ & RP & WT & MS & $f_A$ & $m_b$ & Description \\
\midrule
no-spins      & 0.0 & 1 & 1 & 1 & 0.30 & 2 & zero natal BH spin \\
RP0           & 0.2 & 0 & 1 & 1 & 0.30 & 2 & remnant masses from SEVN code \\
no $\mu$TDEs  & 0.2 & 1 & 0 & 1 & 0.30 & 2 & $\mu$TDE channel disabled \\
no IMBH       & 0.2 & 1 & 1 & 0 & 0.30 & 2 & no IMBH mass seed \\
3\% acc.      & 0.2 & 1 & 1 & 1 & 0.03 & 2 & reduced accretion fraction \\
no-mass-bias  & 0.2 & 1 & 1 & 1 & 0.30 & 0 & no mass bias \\
mb1           & 0.2 & 1 & 1 & 1 & 0.30 & 1 & intermediate mass bias \\
\textbf{default} & \textbf{0.2} & \textbf{1} & \textbf{1} & \textbf{1} & \textbf{0.30} & \textbf{2} & \textbf{fiducial model} \\
fA0.5         & 0.2 & 1 & 1 & 1 & 0.50 & 2 & higher accretion fraction \\
SP0.5         & 0.5 & 1 & 1 & 1 & 0.30 & 2 & higher natal spin \\
\bottomrule
\end{tabular}
\end{table*}
 }

\subsubsection{Flux and detectable fraction}
\label{sec:flux_detectability}

We estimate the observability of a micro-TDE's disk-accretion flare in the $V$~band (central wavelength $\lambda=550\,{\rm nm}$, bandwidth ratio $\Delta\lambda/\lambda=0.2$~\cite{2008gady.book.....B}) by assuming a fraction $f_{\rm band}$ of $L_{\rm disk}$ [Eq.~\eqref{eq:Ldisk}] is radiated into that band, so that the specific flux at Earth is
\begin{equation}
  f_\nu = \frac{f_{\rm band}\,L_{\rm disk}}{4\pi d_L^2\,\Delta\nu},
  \qquad \Delta\nu = \frac{c\,\Delta\lambda}{\lambda^2},
  \label{eq:flux-em}
\end{equation}
with $d_L(z)$ the luminosity distance computed assuming a standard cosmological model with Planck-18 cosmological parameters. The corresponding AB magnitude is~\cite{1974ApJS...27...21O}
\begin{equation}
  m_{\rm AB} = -2.5\log_{10}f_\nu - 48.60,
  \label{eq:AB-mag}
\end{equation}
with $f_\nu$ in ${\rm erg\,s^{-1}\,cm^{-2}\,Hz^{-1}}$. For each Monte Carlo realization (Sec.~\ref{sec:universe_realizations}) and a grid of $f_{\rm band}\in[10^{-7},10^{-1}]$, we evaluate Eq.~\eqref{eq:AB-mag} for every $\mu$TDE in the catalog and report the fraction with $m_{\rm AB}$ below a limiting magnitude ${\rm AB}\in\{22,23,24,25\}$, as representative values bracketing the expected sensitivity of the Vera C. Rubin Observatory, also known as the Large Synoptic Survey Telescope (LSST)~\cite{2019ApJ...873..111I}.

\subsection{Stochastic gravitational-wave background from \texorpdfstring{$\mu$}{u}TDEs}
\label{sec:gwb_methods}

Each micro-TDE emits a burst of gravitational radiation at the pericenter passage $r_p$, following the periastron-encounter formalism of Refs.~\cite{Berry:2010gt,Toscani:2021bzr,Toscani:2025uar}. The characteristic strain of a single event, as a function of observed GW frequency $f$, is
\begin{equation}
  h_c^2(f) = \frac45\,\frac{G^4}{c^8}\,\frac{(1+z)^4}{d_L^2}\,
  \frac{(m_{\rm BH}m_\star)^2}{r_p^2}\,
  \ell\!\left[\frac{(1+z)f}{f_c}\right],
  \label{eq:hc-tde}
\end{equation}
where $f_c=\sqrt{G(m_{\rm BH}+m_\star)/(4\pi^2 r_p^3)}$ is the Keplerian frequency at pericenter and $\ell(x)$ is the dimensionless burst spectral shape function, expressed in terms of modified Bessel functions of the second kind evaluated at $2\sqrt2\,x/3$ (Ref.~\cite{Berry:2010gt}). The stochastic background from the incoherent superposition of all $\mu$TDEs in a given realization (Sec.~\ref{sec:universe_realizations}), observed over a time $T$, is~\cite{Belgacem:2024ohp}
\begin{equation}
  \Omega_{\rm GW}(f) = \frac{4\pi^2}{3H_0^2\,T}\,f^3
  \sum_{\rm events} h_{c,{\rm event}}^2(f),
  \label{eq:omega-gw}
\end{equation}
summed over the $\mu$TDE events retained in that realization's catalog (Sec.~\ref{sec:universe_realizations}).

The YMC formation history follows the cosmic star-formation rate of Ref.~\cite{Madau:2016jbv}, with a fraction 10\% of the total star-formation rate channeled into YMCs, following Ref.~\cite{Mapelli:2021gyv}. The GC formation rate is taken from the digitized model in Fig.~A1 of Ref.~\cite{Rodriguez:2018rmd} based on~\cite{2019MNRAS.482.4528E}, which peaks at a redshift $z\approx3$--$4$.

\section{Results}
\label{sec:Results}

We present the results of our population synthesis study, based on rapid \href{https://github.com/Kkritos/Rapster}{\sc Rapster}~\citep{Kritos:2022ggc} simulations of $10^5$ star clusters per model population.  We explore ten model configurations listed in Table~\ref{tab:populations}, varying the natal BH spin magnitude ($\chi_{1g}$), the IMBH mass-seeding prescription (MS), the accretion fraction during $\mu$TDEs ($f_{A}$), the mass bias parameter ($m_b$), the retention prescription (RP), and the presence or absence of the $\mu$TDE channel (WT). 
For each population we compute statistics from $100$ independent universe realizations, retaining $10$ full event catalogs per formation history (YMC and GC) for catalog-level analysis.  All quoted means and standard deviations refer to the scatter across the 100 realizations unless stated otherwise.  We adopt $M_{\rm TOV}=2.20\,M_\odot$ as the NS-BH threshold throughout, motivated by the APR EOS.
 
\subsection{Merger and \texorpdfstring{$\mu$}{micro}TDE rate densities}
\label{sec:rates}

\subsubsection{Source-frame rate densities}
\label{sec:source_frame_rates}

\begin{figure*}
    \centering
    \includegraphics[width=\linewidth]{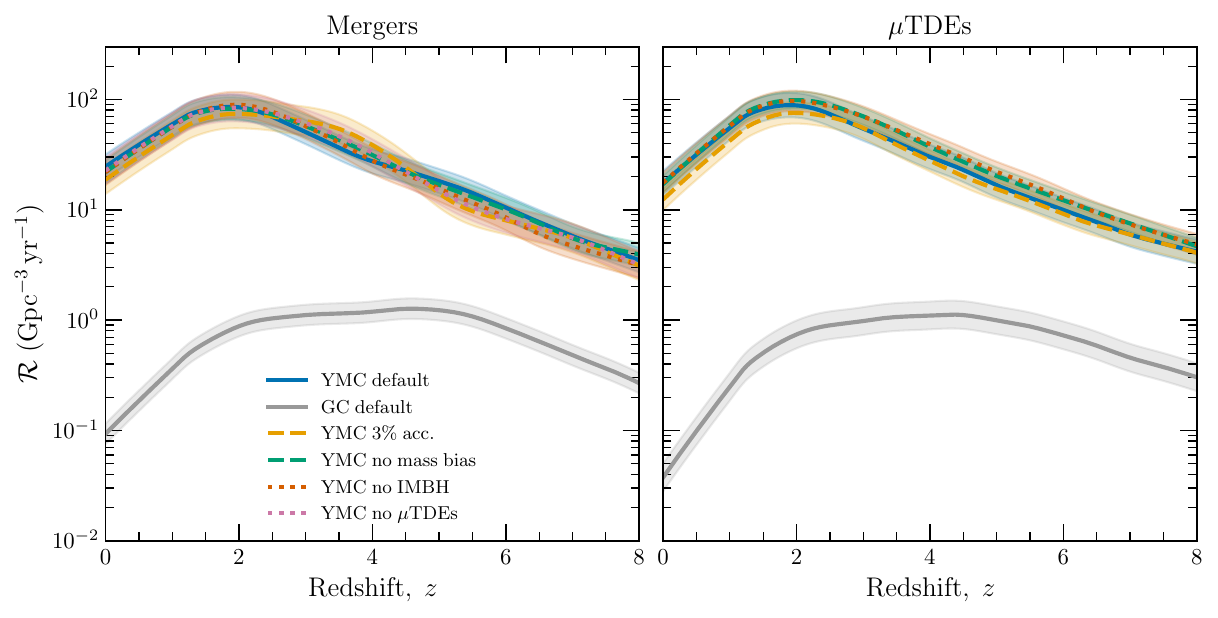}
    \caption{Volumetric source-frame rate density of mergers (left) and $\mu$TDEs (right) across redshift for various population models (described in Table~\ref{tab:populations}) and two star cluster formation histories (YMCs and GCs).}
    \label{fig:rate_density}
\end{figure*}

\begin{figure*}
    \centering
    \includegraphics[width=\linewidth]{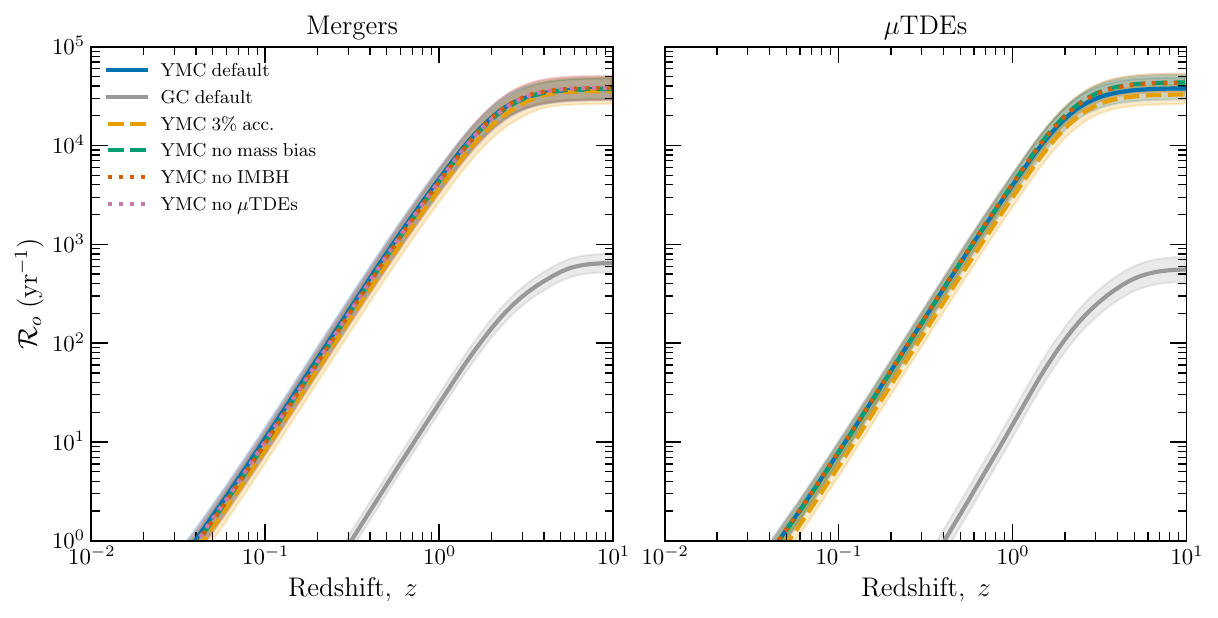}
    \caption{Cumulative observer-frame merger and $\mu$TDE rates out to redshift $z$ for the same model variations as in Fig.~\ref{fig:rate_density}.}
    \label{fig:cumulative_rates}
\end{figure*}

\begin{figure*}
    \centering
    \includegraphics[width=\linewidth]{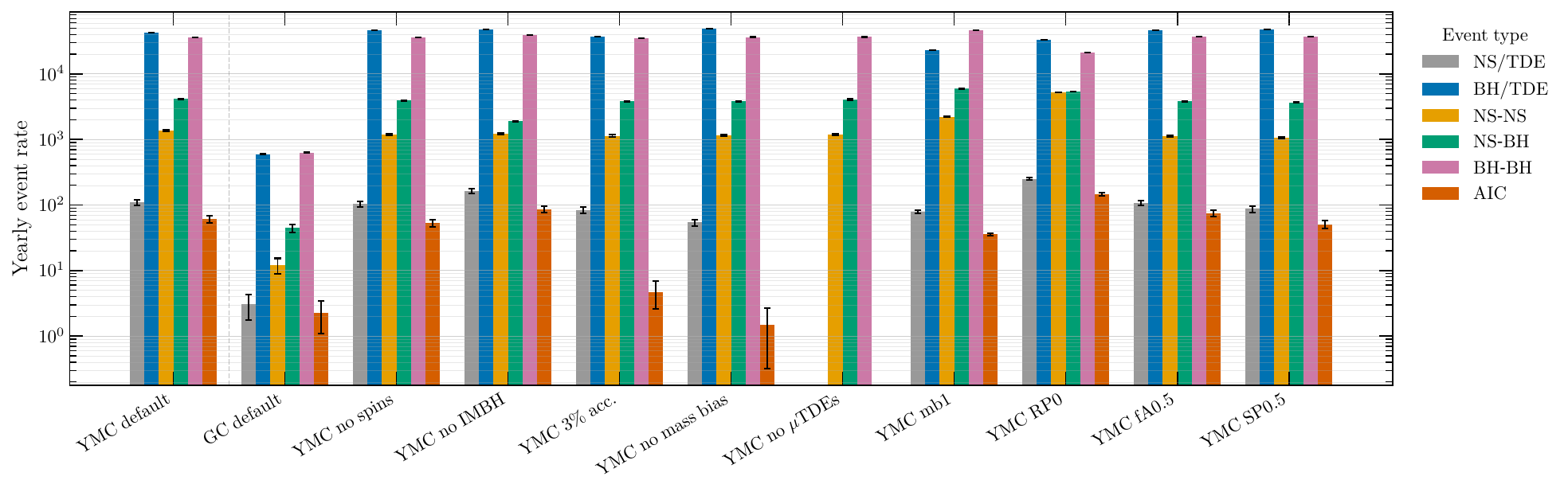}
    \caption{Yearly intrinsic event rates across the ten simulated populations, for transients of various types including mergers (NSNS, NSBH, BHBH), $\mu$TDEs (NS/TDE, BH/TDE), and accretion-induced collapse (AIC).}
    \label{fig:event_statistics}
\end{figure*}

We compute the comoving merger rate density ${\cal R}_{\rm merger}(z)$ and the $\mu$TDE rate density ${\cal R}_{\mu\rm TDE}(z)$ using the importance-sampling estimator described in Sec.~\ref{sec:rate_density}.  Both rate curves are deterministic given the cluster formation history and prior weights; they are computed from the full simulation set and are therefore not subject to realization-to-realization scatter.  We show their central estimates with LOWESS-smoothed uncertainty bands~\cite{Cleveland1979RobustLW} that reflect the dispersion of the rate estimator across the redshift bins.

For the default YMC population (see Fig.~\ref{fig:rate_density}), the BH-BH merger rate density peaks at $z\approx2$--$3$ with a peak value of several tens $\,{\rm Gpc}^{-3}\,{\rm yr}^{-1}$, broadly consistent with the upper end of the LVK inferred rate 
based on GWTC-5.0 (see~\cite{LIGOScientific:2026ctl}, Fig.~5).
The $\mu$TDE rate density follows a similar redshift profile and, at the peak, has the same value as the merger rate. However, at higher redshift the $\mu$TDE rate dominates over the merger rate, and overall, the cumulative number of $\mu$TDEs exceeds the CO binary mergers by a factor of a few, reflecting the larger rate for CO-star encounters relative to CO-CO binary formation and hardening.

The GC channel produces rate densities approximately 10--100 times lower than the YMC rates at any given redshift, peaking $\sim1$ unit of redshift later (that is at $z\approx3$--$4$) due to the later-peaking and narrower GC formation rate~\cite{2019MNRAS.482.4528E}.  Because GC-born BH-BH mergers occur at a higher typical redshift, their contribution to the LIGO-detectable (low-$z$) rate is further suppressed relative to the intrinsic rate density. The ratio ${\cal R}_{\mu\rm TDE}/{\cal R}_{\rm merger}\approx2$--$5$ is approximately preserved between the YMC and GC channels, indicating that the relative efficiency of the two channels within a cluster does not depend strongly on the cluster formation history.

Among the model variations, removing the $\mu$TDE channel (no-$\mu$TDEs) does not appreciably alter ${\cal R}_{\rm merger}$ relative to the default. This confirms that $\mu$TDE-driven mass accretion does not significantly perturb the cluster dynamics or the total merger budget. Removing the IMBH seed (no IMBH) reduces both ${\cal R}_{\rm BHBH}$ and ${\cal R}_{\mu\rm TDE}$ by a factor of roughly 2--3 across all redshifts, because the IMBH serves as a BH retention mechanism that keeps merger products in the cluster and facilitates subsequent interactions.  The 3\% accretion model ($f_{A}=0.03$) has both ${\cal R}_{\rm merger}$ and ${\cal R}_{\mu\rm TDE}$ consistent with the default to within $10\%$, since $f_A$ does not enter the dynamical encounter rate, which is set entirely by the cluster density, velocity dispersion, tidal radius, and CO mass. The effect of reducing $f_A$ is instead manifest in the CO spin and $\mu$TDE luminosity distributions (Secs.~\ref{sec:spin} and~\ref{sec:EM}).  The mass-bias parameter $m_b$ introduces a factor of $\sim2$--$3$ uncertainty in the overall rate normalization without changing the redshift dependence; models with $m_b=0$, 1, and 2 span this range monotonically. Models varying only the natal BH spin (no-spins, SP0.5) produce rates within the scatter of the default, and now shown in the figure to avoid cluttering, confirming that these parameters affect the spin distribution (Sec.~\ref{sec:spin}) but not the overall interaction rates.

\subsubsection{Observer-frame cumulative rates}
\label{sec:cumulative_rates}

The cumulative observer-frame event rate up to redshift $z$,
is shown in Fig.~\ref{fig:cumulative_rates} on a logarithmic redshift axis to resolve the low-$z$ regime and identify where the nearest GW- and EM-detectable events reside.
Notice that the rate computed with Eq.~\eqref{eq:cumulative-rate} is the intrinsic one; we discuss detectability of these signals by current and future observatories in Sec.~\ref{sec:mergers-catalog}.

For the default YMC population, $\mathcal{R}_o^{\rm BHBH}(<z=1)\sim10^3$--$10^4\,{\rm yr}^{-1}$, saturating at $\mathcal{R}_o^{\rm BHBH}(<z=10)\sim{\rm few}\times10^4\,{\rm yr}^{-1}$. The $\mu$TDE cumulative rate consistently exceeds the merger rate by a factor
of a few at all redshifts. The GC default rates are roughly 10--30 times smaller. Thus, in the redshift range $z<1$ the GC formation history channel predicts just a few tens of mergers and a few tens of $\mu$TDEs with a saturation value within $z=10$ that is not above $\sim10^3$ events yr$^{-1}$.
In a particular realization, the closest transients (a merger or a $\mu$TDE) considering the YMC (GC) formation history could be at a redshift of $\approx0.04$--$0.05$ ($\approx0.3$--$0.4$), corresponding to a luminosity distance of $\approx180$--$230\,\rm Mpc$ ($\approx1.6$--$2.2\,\rm Gpc$).

\subsubsection{Yearly event counts}
\label{sec:yearly_counts}

Figure~\ref{fig:event_statistics} shows the mean and standard deviation of the yearly event counts for all channel transients across all ten YMC populations and the GC default model, estimated from 100 universe realizations each. The event count rates are grouped into a bar chart.

The BH-BH and BH-$\mu$TDE rates are the dominant channels in all populations. The dominance of BH-transients reflects the effects of mass segregation and the higher BH abundance in the cluster core compared to the more extended NS population. Moreover, by the time the BH subpopulation has largely evaporated and NSs start to condense toward the center, the cluster has significantly expanded, lowering core densities and limiting the rate of NS transients.

For the default YMC model, the realization-to-realization scatter in the dominant channels is $\lesssim10\%$, indicating that the 100-realization average is well converged.
In addition, the NS-NS and NS-BH rates are suppressed by two to three orders of magnitude relative to the BH channels, because most NSs receive large natal kicks (we have assumed the Hobbs distribution with one-dimensional Maxwellian parameter $\sigma_{\rm kick}=265\,{\rm km\,s^{-1}}$~\cite{Hobbs:2005yx}) and are ejected from the cluster. We find $N_{\rm NSNS}/N_{\rm BHBH}\approx{\rm few}\times10^{-1}$ across populations. The AIC (accretion-induced collapse) count ($N_{\rm AIC}\sim10$--$100\,{\rm yr}^{-1}$) reflects the small fraction of NS-$\mu$TDE events in which the NS exceeds the supramassive limit ($\sim1.2M_{\rm TOV}$) and collapses to a BH.

Removing the IMBH seed (MS0) reduces $N_{\rm BHBH}$ by $\approx30$--$50\%$ and $N_{\rm BH\mu TDE}$ by a similar factor. The WT0 (no-$\mu$TDE) population produces a BH-BH merger count consistent with the default to within $\lesssim10\%$, confirming that $\mu$TDE accretion does not feedback significantly on the merger rate.
NS/TDE transients are particularly rare in the GC default scenario, with less than one event per year on average within redshift $z=10$. 
The RP0 model, which adopts the SEVN remnant mass prescription~\citep{Spera:2015vkd} instead of the SSE/updated-BSE prescription used in RP1 (default), also produces event counts consistent with the default at the $\lesssim10\%$ level, indicating that the choice of remnant mass prescription does not significantly affect the dynamical interaction rates.

\subsubsection{Decomposing the BH-BH rate by mass ratio}
\label{sec:imbh_sink}

\begin{figure*}
    \centering
    \includegraphics[width=\linewidth]{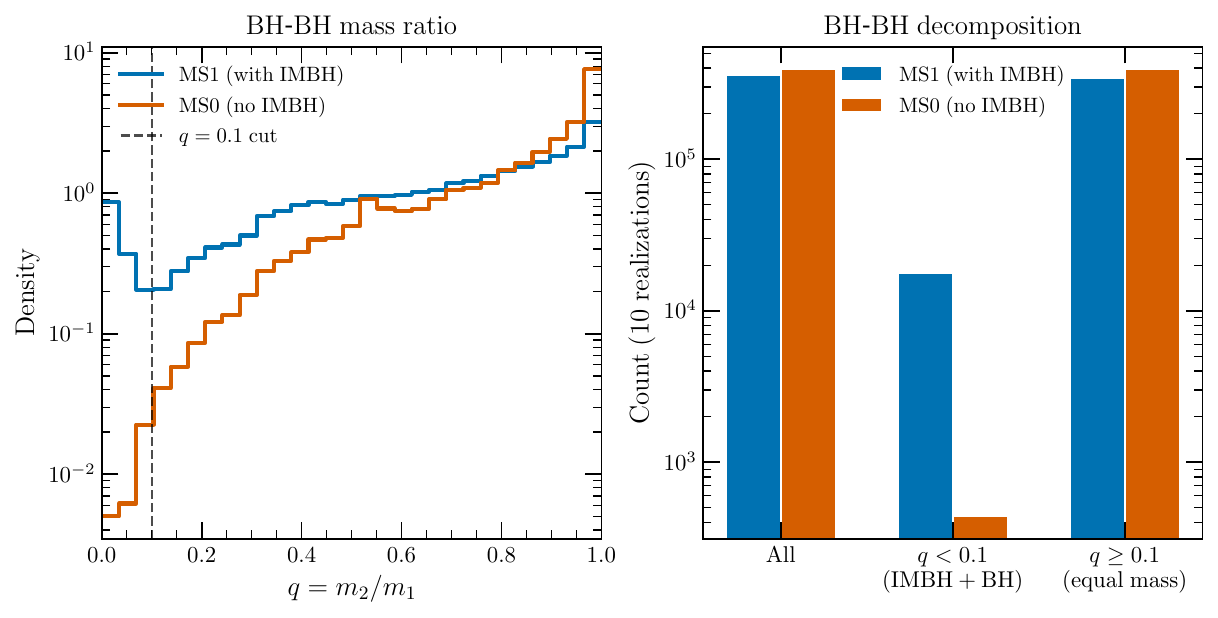}
    \caption{Left: Mass-ratio distribution for mergers under the assumption of no IMBH seed (orange, MS0) versus an IMBH seed with mass $10^{-3}$ of the initial cluster mass placed in the system (blue, MS1). Right: Decomposition of the number of mergers in IMBH+BH ($q<0.1$) and equal-mass ($q\ge0.1$) the two scenarios (MS0 and MS1).}
    \label{fig:IMBH_decomposition}
\end{figure*}

The total $N_{\rm BHBH}$ in the MS1 model includes a contribution from IMBH+BH mergers, which are highly asymmetric ($q=m_2/m_1\ll1$, with $m_2\leq m_1$) and qualitatively distinct from equal-mass BH-BH coalescences in terms of both GW waveform morphology and spin phenomenology. Such IMBH+BH events are also known as intermediate-mass ratio inspirals~\cite{Mandel:2007hi}. To assess the dual role of the IMBH as (i) a BH sink, consuming BHs one by one, depleting the equal-mass merger pool, and (ii) a BH retainer, keeping merger products in the cluster through its deep gravitational potential, we decompose the BH-BH merger catalog by mass ratio at the arbitrary threshold $q=0.1$ (see Fig.~\ref{fig:IMBH_decomposition}).

We find that in the MS1 (with IMBH) model, $3.7\%$ of all BH-BH mergers have $q<0.1$, with a total of $\approx13{,}800$ such events across the 10 retained realizations. By contrast, the MS0 (no IMBH) model has only $0.1\%$ of mergers at $q<0.1$ ($\approx320$ events), confirming that the low-$q$ tail in MS1 is almost entirely attributable to IMBH+BH encounters where the IMBH 
did not necessarily form via repeated mergers.

Disregarding for the moment the IMBH+BH mergers ($q<0.1$), the equal-mass merger count in MS1 is $\approx 362{,}000$ compared to $\approx 381{,}000$ in MS0, giving a ratio of $0.95$.  The equal-mass BH-BH merger rate is therefore $\approx 5\%$ lower in MS1 than in MS0, confirming that the IMBH acts as a net sink for the equal-mass merger population. This is a physically important result: while the presence of an IMBH seed is often assumed to increase hierarchical merger rates, it also removes BHs from the equal-mass merger channel, leading to a mild overall suppression of the equal-mass BH-BH merger rate when the two effects are accounted for separately.

The mass ratio distribution in the left panel of Fig.~\ref{fig:IMBH_decomposition} clearly shows the difference between MS1 and MS0: the MS1 distribution has a distinct excess at $q\lesssim0.1$ absent in MS0, while at $q\gtrsim0.2$ both models converge to the same shape, demonstrating that the IMBH affects only the low-$q$ regime without significantly altering the overall dynamical interaction rates for equal-mass BH pairs.  The total BH-BH merger count (including IMBH+BH mergers) is $\approx376{,}000$ (MS1) vs $\approx381{,}000$ (MS0), a difference of only $1\%$, suggesting that the two effects of the IMBH as a sink and as a retainer nearly cancel when focusing on total numbers.

\subsection{Electromagnetic detectability of \texorpdfstring{$\mu$}{micro}TDEs}
\label{sec:EM}

\subsubsection{Peak central engine luminosity distribution}
\label{sec:peak_luminosity}

\begin{figure}
    \centering
    \includegraphics[width=\linewidth]{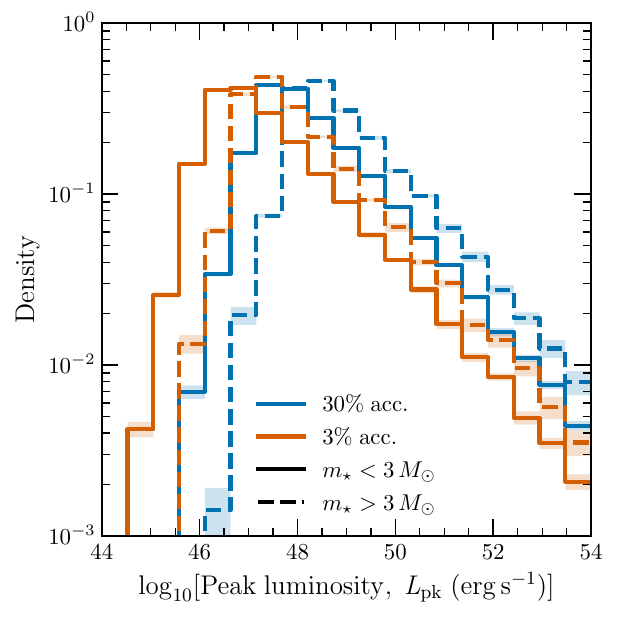}
    \caption{Distribution of peak luminosities of $\mu$TDEs for the default model which assumes accretion of $f_{A}=30\%$ (blue) and the model with $f_{A}=3\%$ (orange). The results are split into stars with mass $<3\,M_\odot$ (low-mass, solid) and $>3\,M_\odot$ (high mass, dashed). The bands correspond to simulation noise over multiple realizations.}
    \label{fig:peak_luminosity}
\end{figure}

Figure~\ref{fig:peak_luminosity} shows the distribution of peak accretion luminosities $L_{\rm pk}$ for the 30\% and 3\% accretion models, split by stellar mass ($m_\star<3\,M_\odot$ and $m_\star>3\,M_\odot$).  For low-mass stars, the peak of the  luminosity distribution lies at $\log_{10}(L_{\rm pk}/{\rm erg\,s^{-1}})\approx47$--$48$ in the 30\% accretion model, shifting down by $\approx1.5\,{\rm dex}$ for the 3\% model. For high-mass stars, the distribution peaks at $\approx49$--$51$ in the 30\% model.  Peak luminosities in the range $10^{47}$--$10^{50}\,{\rm erg\,s^{-1}}$ are generally larger than the luminosities of known FBOTs~\cite{Ho:2021fyb}, supporting the hypothesis that some FBOTs may arise from BH-star $\mu$TDEs in dense stellar environments, where a fraction of the available energy is reprocessed in a wind \cite{Kremer:2023sof}.
A similar class of transients whose energetics fall within the high-tail of our predicted luminosity distribution is that of the ultra-long GRBs, which have been proposed to result from the $\mu$TDE of a main-sequence star by a rapidly spinning BH~\cite{Perets:2016pwr,Neights:2025keq}.

\begin{figure}
    \centering
    \includegraphics[width=\linewidth]{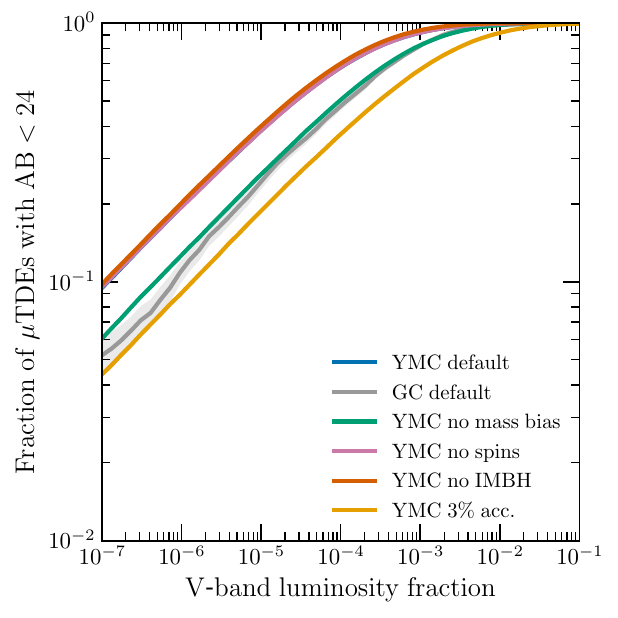}
    \caption{Fraction of $\mu$TDEs with AB magnitude less than 24, as a function of the fraction of the bolometric luminosity in the V-band (central wavelength at $550\,\rm nm$, bandwidth $\Delta \lambda/\lambda=0.2$~\cite{2008gady.book.....B}) across various models.}
    \label{fig:detectability}
\end{figure}

\subsubsection{V-band detectability fraction}
\label{sec:vband}

We compute the fraction $f({\rm AB}<m_{\rm lim})$ of $\mu$TDEs with apparent V-band magnitude brighter than the limiting magnitude $m_{\rm lim}$, as a function of $f_{V}$, which is the assumed fraction of the bolometric luminosity that corresponds to radiation in the V-band (Fig.~\ref{fig:detectability}). For the default YMC model and a limiting magnitude ${\rm AB}=24$ (appropriate for Rubin/LSST in a single pointing), the detectable fraction ranges from $\sim10^{-3}$ at $f_{\rm V}\sim10^{-6}$ to $\sim0.1$ at $f_{\rm V}\sim0.1$.  For ${\rm AB}=25$, the fraction increases by a factor of $\sim2$--$3$ at fixed $f_{\rm V}$. The detectability is systematically highest for the default 30\% accretion model and suppressed by $\sim1$--$1.5\,{\rm dex}$ for the 3\% accretion model, reflecting the direct scaling $L_{\rm pk}\propto f_A$.
The large number of $\mu$TDE transients at redshift $z\sim2$ (cf.~Fig.~\ref{fig:rate_density}) together with their large intrinsic luminosities ($L_{\rm pk}\sim10^{48}\,\rm erg\,s^{-1}$) implies that the majority of such transients should be detectable with Rubin/LSST solely based on the peak flux estimate.
However, we note that the detectability of a $\mu$TDE with LSST
depends not only on whether the peak flux exceeds the detection threshold in a single observation, but also on whether the transient can be characterized with sufficient temporal coverage to distinguish it from other classes of astrophysical transients.
Consequently, our estimates should be interpreted as upper limits on the number of $\mu$TDEs that are detectable based solely on their peak flux, while the number of events that can be confidently identified is expected to be lower by a factor determined by the cadence-dependent recovery efficiency. 

\subsection{Stochastic gravitational-wave background from \texorpdfstring{$\mu$}{u}TDEs}
\label{sec:GWB}

During the closest approach of the star to the BH, GW radiation is emitted with a characteristic strain that depends on the inverse of the pericenter distance [cf.~Eq.~\eqref{eq:hc-tde}]. The peak GW frequency is calculated via Kepler's third law and depends only on the properties of the star, with a value of $f_{\rm GW}\approx4\times10^{-5}\,\beta^{3/2}\,{\rm Hz}$ for a solar star, where $\beta\ge1$ is the penetration parameter defined as the ratio of the tidal radius to the pericenter distance. Therefore, the more a star penetrates, the higher the emitted GW frequency and the stronger the GW strain.

We use the formalism of \citet{Berry:2010gt} for the computation of the GW energy spectrum from a single $\mu$TDE event. This formalism assumes a parabolic orbit and is valid for a point particle; thus, we expect it to break down for events that penetrate beyond a critical value. We make the physical assumption that the pericenter of the orbit be larger than the stellar radius.

\begin{figure*}
    \centering
    \includegraphics[width=0.49\linewidth]{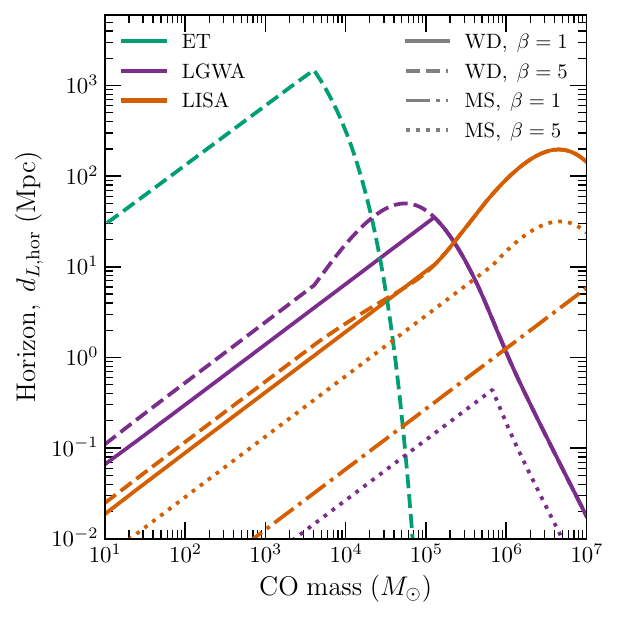}
    \includegraphics[width=0.49\linewidth]{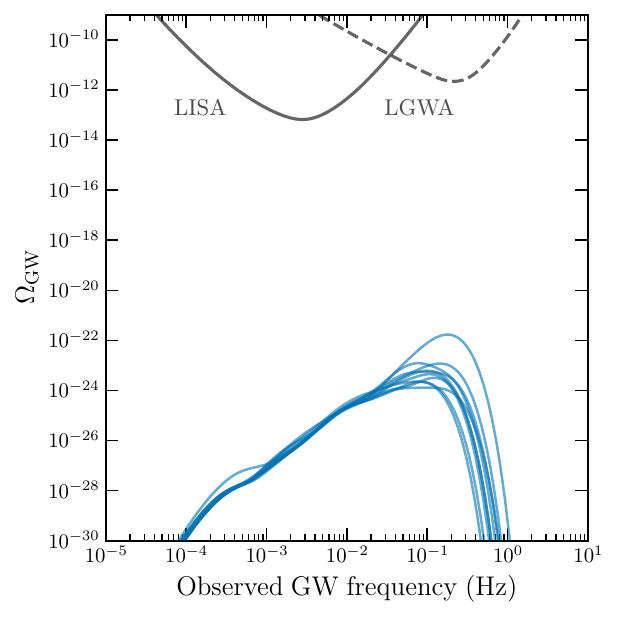}
    \caption{Left: Detector horizon, in terms of luminosity distance, as a function of CO mass for a $\mu$TDE for ET, LGWA, and LISA (see Sec.~\ref{sec:mergers-catalog}). The SNR detectability threshold is set to 8. The WD mass is fixed at $0.6\,M_\odot$ and radius of $5000\,\rm km$. The MS mass is taken $0.1\,M_\odot$ and $0.12\,R_\odot$. Right: Energy density parameter of GWs produced from the default YMC catalog of $\mu$TDEs as a function of the observer-frame GW frequency. The different lines show the resulting GW background across 10 realizations.}
    \label{fig:GWmicroTDEs}
\end{figure*}

The nearest $\mu$TDE involving a star occurs at a redshift $z\approx0.04$--$0.05$ (see Fig.~\ref{fig:cumulative_rates}) or at a luminosity distance $180$--$230\,\rm Mpc$. Based on the left panel of Fig.~\ref{fig:GWmicroTDEs}, 
the GWs emitted even from the nearest predicted $\mu$TDE are too weak to be directly detectable.
Therefore, the energy density of GWs emitted during each $\mu$TDE contributes to the stochastic GW background. The right panel of Fig.~\ref{fig:GWmicroTDEs} shows the GW energy density spectrum $\Omega_{\rm GW}$ as a function of detectable GW frequency and for ten realizations assuming the default YMC model. The spectral amplitude lies multiple orders of magnitude below the power-law sensitivity curves of either LISA or LGWA, making it undetectable. This is consistent with the findings of~\citet{Toscani:2025uar}, despite differences in the population properties of the events.

\subsection{Detectability of mergers}
\label{sec:mergers-catalog}

We now assess the detectability of the simulated catalogs by different detectors. Given the mass spectrum of the sources, we consider the following observatories and networks:
\begin{description}[align=left]
    \item[LIGO--Virgo at design sensitivity] This network comprises the two LIGO interferometers in the US (LIGO-Hanford and LIGO-Livingston)~\cite{LIGOScientific:2014pky} and the Virgo interferometer in Italy~\cite{Virgo:2014yos}. We use noise curves corresponding to the target sensitivity of the detectors~\cite{KAGRA:2013rdx}, and a low-frequency cutoff of 10\,Hz. The sensitivity curves used can be found in Refs.~\cite{noise_curve_Aplusdes}.
    \item[ET] The planned European 3G detector ET~\cite{Punturo:2010zz,Hild:2010id,ET:2025xjr}. We consider two configurations for the instrument, corresponding to the two designs under active consideration: a single-site triangular geometry with 10\,km arms and a network of 2 L-shaped detectors with 15\,km long arms and a relative orientation of $42.5^\circ$ with respect to each other; see  Ref.~\cite{Branchesi:2023mws}. In both cases, we adopt a low frequency cutoff of 3\,Hz and employ the sensitivity curves from Ref.~\cite{Branchesi:2023mws}.
    \item[ET+CE] The same ET configuration as above in combination with the planned US-based CE detector~\cite{Reitze:2019iox,Evans:2021gyd,Evans:2023euw}. Following the recommendation in Ref.~\cite{Evans:2023euw}, for CE we consider a single-site observatory with 40\,km, and we adopt the latest sensitivity curves publicly available in the \href{https://github.com/koustavchandra/gwforge}{\sc gwforge} package~\cite{Chandra:2024dhf} in the most optimistic scenario, which corresponds to a high laser power of $1.5\,{\rm MW}$ and improved coatings~\cite{noise_curves_gwforge}. For this detector, the low frequency cutoff is set to 5\,Hz.
    \item[LGWA] The proposed lunar mission LGWA, which leverages seismometers installed in a lunar crater to detect the response of the Moon to passing GWs~\cite{LGWA:2020mma,Ajith:2024mie}. This detector will be sensitive to GWs with frequencies between 1\,mHz and a few Hz, with the bucket of the sensitivity curve being at $\sim\!0.1$\,Hz. Following Ref.~\cite{Ajith:2024mie}, we consider a single-site observatory with 4 stations at the lunar south pole, and the more optimistic ``Silicon'' sensitivity curve. The assumed mission life time for this experiment is 10 years.
    \item[LISA] The planned space-based mission LISA~\cite{LISA:2017pwj,LISA:2024hlh}, which leverages a constellation of three spacecraft separated by $\sim\!2.5\times10^6\,{\rm km}$ to detect GWs in the frequency range of $10^{-5}\,{\rm Hz}$ to  a few dHz, with the bucket of the sensitivity curve being at $\sim\!4\,$mHz. For the noise curve, we use the same analytical prescription provided in Ref.~\cite{Babak:2021mhe}, further including the contribution of the Galactic White Dwarf confusion noise. For this detector, we assume a mission lifetime of 4 years~\cite{Seoane:2021kkk}.
\end{description}
To compute the SNR in each detector we use the publicly available \href{https://github.com/CosmoStatGW/gwfast}{\sc gwfast} package~\cite{Iacovelli:2022mbg,Iacovelli:2022bbs}, which implements $(i)$ time-evolving pattern functions for ground-based detectors, relevant for ET and CE; $(ii)$ a detailed treatment of the orbital modulations of the Moon motion for LGWA~\cite{Iacovelli:2025kwn,Tissino:2026skt}; $(iii)$ the so-called $AET$ basis~\cite{Prince:2002hp}, with the same expressions detailed in Ref.~\cite{Marsat:2020rtl} for LISA. The waveform is computed using the inspiral-only \href{https://github.com/gmorras/pyEFPE}{\textsc{pyEFPE}} approximant~\cite{Morras:2025nlp}, which models the fundamental emission mode of precessing binaries on generic orbits, and whose computational efficiency allows to compute the waveform for the long inspirals simulated here in a reasonable amount of time. For each detector, we assume a 100\% duty cycle. Considering these different detector concepts allows us to assess the detectability of the simulated objects across an extremely large frequency range, extending from $10^{-5}$\,Hz to a few kHz.

Before computing the SNR, we verify which binaries fall in the frequency band of each detector at formation. For binaries formed with semi-major axes corresponding to peak frequencies \emph{lower} than the assumed minimum frequency of a detector, we evolve the eccentricity up to the frequency corresponding to a time-to-merger equal to the mission life time for LGWA and LISA, and equal to the minimum frequency for ground-based detectors. The peak frequency is estimated using the fit from Ref.~\cite{Hamers:2021eir}. The evolution is performed by numerically finding the solution of $f(e)=f^\star$, where $f^\star$ is the desired peak frequency, and $f(e)$ is obtained through the 2\,PN-accurate expression in Eq.~(2) of Ref.~\cite{Fumagalli:2025asw}. The conversion between orbital frequency and semi-major axis is handled through Peters's equation~\cite{1964PhRv..136.1224P}. To improve computational efficiency, the initial guess for the root finding scheme is obtained from the low-eccentricity  asymptotic Peters relation in Eq.~(5.11) of Ref.~\cite{1964PhRv..136.1224P}. For binaries formed with semi-major axes corresponding to peak frequencies \emph{inside} the frequency band of a detector, we only perform the evolution if the time-to-merger for the binary at that frequency is longer than the mission lifetime for LGWA and LISA, and we do not perform any evolution otherwise. No evolution is performed in this case for ground-based instruments. For binaries formed with semi-major axes corresponding to peak frequencies \emph{higher} than the assumed maximum frequency of a detector, we simply deem the binary undetectable at that detector.

\begin{figure*}
    \centering
    \includegraphics[width=0.49\linewidth]{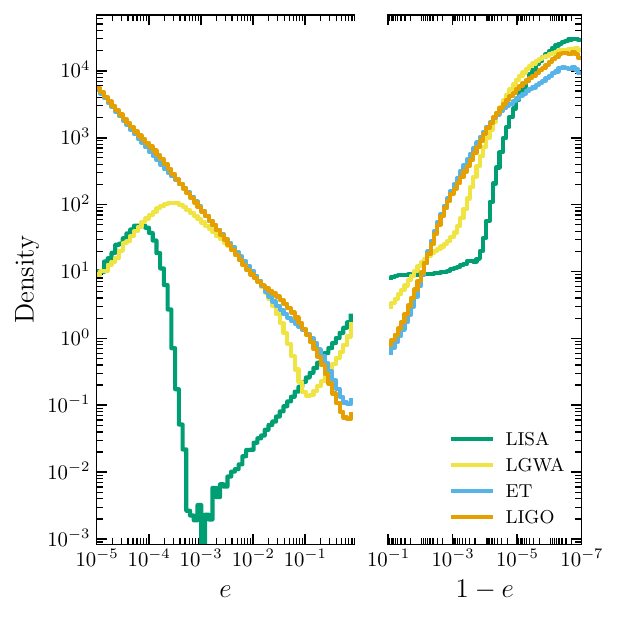}
    \includegraphics[width=0.49\linewidth]{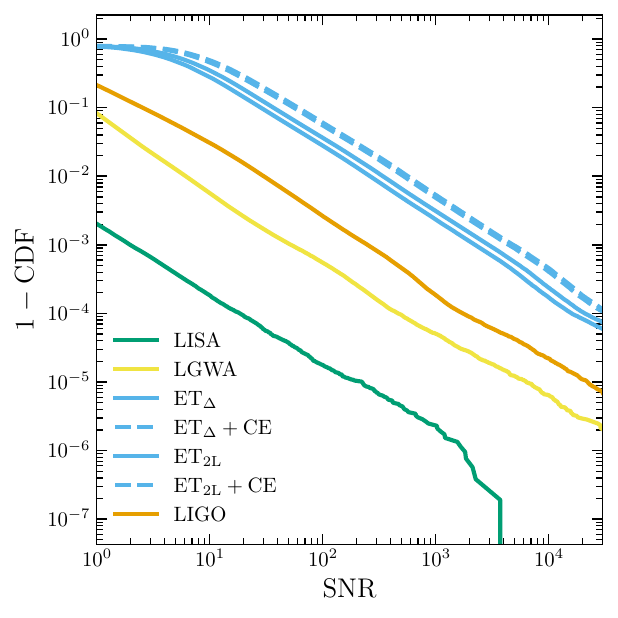}
    \caption{Left: Eccentricity ($e$) density distribution of each compact binary coalescence as it enters each GW detector (LISA, LGWA, ET, and LIGO). The histograms are from stacked data over 10 realizations of the default YMC population model. To resolve the high-eccentricity regime, the eccentricity axis is broken at $e=0.9$, and the range $0.9<e<1$ is plotted as $1-e$ for resolvability. Right: The corresponding complementary cumulative density function ($1-\rm CDF$) of the signal-to-noise ratio (SNR) for each merger is plotted assuming different GW detector networks.}
    \label{fig:mergers}
\end{figure*}

The left panel of Fig.~\ref{fig:mergers} shows the eccentricity density distribution of merging CO binaries across various GW frequency bands obtained through our evolution procedure, and restricting to binaries whose signal has a portion falling inside the band of each instrument, irrespective of its detectability.

As expected, binaries are more eccentric in the LISA band and tend to circularize as they evolve toward the higher-frequency bands of LIGO and ET due to GW emission. The subpopulation of low-eccentricity events in the LISA band (the peak at $e\sim10^{-4}$) corresponds to binaries that got ejected from the cluster due to a strong recoil in a binary-single interaction. The most eccentric events include single-single and binary-single captures, as well as Lidov-Kozai mergers in dynamically formed triple systems of COs.
We find that roughly half of all merging pairs form at a GW frequency too high for LISA to detect, consistent with Ref.~\cite{Samsing:2018isx}. 
In particular, we find that about 17\% (29\%) of all merging binaries have an eccentricity of at least 0.2 when they enter the LIGO (ET) band, at a GW frequency of $10\,\rm Hz$ ($3\,\rm Hz$). This is a significant fraction that could be linked observationally to constrain the branching ratio of dynamical binaries in current and future GW transient catalogs.

The right panel of Fig.~\ref{fig:mergers} shows the fraction of events detectable by each GW detector network: LISA, LGWA, ET/CE, and LIGO. 
The bulk of the emitted GWs originate close to the merger. Since the majority of the merged binaries are stellar mass, the mergers emit most of their GW radiation in ground-based GW detectors above $\sim1\,\rm Hz$. The fraction of events with an $\rm SNR>10$ ($\rm SNR>100$) is $3.1\times10^{-2}$ ($2.6\times10^{-3}$), and $1.8\times10^{-4}$ ($1.8\times10^{-5}$), for LIGO and LISA, respectively, with somewhat intermediate values for LGWA. In particular, since the intrinsic annual event rate is a $\rm few\times10^4$ (several hundreds) assuming the YMC (GC) population, LISA is predicted to observe a few eccentric inspiraling binaries per year (per century), assuming an SNR threshold of 10. A network of next-generation detectors (ET/CE) will detect a few $10\%$ (percent) of all events with $\rm SNR>10$ ($\rm SNR>100$), depending on the configuration.

\subsection{Compact object spin evolution through \texorpdfstring{$\mu$}{micro}TDE
accretion}
\label{sec:spin}

\subsubsection{Individual spin distributions}
\label{sec:spin_dist}

\begin{figure}
    \centering
    \includegraphics[width=\linewidth]{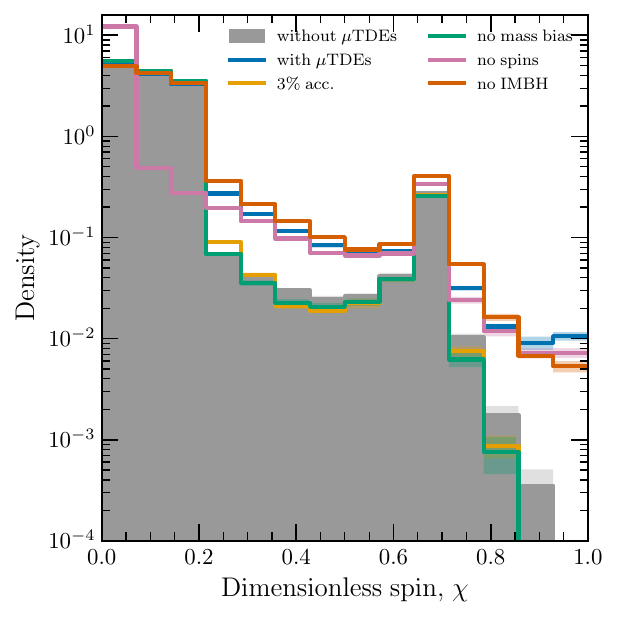}
    \caption{Dimensionless spin magnitude distributions of all merging COs across various models: we compare the effect that $\mu$TDEs have on the spin distribution for models that include $\mu$TDEs (step-colored histograms) against simulations where $\mu$TDEs are ignored (gray-filled histogram). The default model is represented by the ``with $\mu$TDEs'' histogram (solid blue).}
    \label{fig:spin_distribution}
\end{figure}

Figure~\ref{fig:spin_distribution} shows the distribution of individual dimensionless spin magnitudes $\chi$ of the merging COs.  For the no-$\mu$TDE population (WT0, gray filled), the spin distribution is concentrated at low values ($\chi\lesssim0.3$) for first-generation BHs, reflecting the natal spin prior $\chi_{1g}\in[0,0.2]$ in the default model. A secondary population near $\chi\approx0.65$--$0.75$ arises from second-generation merger remnants~\cite{Berti:2008af}.  When $\mu$TDEs are enabled, the distribution acquires a broader high-spin tail extending to $\chi\approx0.5$--$0.9$, driven by spin-up through prograde disk accretion. The broadening is strongly suppressed in the 3\%-accretion and no-mass-bias models because in both cases accretion onto the CO is reduced.

The no-spins model ($\chi_{1g}=0$) provides a null hypothesis: without natal spin, the only source of spin is $\mu$TDE accretion.  Even from $\chi=0$, the spin distribution develops a distinct tail extending to $\chi\approx0.3$--$0.5$, confirming a non-negligible spin-up from $\mu$TDEs.

\subsubsection{CO mass distribution: pre- and post-\texorpdfstring{$\mu$}{u}TDE}
\label{sec:mass_dist}

\begin{figure}
    \centering
    \includegraphics[width=\linewidth]{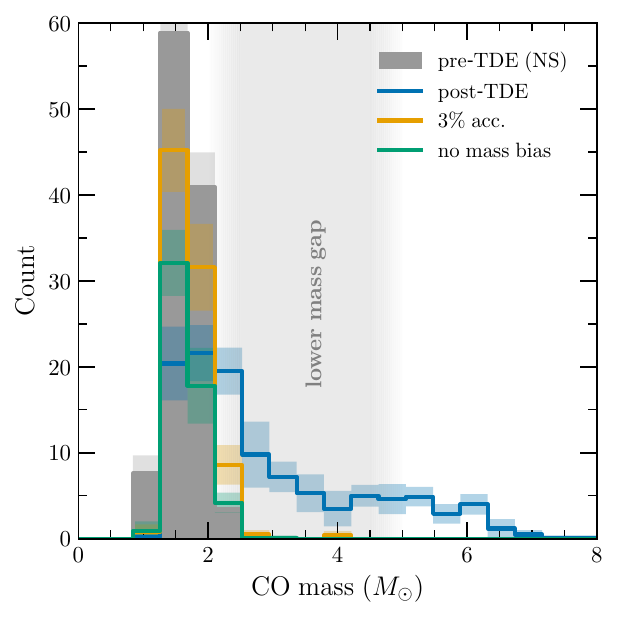}
    \caption{NS mass distribution before (gray) and after (blue, orange, green) a $\mu$TDE episode for three model populations. The shaded region marks the lower mass gap ($\sim2.5$--$\sim5\,M_\odot$); post-TDE accretion pushes a fraction of NSs across $M_{\rm TOV}$, populating it via accretion-induced collapse (AIC).}
    \label{fig:mass_distribution}
\end{figure}

Figure~\ref{fig:mass_distribution} shows the distribution of CO masses before and after a $\mu$TDE episode, restricted to NS progenitors ($m_{\rm CO,ini}<M_{\rm TOV}$). The pre-TDE mass distribution (gray filled) is concentrated below $M_{\rm TOV}\approx2.2\,M_\odot$, peaking near $m\approx1.3$--$1.4\,M_\odot$, consistent with the canonical NS birth mass from core-collapse supernovae~\citep{Lattimer:2012nd}. The distribution falls steeply toward higher masses, with fewer NSs born above $2\,M_\odot$.

\begin{figure*}
    \centering
    \includegraphics[width=0.49\linewidth]{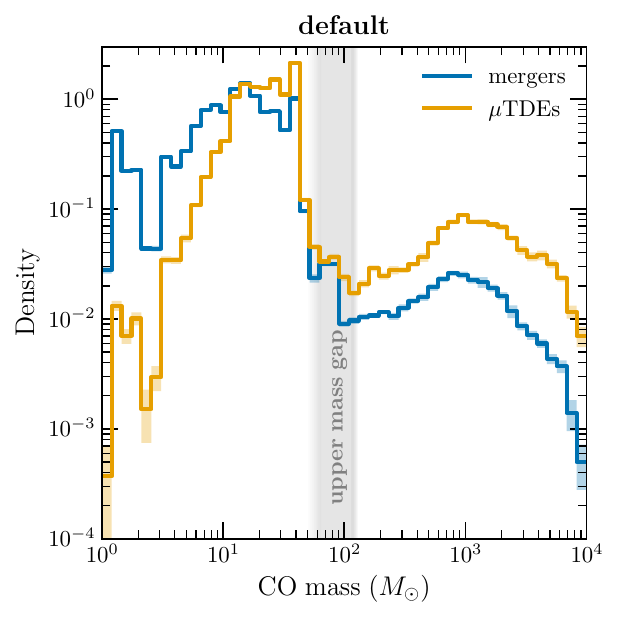}
    \includegraphics[width=0.49\linewidth]{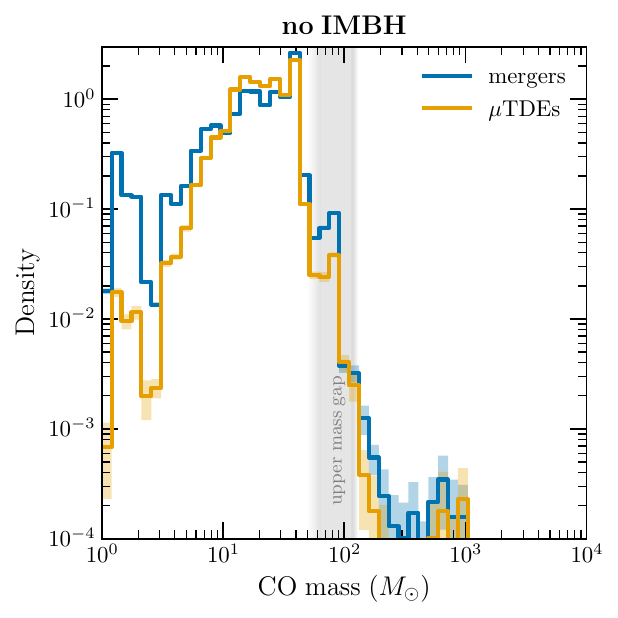}
    \caption{The density distribution of CO mass in merging binaries (either primary or secondary, blue) and in $\mu$TDEs (orange).
    The left panel shows the default model while the right panel the case where no IMBH seeds have been placed initially in the simulations.
    The band has been estimated across 100 realizations of the default YMC model. The upper mass gap is from $\approx50\,M_\odot$ to $\approx130\,M_\odot$.}
    \label{fig:co_mass_distribution}
\end{figure*}

After accreting a fraction $f_A$ of a disrupted star, the post-TDE mass distribution (blue) is shifted to higher values. The shift is most pronounced between $1.5$ and $2.5\,M_\odot$: events that populate this range are NSs that have accreted $\Delta m = f_A m_\star \sim 0.05$--$0.5\,M_\odot$ from a typical low-mass main-sequence star. A fraction of these post-TDE NSs exceed $M_{\rm TOV}$ and eventually collapse to BHs (accretion-induced collapse, AIC), populating the lower mass gap ($2.2$--$5\,M_\odot$, shaded region) from below. The lower mass gap shading highlights the region between the maximum NS mass ($M_{\rm TOV}\approx2.2\,M_\odot$) and the minimum observed BH mass ($\approx5\,M_\odot$)~\citep{Farr:2010tu}, which is predicted to be underpopulated in certain stellar evolution models~\cite{Olejak:2022zee} but, as we show here, can be populated through AIC events driven by $\mu$TDE accretion.

The 3\% accretion model (orange) shows a much smaller post-TDE mass shift: with only $f_A=0.03$, the typical accreted mass is $\Delta m\approx0.015\,M_\odot$, insufficient to push most NSs significantly above their birth mass or across $M_{\rm TOV}$. The AIC contribution to the lower mass gap is correspondingly suppressed. The no-mass-bias model (green) shows a post-TDE distribution similar to the default, confirming that the mass bias parameter $m_b$ affects the CO encounter rate but not the mass evolution per event.

The lower mass gap population produced by $\mu$TDE-driven AIC events is a distinctive prediction of our model. In the absence of $\mu$TDE accretion, NSs in dense clusters cannot grow significantly in mass through dynamical interactions alone. The presence of a population of compact objects with masses $2.2$--$5\,M_\odot$ in GW merger catalogs --- if confirmed --- would therefore constitute evidence for accretion-driven mass growth, with $\mu$TDEs in dense stellar environments as a natural channel. This is particularly relevant for LVK events such as GW190814~\citep{LIGOScientific:2020zkf}, where the secondary component has a mass of $\approx2.6\,M_\odot$, which could be consistent with an AIC product from NS-$\mu$TDE accretion.

In Fig.~\ref{fig:co_mass_distribution} we show the mass distribution of merging COs and COs that cause $\mu$TDEs under the ``default'' and ``no IMBH'' models. We assume the ``default'' (left panel) and ``no IMBH'' (right) YMC catalog. We conclude that $\mu$TDEs probe better the IMBH mass range in the default model, as the mass distribution has more probability weight for heavier COs. This is because the tidal disruption cross section scales as $M\substack{4/3\\{\rm CO}}$. We observe that the no heavy seeds model lacks the long IMBH tail, while in both scenarios the upper mass gap is filled by hierarchical mergers and $\mu$TDEs. Finally, NS transients are dominated by mergers.

\subsubsection{Effective spin distribution}
\label{sec:chieff_dist}

\begin{figure}
    \centering
    \includegraphics[width=\linewidth]{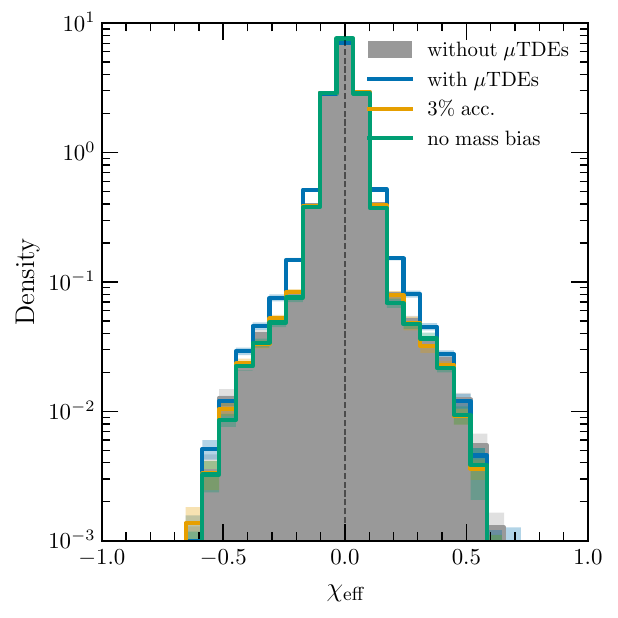}
    \caption{Same as Fig.~\ref{fig:spin_distribution} but with fewer models and showing the dimensionless effective spin parameter of merging CO pairs.}
    \label{fig:chieff_with_without}
\end{figure}

Figure~\ref{fig:chieff_with_without} shows the $\chi_{\rm eff}$ distribution for BH-BH mergers across four model populations: without $\mu$TDEs (gray), with $\mu$TDEs at 30\% accretion (blue), at 3\% accretion (orange), and without mass bias (green). The four distributions are nearly indistinguishable across the full $\chi_{\rm eff}\in[-1,1]$ range, indicating that neither the presence of $\mu$TDEs nor the choice of accretion fraction or mass bias parameter produces a detectable shift in the shape of the $\chi_{\rm eff}$ distribution at the population level. The distribution is approximately symmetric around zero for all models, with a width of $\sigma_{\chi_{\rm eff}}\approx0.2$--$0.3$ (FWHM). 
This is consistent with the $\mu$TDE spin-up signal being diluted at the population level by the large fraction ($\approx70\%$, corresponding to the $h=0$ bin in Fig.~\ref{fig:chieff_conditioned_h}) of merging BHs that have not experienced any prior $\mu$TDE episode.

\subsubsection{Separating the \texorpdfstring{$\mu$}{u}TDE and hierarchical contributions to the \texorpdfstring{$\chi_{\rm eff}$}{chi\_eff} distribution}
\label{sec:chieff_separation}

Hierarchical mergers are expected to broaden the $\chi_{\rm eff}$ distribution at high total mass, since merger remnants are more massive than their progenitors and tend to carry high remnant spins ($\chi_{\rm rem}\approx0.7$, e.g., Ref.~\cite{Berti:2008af}). On the other hand, $\mu$TDE spin-up is expected to operate primarily for low-mass COs where $\Delta\chi\approx \sqrt{2/3}f_A\,m_\star/m_{\rm BH}$ is largest. This follows from the Bardeen accretion formula~\citep{1970Natur.226...64B} in the limit $\Delta m = f_A m_\star \ll m_{\rm BH}$: the spin increment is $\Delta\chi \approx \sqrt{2/3}\,\Delta m/m_{\rm BH}$, where $\sqrt{2/3}$ is the specific angular momentum of the ISCO in the Schwarzschild limit. The two effects are therefore expected to occupy different regions of the $(M_{\rm tot},\,\chi_{\rm eff})$ plane.

\begin{figure*}
    \centering
    \includegraphics[width=\linewidth]{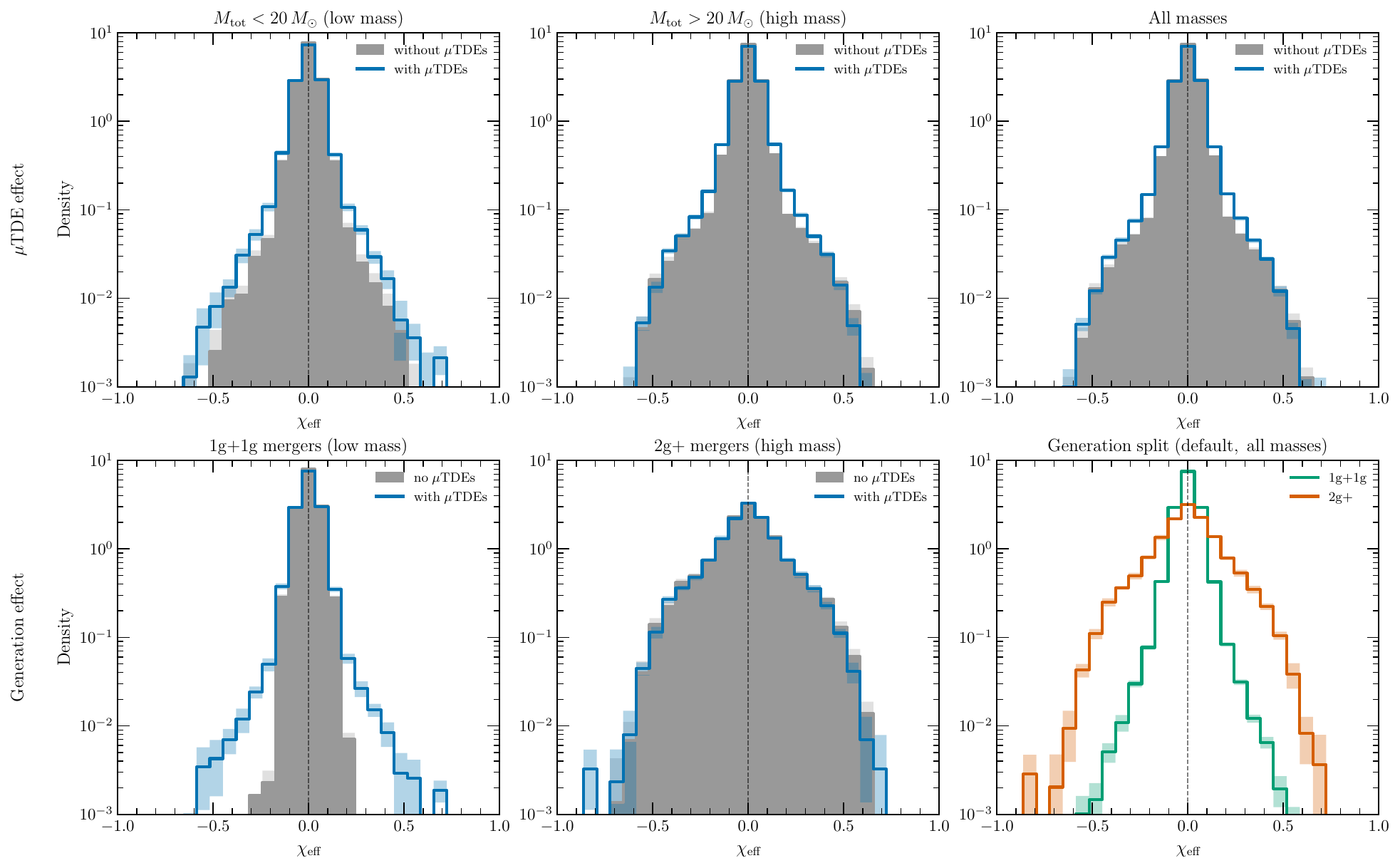}
    \caption{Distribution of the effective spin parameter $\chi_{\rm eff}$ of merged CO binaries. Top row shows the effect of $\mu$TDEs on the distribution, while the lower row the effect of hierarchical mergers. Low-mass (total mass $M_{\rm tot}<20\,M_\odot$) mergers are shown in the left column, high-mass ($M_{\rm tot}>20\,M_\odot$) shown in the middle, and all masses in the right column. Blue histograms correspond to the default model, while the solid gray histogram to the no-$\mu$TDEs scenario (WT0). Finally, the lower right panel shows the $\chi_{\rm eff}$ distribution as split based on BH generation (first generation 1g, second or higher generation 2g+).}
    \label{fig:chieff_muTDE_vs_hierarchical}
\end{figure*}

We test this hypothesis with Fig.~\ref{fig:chieff_muTDE_vs_hierarchical}.
{We find that coalescing binaries with first-generation (1g) members, which we call ``1g+1g mergers'', contain high-$\chi_{\rm eff}$ wings that extend up to a magnitude value of $0.5$. This wing does not appear when we turn $\mu$TDEs off (model WT0, see lower left panel of Fig.~\ref{fig:chieff_muTDE_vs_hierarchical}). Moreover, we find no difference in the $\chi_{\rm eff}$ distribution between the WT0 and WT1 models. In the lower panels of Fig.~\ref{fig:chieff_muTDE_vs_hierarchical} we show that hierarchical mergers show a stronger $\chi_{\rm eff}$ wing than $\mu$TDEs, which extends in magnitude to a value of $\approx0.7$, and we demonstrate that the wing appears for high-mass (total mass $>20\,M_\odot$) mergers. The lower-right panel of Fig.~\ref{fig:chieff_muTDE_vs_hierarchical} proves that the high-mass strong wing is associated with hierarchical mergers.}

\subsubsection{Effective spin conditioned on prior \texorpdfstring{$\mu$}{u}TDE count}
\label{sec:chieff_vs_h}

\begin{figure}
    \centering
    \includegraphics[width=\linewidth]{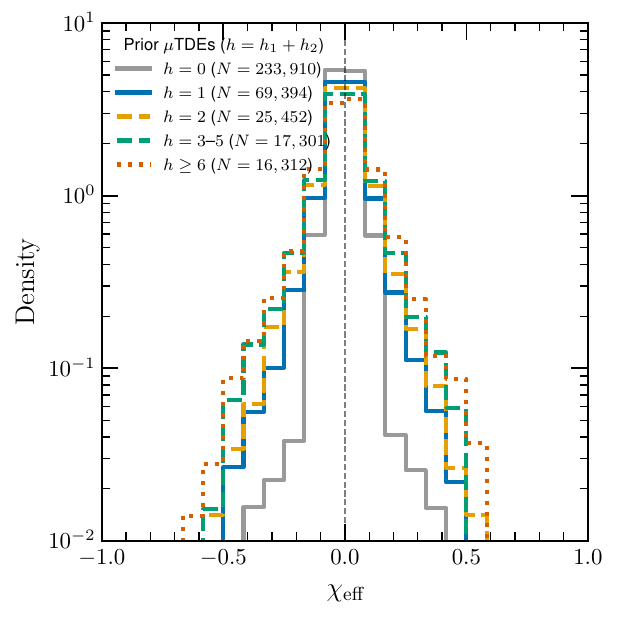}
    \caption{Density distribution of the dimensionless effective spin parameter of merging CO binaries conditioned on $h$, the $\mu$TDE count experienced by either CO progenitor prior to the merger.}
    \label{fig:chieff_conditioned_h}
\end{figure}

We define the total prior $\mu$TDE count $h = h_1+h_2$, where $h_i$ is the number of $\mu$TDE episodes experienced by the $i$-th BH ($i=1$, the primary, or $i=2$, the secondary member of the binary) before the final merger. Physically, $h$ denotes the total number of $\mu$TDEs the COs have collectively experienced prior to their GW merger. Figure~\ref{fig:chieff_conditioned_h} shows the $\chi_{\rm eff}$ distribution for BH-BH mergers conditioned on the value of $h$.

The $h=0$ subsample dominates the statistics ($N=265{,}127$) and is approximately symmetric around zero, with a sharp peak near $\chi_{\rm eff}\approx0$ and a full width at half maximum of $\sigma_{\chi_{\rm eff}}\approx0.2$--$0.3$. As $h$ increases the distributions develop progressively broader tails while retaining a peak near $\chi_{\rm eff}\approx0$: 
at $h=2$ ($N=20{,}742$) and $h=3$--$5$ ($N=11{,}898$) the enhancement extends to $\chi_{\rm eff}\approx0.4$--$0.5$; and for $h\geq6$ ($N=11{,}600$) a clear excess at $\chi_{\rm eff}\gtrsim0.2$ is present, with a tail reaching $\chi_{\rm eff}\approx0.5$--$0.6$. The overall trend is a monotonic broadening of the distribution in both directions with increasing $h$, with the positive wing extending somewhat further than the negative, consistent with a mild prograde preference in the $\mu$TDE spin-up, but without a clear suppression of the negative tail. The peak position remains near zero for all values of $h$, indicating that $\mu$TDE spin-up acts as a perturbation on the underlying isotropic spin distribution rather than fully dominating it.

\subsubsection{Mean effective spin versus prior \texorpdfstring{$\mu$}{u}TDE count}
\label{sec:chieff_mean_vs_h}

\begin{figure}
    \centering
    \includegraphics[width=\linewidth]{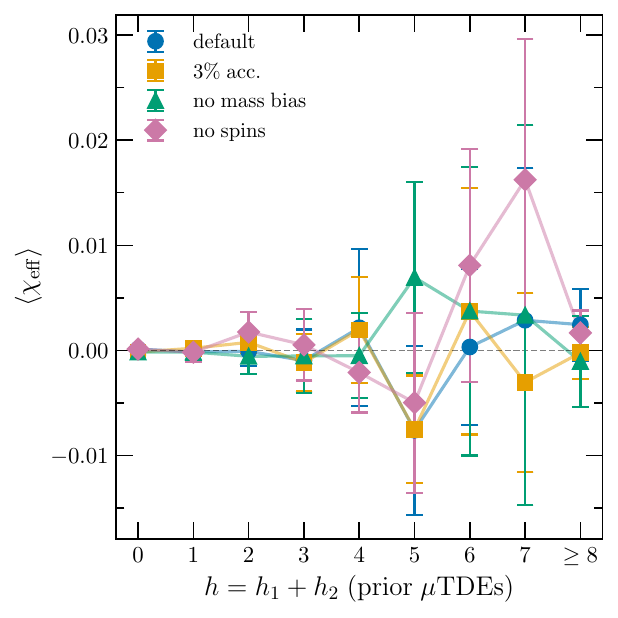}
    \caption{The mean dimensionless effective spin parameter of merging CO binaries as a function of $h$, the total number of prior $\mu$TDEs experienced by either CO.}
    \label{fig:chieff_mean_vs_h}
\end{figure}

Figure~\ref{fig:chieff_mean_vs_h} shows $\langle\chi_{\rm eff}\rangle$ as a function of $h$ for four model populations. At $h=0$--$2$, all models cluster tightly around $\langle\chi_{\rm eff}\rangle\approx0$, with statistical uncertainties (mean and variance) of $\lesssim0.005$. At higher $h$ ($h\geq3$), the error bars grow substantially as the number of events per bin decreases, and the mean $\langle\chi_{\rm eff}\rangle$ begins to scatter around zero across all models. No clear monotonic trend is established in the current simulation output: the default model reaches $\langle\chi_{\rm eff}\rangle\approx0.01$--$0.015$ at $h=6$--$7$ but returns toward zero at $h\geq8$, while the 3\% accretion and no-mass-bias models show comparable scatter without a consistent positive offset.  The no-spins model ($\chi_{1g}=0$) oscillates between $\approx-0.01$ and $\approx+0.01$ at $h\geq3$, consistent with simulation noise.

The absence of a statistically significant monotonic trend at high $h$ is primarily a consequence of the limited number of events in those bins: at $h\geq6$, only $\sim10^3$--$10^4$ mergers contribute across all 10 realizations, giving per-realization scatter that dominates the mean. The low-$h$ regime ($h=0$--$2$), which is well-sampled, shows $\langle\chi_{\rm eff}\rangle$ consistent with zero within the uncertainties for all models, indicating that the cumulative $\mu$TDE spin-up imprint on $\langle\chi_{\rm eff}\rangle$ is below our current statistical sensitivity.

\subsubsection{Spin increment per \texorpdfstring{$\mu$}{u}TDE episode}
\label{sec:delta_chi}

Figure~\ref{fig:delta_chi_vs_mass} shows the per-event spin magnitude increment $\Delta\chi = \chi_{\rm new} - \chi_{\rm old}$ as a function of initial CO mass, split by type (NS-$\mu$TDE, BH-$\mu$TDE prograde, and BH-$\mu$TDE retrograde). 

For NS-$\mu$TDE events, $\Delta\chi$ spans a broad range from $\sim0.01$ to $\sim1$, concentrated at $\Delta\chi\approx0.2$--$0.4$ for typical NS masses. This overdensity is associated with NSs that accrete enough mass to reach the mass-shedding limit, but not enough to collapse into BHs. Another concentration is found at $\Delta\chi\approx1$, corresponding to NSs that disrupt massive stars ($m_*\gtrsim10\,M_\odot$). 
In these systems, the accreting NS undergoes AIC, while the remaining debris is still sufficient to spin up the newly formed BH to nearly maximal rotation. 
This is because a typical NS with mass $1.4\,M_\odot$ will have to accrete about $1.5\,M_\odot$ to reach the supramassive limit of $1.2M_{\rm TOV}\approx2.6\,M_\odot$, then collapse into a BH (AIC), and further accrete $\sim2\,M_\odot$ of matter coherently to spin up from $\sim0.4$ to unity. In total, the NS shall accrete $\sim3\,M_\odot$ for the resulting BH to reach the Thorne limit after the $\mu$TDE episode, which assuming an accretion fraction of $f_{A}=0.30$ requires a star of at least $\sim10\,M_\odot$. In the default scenario, every star with a mass of $>10\,M_\odot$ disrupted by a NS will result in a rapidly spinning BH remnant ($\chi\approx0.998$) of mass $\gtrsim6\,M_\odot$. Since we assume NSs have zero spin, every accretion spin-up episode is necessarily prograde. 

For prograde BH-$\mu$TDE events, $|\Delta\chi|$ is largest at low BH masses: up to unity at $m_{\rm BH}\approx3\,M_\odot$, dropping to $\approx0.0$--$0.8$ at $m_{\rm BH}\sim10\,M_\odot$, and becoming negligible ($|\Delta\chi|<0.1$) at IMBH masses ($m_{\rm BH}\gtrsim100\,M_\odot$). This confirms the mass-dependence identified in Sec.~\ref{sec:chieff_separation}: the $\mu$TDE spin-up effect is concentrated at low CO masses, validating the hypothesis that the $\mu$TDE and hierarchical-merger contributions to the $\chi_{\rm eff}$ broadening are mass-separated. 

Retrograde BH/$\mu$TDE events produce $\Delta\chi<0$ at all BH masses, with $|\Delta\chi|$ slightly smaller than for prograde events at the same mass, as expected from the Bardeen prescription~\cite{1970Natur.226...64B}.

\subsubsection{The mass-spin parameter space of merging COs}
\label{sec:m-chi}

Figure~\ref{fig:chi_vs_mass} shows the spin parameter of all COs merging in binaries (NS-NS, NS-BH, and BH-BH) as a function of CO mass. 
Since the default YMC model has been assumed, the BH-BH spin parameter is uniform in the range $0.0<\chi<0.2$ at fixed mass $m<45\,M_\odot$. The overdensity at $\chi\approx0.7$ corresponds to second-generation BHs. Some Second-generation BHs have formed out of previous NS-NS mergers, giving rise to the overdensity at $(m,\chi)\approx(2.8\,M_\odot,0.7)$ since the typical NS mass is around $1.4\,M_\odot$ (the primary mode in the NS mass distribution assumed from Ref.~\cite{Rocha:2023xwp}). 

In the IMBH mass range ($m>100\,M_\odot$), and assuming isotropy, the remnants maintain modestly low spins ($\chi<0.3$), while a smaller fraction has spins clustering around $\chi\approx0.7$ with a long low-$\chi$ tail, corresponding to merger remnants from relatively symmetric progenitors. 

\begin{figure*}
    \centering
    \includegraphics[width=\linewidth]{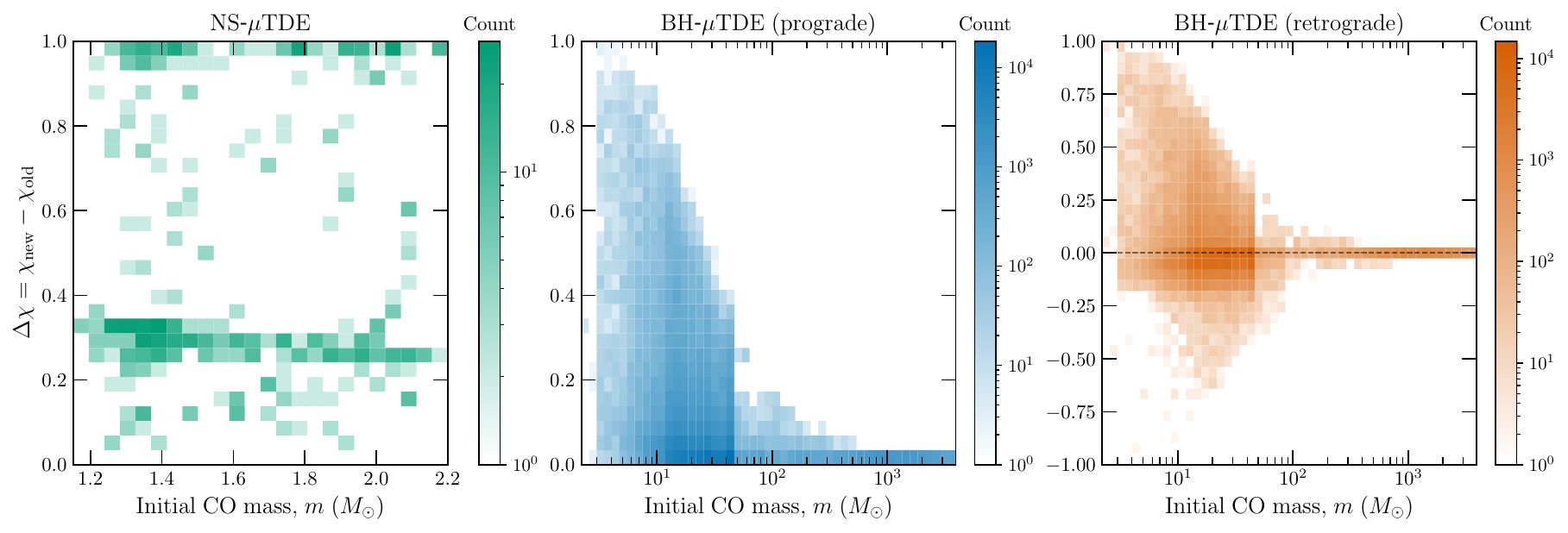}
    \caption{Change in the CO dimensionless spin magnitude in $\mu$TDE accretion episodes, decomposed to NS-$\mu$TDE (left, always prograde because NSs are assumed originally non-spinning), BH-$\mu$TDE (prograde, middle), and BH-$\mu$TDE (retrograde, right). Negative values, $\Delta\chi<0$, correspond to cases where the CO spins down due to accretion and the final spin magnitude is smaller than the initial one.}
    \label{fig:delta_chi_vs_mass}
\end{figure*}

\begin{figure*}
    \centering
    \includegraphics[width=\linewidth]{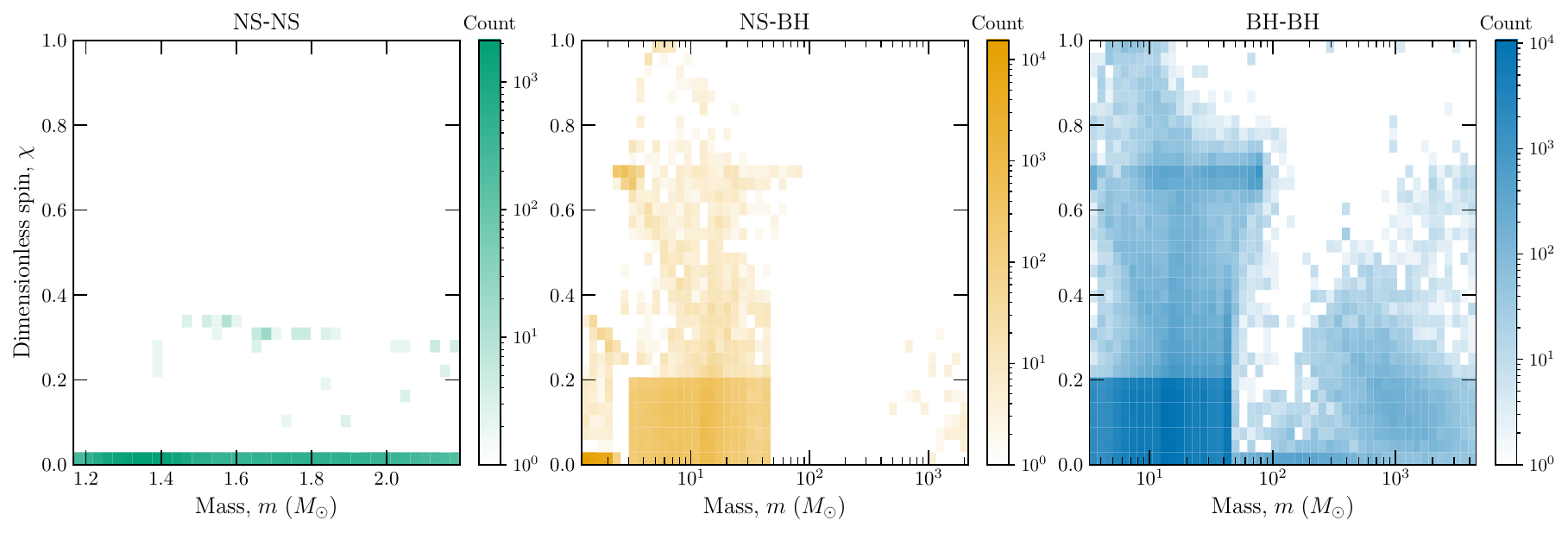}
    \caption{Dimensionless spin vs. mass of COs merging in pairs, partitioned based on pair flavor: NS-NS (left), NS-BH (middle), and BH-BH (right). For the default YMC model.}
    \label{fig:chi_vs_mass}
\end{figure*}

Moreover, since low-mass COs can vary their spin magnitude significantly (cf.~Fig.~\ref{fig:delta_chi_vs_mass}), a population of highly spinning COs ($\chi>0.8$) appears with masses $m\lesssim15\,M_\odot$. 
Finally, while NSs are assumed to have zero spin and most NS mergers have zero spins, some of them have undergone past $\mu$TDE episodes with low-mass ($<\,M_\odot$) MS stars spinning up to or below their mass shedding limit ($\chi\approx0.3$--$0.4$).

\subsection{\texorpdfstring{$\mu$}{u}TDE dynamical channel properties}
\label{sec:TDE_channels}

\subsubsection{Dynamical channel branching ratios}
\label{sec:branching}

\begin{figure}
    \centering
    \includegraphics[width=\linewidth]{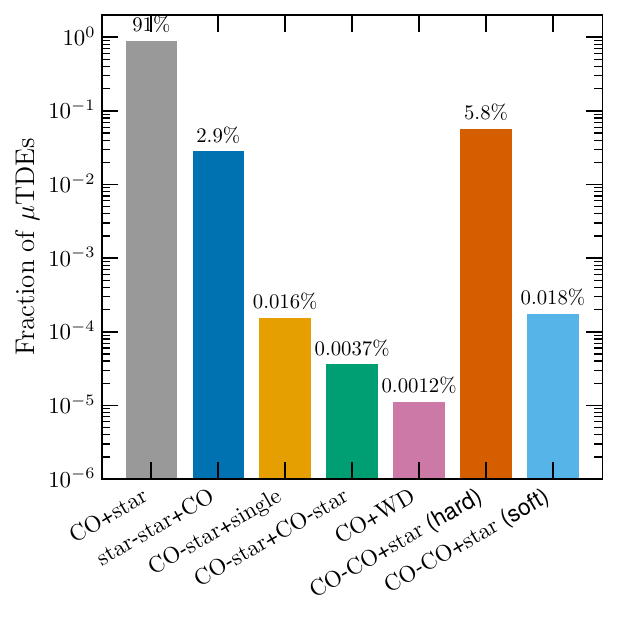}
    \caption{Branching ratio of various $\mu$TDE subchannels across all events in a single realization of the default catalog.}
    \label{fig:tde_branching}
\end{figure}

Figure~\ref{fig:tde_branching} shows the fraction of $\mu$TDEs in each dynamical channel for the default YMC population. The dominant channel is CO+star,
contributing $\approx90\%$ of all $\mu$TDE events, consistent with the expectation that single BHs and NSs are more numerous than CO-star binaries, which individually have larger TDE cross sections. 

About $8\%$ of all $\mu$TDEs are produced during hard binary CO - single star interactions.
We remind the reader that, by ``hard interaction'', we refer to a resonant CO-CO-star encounter in which
 the star is tidally disrupted by a single CO member of the binary. In contrast, CO-CO+star (soft)-induced $\mu$TDEs are much rarer because this channel requires very hard CO binaries. Such binaries have smaller interaction cross-sections, leading to lower encounter rates. Moreover, such pairs do not survive in the cluster for long:  they either dynamically decouple rapidly entering the GW-driven evolution until merger, or are ejected  from the cluster following the strong recoil imparted in the more frequent CO-CO+CO encounters, a consequence of their large binding energies and mass segregation. These combined effects make CO-CO+star soft $\mu$TDEs the rarest of all subchannels considered here. 

The star-star+CO channel (type 2) contributes $\approx4\%$ and is the third most popular subchannel on the list. The remaining subchannels each contribute $\lesssim0.04\%$.
The combined contribution of types 21 and 22 ($\lesssim0.02\%$) confirms that BBH-star TDEs are subdominant when primordial binary fractions are not inflated, consistent with~\citet{2026A&A...707A.217R}.

\subsubsection{CO generation distribution and prior \texorpdfstring{$\mu$}{u}TDE counts}
\label{sec:co_gen}

\begin{figure}
    \centering
    \includegraphics[width=\linewidth]{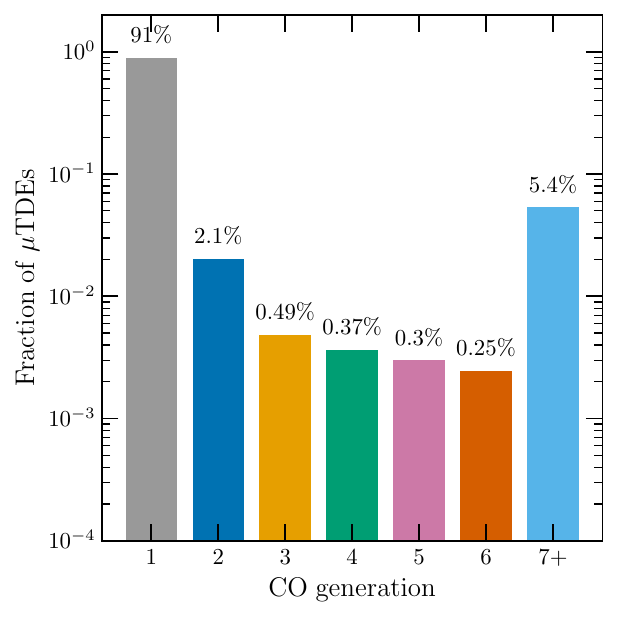}
    \caption{Fraction of $\mu$TDEs as a function of the CO merger-generation in a single realization of the default catalog. The last bin ``7+'' denotes all COs (all BHs) whose merger generation is at least 7, i.e., the BH has undergone at least 7 previous mergers with other COs in the cluster prior to undergoing a $\mu$TDE.}
    \label{fig:co_generation}
\end{figure}

Figure~\ref{fig:co_generation} shows the generation distribution of $\mu$TDE
progenitors: $\approx92\%$, the majority, are first-generation (1g) COs, $\approx2\%$ are 2g, and $\approx1.2\%$ are 3g--6g. The $\approx4.4\%$ in the class ``7+'' corresponds to IMBHs which have undergone multiple (at least 7) past mergers with other COs, and due to their large mass and cross-section have a relatively enhanced probability for undergoing a $\mu$TDE episode. Nevertheless, 1g COs dominate the number of $\mu$TDEs due to the large number of 1g BHs in the cluster.

\begin{figure}
    \centering
    \includegraphics[width=\linewidth]{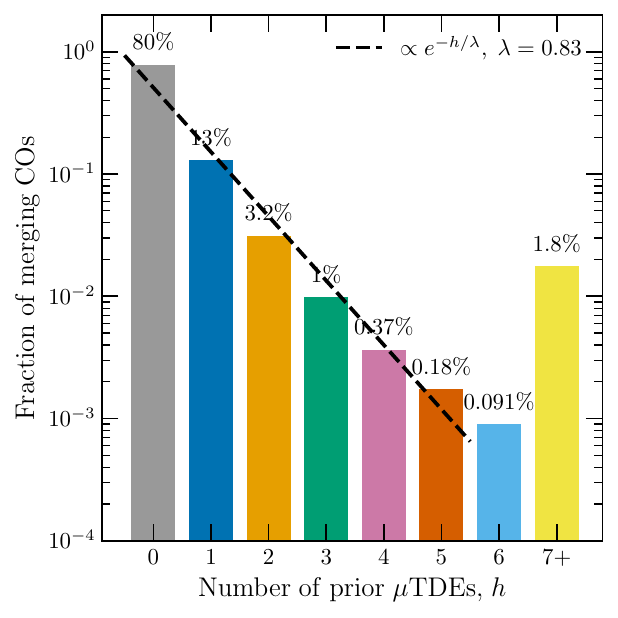}
    \caption{Fraction of CO-CO mergers as a function of the number of prior $\mu$TDEs in a single realization of the default catalog. The black dashed line corresponds to the best-fit exponential curve. The last bin ``7+'' compresses all merging COs that had at least 7 previous $\mu$TDEs. The parameter $\lambda$ is fit to the data and represents the exponential cutoff scale for $h$.}
    \label{fig:h_prior_tdes}
\end{figure}

Figure~\ref{fig:h_prior_tdes} shows the distribution of prior $\mu$TDE count $h=h_1+h_2$. We find that $\approx84\%$ of merging COs have $h=0$, while $\approx12\%$ have $h=1$, and $\approx4\%$ have $h\geq2$. The decay is well represented by an exponential $e^{-h/\lambda}$ with best-fit parameter $\lambda=0.76$, consistent with $\mu$TDEs being a Poisson process. The physical interpretation of $\lambda$ is that it is the average number of prior $\mu$TDEs, $\lambda=\langle h\rangle\approx0.76$, i.e., on average a CO undergoes 4 $\mu$TDEs for every 5 CO-CO mergers, assuming it is retained in the cluster.

\subsection{GW--\texorpdfstring{$\mu$}{u}TDE cross-correlation}
\label{sec:cross_corr}

\subsubsection{Rate correlation across populations}
\label{sec:rate_corr}

Table~\ref{tab:event_rates} shows the yearly BH-BH merger and BH-$\mu$TDE rates across all YMC and GC model populations. The ratio $r = N_{\rm BH\mu TDE}/N_{\rm BHBH}$ is stable at $r\approx1.28$--$1.30$ (YMC) and $r\approx1.25$--$1.27$ (GC) for the default, no-spins, and no-mass-bias models, confirming that neither natal spin nor mass-bias parameter affects the relative efficiency of the two channels. The no-IMBH model gives a slightly lower ratio ($r\approx1.25$ YMC, $r\approx1.22$ GC), indicating that the IMBH boosts the $\mu$TDE rate proportionally more than the merger rate. The 3\% accretion model shows a more significant reduction ($r\approx1.04$ YMC, $r\approx0.75$ GC), well outside the realization scatter, suggesting an indirect effect of reduced BH mass growth on subsequent encounter rates. The overall correlation has a clear physical origin: both rates scale as $\Gamma\propto n_{\rm BH}$, so any process modulating the BH density shifts both rates simultaneously, largely preserving $r$.

{
\setlength{\tabcolsep}{7.5pt}
\begin{table*}
\centering
\caption{Mean yearly event counts $\langle N \rangle \pm \sigma$ for BH-BH mergers and BH-$\mu$TDEs across model populations, for both the YMC and GC formation histories, together with the ratio $r = N_{\rm BH\mu TDE}/N_{\rm BHBH}$. Uncertainties are standard deviations across 100 universe realizations. The no-$\mu$TDE model (WT0) has no BH-$\mu$TDE channel by construction. The error on the ratio $r$ is computed via standard error propagation: $\sigma_r = r\sqrt{(\sigma_{N_{\rm BH\mu TDE}}/N_{\rm BH\mu TDE})^2 +
(\sigma_{N_{\rm BHBH}}/N_{\rm BHBH})^2}$.}
\label{tab:event_rates}
\begin{tabular}{lcccccc}
\toprule\midrule
Population &
$N_{\rm BHBH}$ (YMC) &
$N_{\rm BH\mu TDE}$ (YMC) &
$r$ (YMC) &
$N_{\rm BHBH}$ (GC) &
$N_{\rm BH\mu TDE}$ (GC) &
$r$ (GC) \\
\midrule
default      & $36309\pm68$ & $42931\pm10$ & $1.182\pm0.002$ & $633\pm7$ & $598\pm1$ & $0.944\pm0.011$ \\
no-spins     & $36042\pm61$ & $46404\pm9$  & $1.288\pm0.002$ & $651\pm7$ & $795\pm1$ & $1.221\pm0.013$ \\
no IMBH      & $39373\pm45$ & $48101\pm14$ & $1.222\pm0.001$ & $703\pm5$ & $826\pm1$ & $1.174\pm0.008$ \\
3\% acc.     & $35120\pm54$ & $37175\pm9$  & $1.059\pm0.002$ & $649\pm6$ & $551\pm1$ & $0.849\pm0.008$ \\
no mass bias & $36519\pm68$ & $48629\pm6$  & $1.332\pm0.002$ & $638\pm7$ & $834\pm1$ & $1.307\pm0.014$ \\
no $\mu$TDEs & $36683\pm68$ & ---          & ---             & $668\pm7$ & ---       & ---             \\
\bottomrule
\end{tabular}
\end{table*}
}

\subsubsection{Cluster properties driving both channels}
\label{sec:cluster_props}

\begin{figure*}
    \centering
    \includegraphics[width=\textwidth]{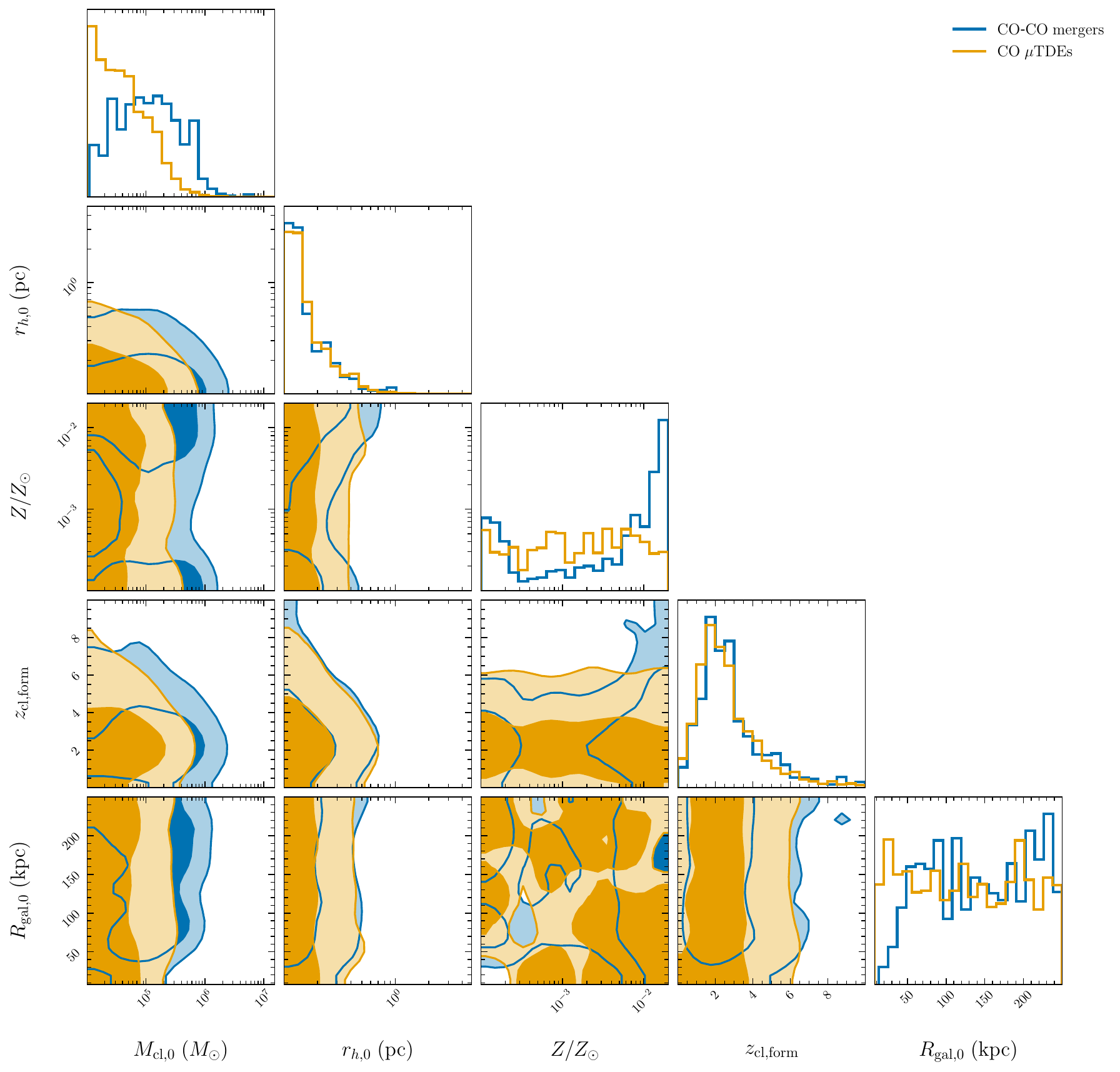}
    \caption{Corner plot of host cluster properties for CO-CO mergers (blue) and CO $\mu$TDEs (orange) in the default YMC population. Diagonal panels show the marginalized weighted probability density of each cluster property; off-diagonal panels show the joint distribution as $68\%$ and $95\%$ credible contours. Each distribution is weighted by the per-cluster event rate assuming the cluster astrophysical
prior, showing which cluster environments contribute most to the total merger and $\mu$TDE budgets.}
    \label{fig:cluster_props}
\end{figure*}

Figure~\ref{fig:cluster_props} shows the astrophysically weighted distributions of initial cluster mass $M_{\rm cl,0}$, half-mass radius $r_{h,0}$, metallicity $Z$, formation redshift $z_{\rm cl,form}$, and galactocentric radius $R_{\rm gal,0}$ for CO-CO mergers and CO-$\mu$TDEs separately. Both channels are dominated by the same cluster class in all dimensions, reflecting their common physical origin.

The weighted cluster mass function peaks at $M_{\rm cl,0}\sim10^4\,M_\odot$ and declines steeply toward higher masses, with $\mu$TDEs slightly more concentrated at low-mass clusters than mergers. This is because COs need to pair up, harden, and eventually merge in a process that depends on CO number density. On the other hand, $\mu$TDEs can also occur in a system with fewer COs, given that the disruption occurs along a single hyperbolic passage of a star by a CO. While the more massive the cluster, the larger the number of transients, there is less weight at the population level because there a fewer heavier clusters forming with masses $M_{\rm cl,0}>10^6\,M_\odot$ than lighter ones.

The half-mass radius distribution strongly favors compact clusters with $r_{h,0}\lesssim0.3\,{\rm pc}$, with a sharp cutoff above $\sim1\,{\rm pc}$; the two channels are nearly indistinguishable in this dimension. 
The metallicity distribution is consistent with a preference for low-metallicity clusters ($Z\lesssim0.1\,Z_\odot$), though the discrete metallicity grid used introduces sampling noise that prevents a precise characterization of the metallicity dependence. 
The formation redshift peaks at $z_{\rm cl,form}\approx2$--$3$, consistent with the adopted YMC formation rate. 

Finally, the galactocentric radius distribution is approximately flat across $R_{\rm gal,0}\sim50$--$250\,{\rm kpc}$, with no strong preference for either central or halo clusters, and the two channels are again nearly indistinguishable, with the exception of clusters near the galactic center. Those clusters have shorter lifetimes because strong galactic tides in the simulation enhance mass loss, and while the merger rate peaks with an average delay -- formation and hardening timescales -- of $\approx30\,\rm Myr$ from cluster formation, $\mu$TDEs show no delay once the CO population has assembled because they occur within a crossing time which is typically $\lesssim1\,\rm Myr$ for young clusters~\cite{2010ARA&A..48..431P}. Thus, once these clusters with $R_{\rm gal,0}\lesssim20\,\rm kpc$ evaporate completely within $30\,\rm Myr$, the probability of a CO binary merger drops significantly.

\section{Comparison with previous literature}
\label{sec:comparison}

In this Section we compare our results with other studies and highlight the differences and similarities. We focus on studies that estimate $\mu$TDE rates.

\begin{description}[align=left,leftmargin=5pt]

    \item[Perets et al.\ (2016)] \citet{Perets:2016pwr} introduced the concept of $\mu$TDEs and provided the first analytical rate estimates for the GC channel, predicting $\sim10^{-6}\,{\rm yr}^{-1}$ per Milky-Way-like galaxy, equivalent to $\sim{\rm few}\,{\rm Gpc}^{-3}\,{\rm yr}^{-1}$ in the local Universe assuming a local galactic number density of $0.01\,\rm Mpc^{-1}$~\cite{2016ApJ...830...83C}. Our GC-channel local rate of $\sim0.04\,{\rm Gpc}^{-3}\,{\rm yr}^{-1}$ is a factor of $\approx25$ smaller.

    Moreover, their NS $\mu$TDE rate is suppressed by a factor of $\approx10$ compared to their BH $\mu$TDE rate, while our NS rate is suppressed by a much larger factor. We attribute this to mass segregation, which is not accounted for in the estimate of~\cite{Perets:2016pwr}.
    \citet{Perets:2016pwr} also predicted that the typical viscous timescale of the debris disk is $\lesssim\rm few$~days [their Eq.~(1)], implying a fast-transient signature, which our model also adopts for computing the accretion luminosity.

    \item[Kremer et al.\ (2019, 2020)]  \citet{Kremer:2019zql,Kremer:2020cne} performed CMC simulations of $\mu$TDEs in GCs and YSCs, respectively. For GCs, they predicted a BH-MS TDE rate of approximately $2$--$5\,{\rm Gpc}^{-3}\,{\rm yr}^{-1}$ in the local Universe, with a cosmological peak of $\sim5$--$40\,{\rm Gpc}^{-3}\,{\rm yr}^{-1}$ at $z\approx3$. For YSCs, \citet{Kremer:2020cne} found that single-encounter BH-MS TDEs contribute $\sim1$--$30\,{\rm Gpc}^{-3}\,{\rm yr}^{-1}$, while binary-mediated (BBH+MS) encounters contribute $\sim20$--$160\,\rm Gpc^{-3}\,yr^{-1}$.

    Our GC-channel local rate of $\sim{\rm few}\times10^{-2}\,{\rm Gpc}^{-3}\,{\rm yr}^{-1}$ is two orders below their nominal estimate of $3\,{\rm Gpc}^{-3}\,{\rm yr}^{-1}$. For the YMC channel, our total rate of $\sim20\,{\rm Gpc}^{-3}\,{\rm yr}^{-1}$ (dominated by single CO-star encounters) is consistent with \citet{Kremer:2020cne} single-encounter estimate of $\sim1$--$30\,{\rm Gpc}^{-3}\,{\rm yr}^{-1}$ but below their total (single+binary) upper limit of $\sim20$--$200\,{\rm Gpc}^{-3}\,{\rm yr}^{-1}$. 

    The key difference in the channel decomposition is more striking: \citet{Kremer:2020cne} find binary-mediated (BBH+MS) events dominate over single-encounter events by a factor of $\sim5$--$10$, while we find $\approx90\%$ of $\mu$TDEs arise from single CO-star encounters (our type 1) with binary-mediated channels contributing at $\lesssim10\%$. Notably,~\citet{Kremer:2019zql} finds that single-single $\mu$TDEs dominate for clusters with mass $>10^6\,M_\odot$.

    \item[Rastello et al.\ (2026)] The most fundamental difference between our work and \citet{2026A&A...707A.217R} is the dynamical channel hierarchy. They find $\approx90\%$ of $\mu$TDEs arise from the ``multiples'' channel (hierarchical triples and quadruples, plus chaotic few-body resonant encounters), $\approx7\%$ from the binary channel (including SN-kick triggers), and only $\approx3\%$ from the single-encounter channel. Moreover, $\approx91\%$ of the multiple-channel events involve exchanged BH-star binaries, i.e., pairs formed dynamically but seeded by original (primordial) binaries. Our results are essentially inverted: $\approx90\%$ single CO-star encounters, $\approx4\%$ star-star+CO, and $\lesssim8\%$ BBH-star channels. This inversion is entirely attributable to the presence or absence of primordial binary stars as well as our negligence of multiples beyond binaries. As \citet{2026A&A...707A.217R} themselves note, original binaries segregate toward the cluster core and enhance few-body interactions, dramatically boosting the multiples channel. Our result is therefore the dynamical-assembly-only lower limit on the $\mu$TDE rate.

    Lacking the multiples channel --- because we do not model quadrupoles or $\mu$TDEs in triples --- which accounts for $\approx90\%$ of events in \citet{2026A&A...707A.217R}, our total number of $\mu$TDEs per cluster mass, is expected to be substantially lower than their $\approx5\times10^{-5}\,M_\odot^{-1}$, by up to a factor of $\sim30$ if our single-encounter efficiency is comparable to their $\approx1.7\times10^{-6}\,M_\odot^{-1}$.

    \citet{2026A&A...707A.217R} predict $\approx350$--$450\,{\rm Gpc}^{-3}\,{\rm yr}^{-1}$ at $z=0$, rising to $\approx2000$--$3000\,{\rm Gpc}^{-3}\,{\rm yr}^{-1}$ at $z=2$, adopting $f_{\rm SF,SC}=0.8$ (the fraction of total star formation in star cluster environments). Our YMC rate at $z=0$ is $\sim20\,{\rm Gpc}^{-3}\,{\rm yr}^{-1}$ and peaks at $z\approx2$ with value $\sim100\,{\rm Gpc}^{-3}\,\rm yr^{-1}$ for $f_{\rm YMC}=0.1$. Scaling our rates to the same $f_{\rm SF,SC}=0.8$ would increase them by a factor of eight, yielding ${\cal R}(z\approx0)\sim160\,{\rm Gpc}^{-3}\,{\rm yr}^{-1}$ and ${\cal R}(z\approx2)\sim800\,{\rm Gpc}^{-3}\,{\rm yr}^{-1}$. For comparison, if we exclude the multiples but include binaries and singles, \citet{2026A&A...707A.217R} find ${\cal R}(z=0)\approx20$--$30\,\rm Gpc^{-3}\,yr^{-1}$ and ${\cal R}(z=2)\approx200$--$400\,\rm Gpc^{-3}\,yr^{-1}$, respectively.

    \citet{2026A&A...707A.217R} find that $\eta$ is largely independent of metallicity across the single and multiples channels, but the binary channel efficiency drops from $\eta_{\rm bin}\approx6\times10^{-6}$ at $Z=0.0002$ to $\approx2\times10^{-6}\,M_\odot^{-1}$ at $Z=Z_\odot$, because larger stellar radii at high metallicity cause premature stellar mergers that prevent BH-star binary formation. Our results are consistent with this trend: both our BH-BH merger and $\mu$TDE rates are dominated by low-metallicity clusters ($Z\lesssim0.01\,Z_\odot$; Fig.~\ref{fig:cluster_props}), and the metallicity dependence is somewhat steeper for the merger channel than for the $\mu$TDE channel.

    \citet{2026A&A...707A.217R} find $\approx20\%$ of their $\mu$TDEs involve NSs, with an efficiency $\eta_{\rm NS}\approx1.4\times10^{-5}\,M_\odot^{-1}$, which is approximately one-fifth of the BH efficiency, and with $\approx91\%$ of NS-$\mu$TDE events in original (primordial) binaries. In our model, without primordial binaries, NS-$\mu$TDE events occur primarily through single CO-star parabolic encounters, and their rate is suppressed relative to \citet{2026A&A...707A.217R} by the same factor as for BH-$\mu$TDEs. Nevertheless, our NS-$\mu$TDE count is non-negligible. \citet{Kremer:2019zql} found a very low NS-$\mu$TDE rate of $\sim10^{-8}\,{\rm yr}^{-1}$ per MW-like galaxy in GCs (two orders of magnitude below their BH rate), while \citet{2026A&A...707A.217R} predict $\sim1$--$3\times10^{-5}\,{\rm yr}^{-1}$ for YSCs where primordial binaries dominate the NS channel. Our model, lacking primordial binaries, produces NS-$\mu$TDE rates closer to the \citet{Kremer:2019zql} GC estimate.

    \item[Kıro\u{g}lu et al.\ (2025a)]  \citet{Kiroglu:2024xpc} used CMC simulations to study BH spin evolution through accretion following BH-star collisions in dense clusters, the same physical process as our $\mu$TDEs. They find that 10\%--60\% of merging BBHs have at least one component spun up through a prior stellar collision, which compares with our $\approx30\%$ ($h\geq1$). Our single-encounter dominance (type 1, $\approx88\%$) is consistent with their finding that $\approx90\%$ of spun-up BBHs are dynamically assembled post-collision with randomized spin-orbit orientations.

\end{description}

\section{Conclusions}
\label{sec:conclusions}

In this work we have used the rapid population-synthesis code \href{https://github.com/Kkritos/Rapster}{\sc Rapster} to simulate $10^5$ star clusters per model population across ten model configurations, jointly tracking CO mergers and $\mu$TDEs. By computing merger and $\mu$TDE rate densities, spin evolution, electromagnetic signatures, and GW emission across cosmic time, we have established a comprehensive picture of the multi-messenger connection between these two channels in dense stellar environments. Our main findings are summarized below.

\begin{itemize}[align=left,labelsep=-4pt]

    \item \textbf{Correlated merger and} $\bm{\mu}$\textbf{TDE rates.} The intrinsic source-frame BH-BH merger and BH-$\mu$TDE rate densities are tightly correlated across all model variations, with a ratio $N_{\rm BH\mu TDE}/N_{\rm BHBH}\approx1.25$--$1.30$ that is approximately constant across redshift. This reflects the common physical origin of both channels: both rates scale with the local BH number density, so any process modulating the BH population shifts both simultaneously.

    \item \textbf{Single-encounter channel dominates.} The vast majority ($\approx88\%$) of $\mu$TDEs arise from single CO--star encounters in our simulations, while binary-mediated (star-star+CO and CO-CO+star) encounters contribute $\approx12\%$ ($\approx4\%$ and $\approx8\%$). This is in contrast with studies that include primordial binaries, where multiple- and binary-mediated channels dominate. Our results represent a conservative lower limit on the total $\mu$TDE rate.

    \item \textbf{Lower rates than the broader literature.} Compared to $N$-body and Monte Carlo studies that include primordial binaries, our predicted rates are lower by up to a factor of $\sim30$, reflecting the absence of primordial binaries and multiple-mediated channels. Within the single-encounter channel, our rates are consistent with the literature.

    \item \textbf{Metal-poor, compact, low-mass clusters dominate.} The cluster environments that contribute most to both the merger and $\mu$TDE budgets are metal-poor, low-mass, and compact clusters that formed at $z\sim2$--$3$. These clusters have the highest BH retention fractions and stellar densities, maximizing the rate of both channels.

    \item $\bm\mu$\textbf{TDEs are electromagnetically luminous.} The peak disk luminosity of $\mu$TDE events ranges between $\sim10^{46}$ and $\sim10^{48}\,{\rm erg\,s}^{-1}$, depending on the accreted mass fraction. Assuming a V-band luminosity fraction of at least $10^{-4}$, we find that more than 10\% of $\mu$TDEs have apparent $V$-band magnitude ${\rm AB}<24$ regardless of model, because the events are intrinsically luminous and most occur at $z\sim2$--$3$, where the cluster formation rate is hypothesized to peak.

    \item \textbf{NS-}$\bm\mu$\textbf{TDEs lead to AIC and populate the lower mass gap.} A significant fraction of NS-$\mu$TDE events result in AIC, in which the accreted mass pushes the NS across $M_{\rm TOV}$, forming a low-mass BH in the $2.2$--$5\,M_\odot$ lower mass gap. This provides a dynamical formation channel for compact objects in a mass range otherwise underpopulated by some stellar evolution models.

    \item $\bm\mu$\textbf{TDEs spin up low-mass COs most efficiently.} $\mu$TDE accretion primarily modifies the spins of low-mass COs. NSs are spun up toward their mass-shedding limit; BHs in the lower mass gap ($M_{\rm TOV}$--$5\,M_\odot$) experience spin increments approaching the maximal value; BHs up to $\approx30\,M_\odot$ reach moderate spin increments of $\Delta\chi\lesssim0.5$; and IMBHs ($m\gtrsim100\,M_\odot$) are affected by less than $\Delta\chi\sim0.1$. In contrast, hierarchical mergers alone rarely bring the spin magnitude above $\chi\approx0.8$ in the population due to the assumption of isotropic spin orientations.

    \item \textbf{About 30\% of merging COs have experienced at least one prior} $\bm\mu$\textbf{TDE.} The distribution of prior $\mu$TDE counts $h$ per merging CO follows an approximately exponential decay $\propto e^{-h/0.76}$, with $\approx70\%$ of merging BHs having $h=0$ and $\approx16\%$ having $h\geq1$.

    \item $\bm\mu$\textbf{TDEs do not detectably broaden the population-level $\chi_{\rm eff}$ distribution.} Despite individual spin-up events, the $\chi_{\rm eff}$ distribution of merging CO binaries is not significantly altered at the population level. The broadening of $\chi_{\rm eff}$ is dominated by hierarchical mergers, which produce systematically more massive remnants for which the per-event $\mu$TDE spin increment is negligible. The $\mu$TDE spin-up signal is diluted by the large fraction ($\approx70\%$) of merging BHs with no prior $\mu$TDE history.

    \item \textbf{WD} $\bm\mu$\textbf{TDEs are absent from our catalogs.} WD $\mu$TDEs are extremely rare in our simulations: WDs have an extended formation history and their population builds up at late times, when most BHs have been ejected and the cluster has expanded, making close encounters inefficient. No WD $\mu$TDE events appear in our retained catalogs.

    \item \textbf{The stochastic GW background from} $\bm\mu$\textbf{TDEs is undetectable.} The GW energy density $\Omega_{\rm GW}$ from the incoherent superposition of $\mu$TDE events lies outside the band of ground-based detectors, and several orders of magnitude below the sensitivity curves of LISA and LGWA at all frequencies.

    \item \textbf{Significant fractions of merging events eccentric and detectable.} A network of next-generation ground-based detectors (ET/CE) will be able to detect more than $20\%$ ($>{\rm few}\%$) of all dynamical events in our generated catalogs with an $\rm SNR>10$ ($\rm SNR>100$), while LISA (LGWA) is predicted to detect $>0.02\%$ ($>0.5\%$) with an $\rm SNR>10$. Approximately half of all events form with an observer-frame GW frequency of $\gtrsim1\,\rm mHz$ and about $17\%$ of the events have an eccentricity of at least 0.2 at $10\,\rm Hz$.

\end{itemize}

Together, these results establish $\mu$TDEs in dense star clusters as a distinctive multi-messenger phenomenon: electromagnetically luminous transients whose host environments also produce GW-detectable BH-BH mergers at a correlated rate. The tight $N_{\rm BH\mu TDE}/N_{\rm BHBH}$ ratio across all model populations makes joint observations of the same sky regions -- by Rubin/LSST and by GW detectors spanning different bands, from LISA to LVK -- a direct test of our predictions. Future work extending these simulations to include primordial binaries, triples, and quadruples, which would boost the $\mu$TDE rate by up to a factor of $\sim30$ and shift the channel decomposition toward binary- and multiple-mediated encounters, will be essential for a complete census of the $\mu$TDE contribution to multi-messenger astrophysics from dense stellar environments.

\begin{acknowledgments}
K.K. thanks K.~K.Y.~Ng for past discussions related to merger rates, and D.~D'Orazio for discussions.
Support for this work was provided by NASA through the NASA Hubble Fellowship Grant No.~HST-HF2-51608.001-A awarded by the Space Telescope Science Institute, which is operated by the Association of Universities for Research in Astronomy, Inc., for NASA, under contract NAS5-26555.
K.K., F.I., and E.B. are supported by NSF Grants No.~AST-2307146, No.~PHY-2513337, No.~PHY-090003, and No.~PHY-20043, by NASA Grant No.~21-ATP21-0010, by John Templeton Foundation Grant No.~62840, by the Simons Foundation [MPS-SIP-00001698, E.B.], by the Simons Foundation International [SFI-MPS-BH-00012593-02], and by Italian Ministry of Foreign Affairs and International Cooperation Grant No.~PGR01167. 
K.K. is also supported by the Onassis Foundation - Scholarship ID: F ZT 041- 1/2023-2024.
The work of F.I. is supported by a Miller Postdoctoral Fellowship.
This work was carried out at the Advanced Research Computing at Hopkins (ARCH) core facility~\cite{rockfish}, which is supported by the NSF Grant No.~OAC-1920103.
\end{acknowledgments}
\appendix

\bibliography{references}

@ARTICLE{Branchesi:2023mws,
       author = {{Branchesi}, Marica and {Maggiore}, Michele and others},
        title = "{Science with the Einstein Telescope: a comparison of different designs}",
      journal = {\jcap},
         year = 2023,
        month = jul,
       volume = {2023},
       number = {7},
          eid = {068},
        pages = {068},
          doi = {10.1088/1475-7516/2023/07/068},
archivePrefix = {arXiv},
       eprint = {2303.15923},
 primaryClass = {gr-qc},
       adsurl = {https://ui.adsabs.harvard.edu/abs/2023JCAP...07..068B}
}

@article{Chia:2021mxq,
    author = "Chia, Horng Sheng and Olsen, Seth and Roulet, Javier and Dai, Liang and Venumadhav, Tejaswi and Zackay, Barak and Zaldarriaga, Matias",
    title = "{Signs of higher multipoles and orbital precession in GW151226}",
    eprint = "2105.06486",
    archivePrefix = "arXiv",
    primaryClass = "astro-ph.HE",
    doi = "10.1103/PhysRevD.106.024009",
    journal = "Phys. Rev. D",
    volume = "106",
    number = "2",
    pages = "024009",
    year = "2022"
}

@article{Seoane:2021kkk,
    author = "Seoane, Pau Amaro and others",
    title = "{The effect of mission duration on LISA science objectives}",
    eprint = "2107.09665",
    archivePrefix = "arXiv",
    primaryClass = "astro-ph.IM",
    doi = "10.1007/s10714-021-02889-x",
    journal = "Gen. Rel. Grav.",
    volume = "54",
    number = "1",
    pages = "3",
    year = "2022"
}

@article{Cleveland1979RobustLW,
  title={Robust Locally Weighted Regression and Smoothing Scatterplots},
  author={William S. Cleveland},
  journal={Journal of the American Statistical Association},
  year={1979},
  volume={74},
  pages={829-836},
  url={https://api.semanticscholar.org/CorpusID:31665444}
}

@ARTICLE{2016ApJ...830...83C,
       author = {{Conselice}, Christopher J. and {Wilkinson}, Aaron and {Duncan}, Kenneth and {Mortlock}, Alice},
        title = "{The Evolution of Galaxy Number Density at z < 8 and Its Implications}",
      journal = {\apj},
         year = 2016,
        month = oct,
       volume = {830},
       number = {2},
          eid = {83},
        pages = {83},
          doi = {10.3847/0004-637X/830/2/83},
archivePrefix = {arXiv},
       eprint = {1607.03909},
 primaryClass = {astro-ph.GA},
       adsurl = {https://ui.adsabs.harvard.edu/abs/2016ApJ...830...83C}
}

@ARTICLE{2019ApJ...873..111I,
       author = {{Ivezi{\'c}}, {\v{Z}}eljko and {Kahn}, Steven M. and {Tyson}, J. Anthony and others},
        title = "{LSST: From Science Drivers to Reference Design and Anticipated Data Products}",
      journal = {\apj},
         year = 2019,
        month = mar,
       volume = {873},
       number = {2},
          eid = {111},
        pages = {111},
          doi = {10.3847/1538-4357/ab042c},
archivePrefix = {arXiv},
       eprint = {0805.2366},
 primaryClass = {astro-ph},
       adsurl = {https://ui.adsabs.harvard.edu/abs/2019ApJ...873..111I}
}

@article{Gerosa:2020xly,
    author = "Gerosa, Davide and Rosotti, Giovanni and Barbieri, Riccardo",
    title = "{The Bardeen{\textendash}Petterson effect in accreting supermassive black hole binaries: a systematic approach}",
    eprint = "2004.02894",
    archivePrefix = "arXiv",
    primaryClass = "astro-ph.GA",
    doi = "10.1093/mnras/staa1693",
    journal = "Mon. Not. Roy. Astron. Soc.",
    volume = "496",
    number = "3",
    pages = "3060--3075",
    year = "2020"
}

@article{Galaudage:2021rkt,
    author = "Galaudage, Shanika and others",
    title = "{Building Better Spin Models for Merging Binary Black Holes: Evidence for Nonspinning and Rapidly Spinning Nearly Aligned Subpopulations}",
    eprint = "2109.02424",
    archivePrefix = "arXiv",
    primaryClass = "gr-qc",
    doi = "10.3847/2041-8213/ac2f3c",
    journal = "Astrophys. J. Lett.",
    volume = "921",
    number = "1",
    pages = "L15",
    year = "2021",
    note = "[Erratum: Astrophys.J.Lett. 936, L18 (2022), Erratum: Astrophys.J. 936, L18 (2022)]"
}

@article{Hotokezaka:2017esv,
    author = "Hotokezaka, Kenta and Piran, Tsvi",
    title = "{Implications of the low binary black hole aligned spins observed by LIGO}",
    eprint = "1702.03952",
    archivePrefix = "arXiv",
    primaryClass = "astro-ph.HE",
    doi = "10.3847/1538-4357/aa6f61",
    journal = "Astrophys. J.",
    volume = "842",
    number = "2",
    pages = "111",
    year = "2017"
}

@inbook{Piran:2018bbt,
    author = "Piran, Tsvi and Hotekezaka, Kenta",
    editor = "Brink, Lars and Mukhanov, Viatcheslav and Rabinovici, Eliezer and Phua, K. K.",
    title = "{Who Ordered That? On the Origin of LIGO{\textquoteright}s Merging Binary Black Holes}",
    booktitle = "{Jacob Bekenstein}: {The Conservative Revolutionary}",
    eprint = "1807.01336",
    archivePrefix = "arXiv",
    primaryClass = "astro-ph.HE",
    doi = "10.1142/9789811203961_0019",
    publisher = "World Scientific",
    address = "Singapur",
    pages = "243--257",
    year = "2020"
}

@article{Kapil:2026hyn,
    author = "Kapil, Veome and Mandel, Ilya and Riley, Jeff and Grishin, Evgeni and Fuller, Jim and Berti, Emanuele",
    title = "{Modern tidal interaction models for rapid binary population synthesis: II. Binary black hole formation, mergers, and spins}",
    eprint = "2606.23773",
    archivePrefix = "arXiv",
    primaryClass = "astro-ph.HE",
    month = "6",
    year = "2026",
    journal = "",
}

@ARTICLE{Planck2020,
        collaboration = {Planck},
       author = {{Aghanim}, N. and {Akrami}, Y. and {Ashdown}, M. and others},
        title = "{Planck 2018 results. VI. Cosmological parameters}",
      journal = {\aap},
         year = 2020,
        month = sep,
       volume = {641},
          eid = {A6},
        pages = {A6},
          doi = {10.1051/0004-6361/201833910},
archivePrefix = {arXiv},
       eprint = {1807.06209},
 primaryClass = {astro-ph.CO},
       adsurl = {https://ui.adsabs.harvard.edu/abs/2020A&A...641A...6P}
}

@ARTICLE{Lattimer2007,
       author = {{Lattimer}, James M. and {Prakash}, Madappa},
        title = "{Neutron star observations: Prognosis for equation of state constraints}",
      journal = {\physrep},
         year = 2007,
        month = apr,
       volume = {442},
       number = {1-6},
        pages = {109-165},
          doi = {10.1016/j.physrep.2007.02.003},
archivePrefix = {arXiv},
       eprint = {astro-ph/0612440},
 primaryClass = {astro-ph},
       adsurl = {https://ui.adsabs.harvard.edu/abs/2007PhR...442..109L}
}

@ARTICLE{Peters1963,
       author = {{Peters}, P.~C. and {Mathews}, J.},
        title = "{Gravitational Radiation from Point Masses in a Keplerian Orbit}",
      journal = {Physical Review},
         year = 1963,
        month = jul,
       volume = {131},
       number = {1},
        pages = {435-440},
          doi = {10.1103/PhysRev.131.435},
       adsurl = {https://ui.adsabs.harvard.edu/abs/1963PhRv..131..435P}
}

@ARTICLE{Rein2012REBOUND,
       author = {{Rein}, H. and {Liu}, S.-F.},
        title = "{REBOUND: an open-source multi-purpose N-body code for collisional dynamics}",
      journal = {\aap},
         year = 2012,
        month = jan,
       volume = {537},
          eid = {A128},
        pages = {A128},
          doi = {10.1051/0004-6361/201118085},
archivePrefix = {arXiv},
       eprint = {1110.4876},
 primaryClass = {astro-ph.EP},
       adsurl = {https://ui.adsabs.harvard.edu/abs/2012A&A...537A.128R}
}

@INPROCEEDINGS{TraniTsunami,
       author = {{Trani}, Alessandro A. and {Spera}, Mario},
        title = "{Modeling gravitational few-body problems with tsunami and okinami}",
    booktitle = {The Predictive Power of Computational Astrophysics as a Discover Tool},
         year = 2023,
       editor = {{Bisikalo}, Dmitry and {Wiebe}, Dmitri and {Boily}, Christian},
       series = {IAU Symposium},
       volume = {362},
        month = jan,
        pages = {404-409},
          doi = {10.1017/S1743921322001818},
archivePrefix = {arXiv},
       eprint = {2206.10583},
 primaryClass = {astro-ph.HE},
       adsurl = {https://ui.adsabs.harvard.edu/abs/2023IAUS..362..404T}
}

@ARTICLE{Perna:2025,
       author = {{Perna}, Rosalba and {Gottlieb}, Ore and {Shukla}, Estuti and {Radice}, David},
        title = "{Connecting GRBs from binary neutron star mergers to nuclear properties of neutron stars}",
      journal = {\prd},
         year = 2025,
        month = mar,
       volume = {111},
       number = {6},
          eid = {063015},
        pages = {063015},
          doi = {10.1103/PhysRevD.111.063015},
archivePrefix = {arXiv},
       eprint = {2412.07846},
 primaryClass = {astro-ph.HE},
       adsurl = {https://ui.adsabs.harvard.edu/abs/2025PhRvD.111f3015P}
}

@ARTICLE{1976A&A....53..259A,
       author = {{Aarseth}, S.~J. and {Heggie}, D.~C.},
        title = "{The probability of binary formation by three-body encounters.}",
      journal = {\aap},
         year = 1976,
        month = dec,
       volume = {53},
        pages = {259-265},
       adsurl = {https://ui.adsabs.harvard.edu/abs/1976A&A....53..259A}
}

@article{Antonini:2016gqe,
    author = "Antonini, Fabio and Rasio, Frederic A.",
    title = "{Merging black hole binaries in galactic nuclei: implications for advanced-LIGO detections}",
    eprint = "1606.04889",
    archivePrefix = "arXiv",
    primaryClass = "astro-ph.HE",
    doi = "10.3847/0004-637X/831/2/187",
    journal = "Astrophys. J.",
    volume = "831",
    number = "2",
    pages = "187",
    year = "2016"
}

@article{Quinlan:1996vp,
    author = "Quinlan, Gerald D.",
    title = "{The dynamical evolution of massive black hole binaries - I. hardening in a fixed stellar background}",
    eprint = "astro-ph/9601092",
    archivePrefix = "arXiv",
    reportNumber = "RUTGERS-ASTROPHYSICS-PREPRINT-SERIES-NO-187",
    doi = "10.1016/S1384-1076(96)00003-6",
    journal = "New Astron.",
    volume = "1",
    pages = "35--56",
    year = "1996"
}

@article{Morscher:2014doa,
    author = "Morscher, Meagan and Pattabiraman, Bharath and Rodriguez, Carl and Rasio, Frederic A. and Umbreit, Stefan",
    title = "{The Dynamical Evolution of Stellar Black Holes in Globular Clusters}",
    eprint = "1409.0866",
    archivePrefix = "arXiv",
    primaryClass = "astro-ph.GA",
    doi = "10.1088/0004-637X/800/1/9",
    journal = "Astrophys. J.",
    volume = "800",
    number = "1",
    pages = "9",
    year = "2015"
}

@article{Wiringa:1984tg,
    author    = {Wiringa, R. B. and Smith, R. A. and Ainsworth, T. L.},
    title     = {Nucleon-nucleon potentials with and without delta (1232) degrees of freedom},
    journal   = {Phys. Rev. C},
    volume    = {29},
    pages     = {1207--1221},
    year      = {1984},
    doi       = {10.1103/PhysRevC.29.1207}
}

@ARTICLE{BerryGair2010,
       author = {{Berry}, Christopher P.~L. and {Gair}, Jonathan R.},
        title = "{Gravitational wave energy spectrum of a parabolic encounter}",
      journal = {\prd},
         year = 2010,
        month = nov,
       volume = {82},
       number = {10},
          eid = {107501},
        pages = {107501},
          doi = {10.1103/PhysRevD.82.107501},
archivePrefix = {arXiv},
       eprint = {1010.3865},
 primaryClass = {gr-qc},
       adsurl = {https://ui.adsabs.harvard.edu/abs/2010PhRvD..82j7501B}
}

@ARTICLE{Ho:2023,
       author = {{Ho}, Anna Y.~Q. and {Perley}, Daniel A. and {Gal-Yam}, Avishay and others},
        title = "{A Search for Extragalactic Fast Blue Optical Transients in ZTF and the Rate of AT2018cow-like Transients}",
      journal = {\apj},
         year = 2023,
        month = jun,
       volume = {949},
       number = {2},
          eid = {120},
        pages = {120},
          doi = {10.3847/1538-4357/acc533},
archivePrefix = {arXiv},
       eprint = {2105.08811},
 primaryClass = {astro-ph.HE},
       adsurl = {https://ui.adsabs.harvard.edu/abs/2023ApJ...949..120H}
}

@ARTICLE{Liu:2025,
       author = {{Liu}, Bin and {Lai}, Dong},
        title = "{Hierarchical Black Hole Mergers in Nuclear Star Clusters: A Combined Dynamical-Secular Channel for GW231123-like Events}",
      journal = {arXiv e-prints},
         year = 2025,
        month = nov,
          eid = {arXiv:2511.13820},
        pages = {arXiv:2511.13820},
          doi = {10.48550/arXiv.2511.13820},
archivePrefix = {arXiv},
       eprint = {2511.13820},
 primaryClass = {astro-ph.HE},
       adsurl = {https://ui.adsabs.harvard.edu/abs/2025arXiv251113820L}
}

@ARTICLE{Yang:2026,
       author = {{Yang}, William Y.~W. and {Kremer}, Kyle and {Lombardi}, Jr., James C. and {Dage}, Kristen C.},
        title = "{Formation of Black Hole-White Dwarf X-ray Binaries in Globular Clusters}",
      journal = {arXiv e-prints},
         year = 2026,
        month = jun,
          eid = {arXiv:2606.25030},
        pages = {arXiv:2606.25030},
          doi = {10.48550/arXiv.2606.25030},
archivePrefix = {arXiv},
       eprint = {2606.25030},
 primaryClass = {astro-ph.HE},
       adsurl = {https://ui.adsabs.harvard.edu/abs/2026arXiv260625030Y}
}

@ARTICLE{Ryu:2024,
       author = {{Ryu}, Taeho and {de Mink}, Selma E. and {Farmer}, Rob and {Pakmor}, R{\"u}diger and {Perna}, Rosalba and {Springel}, Volker},
        title = "{Close encounters of star-black hole binaries with single stars}",
      journal = {\mnras},
         year = 2024,
        month = jan,
       volume = {527},
       number = {2},
        pages = {2734-2749},
          doi = {10.1093/mnras/stad3082},
archivePrefix = {arXiv},
       eprint = {2307.03097},
 primaryClass = {astro-ph.HE},
       adsurl = {https://ui.adsabs.harvard.edu/abs/2024MNRAS.527.2734R}
}

@article{Ryu:2020gxf,
    author = "Ryu, Taeho and Krolik, Julian and Piran, Tsvi",
    title = "{Measuring stellar and black hole masses of tidal disruption events}",
    eprint = "2007.13765",
    archivePrefix = "arXiv",
    primaryClass = "astro-ph.HE",
    doi = "10.3847/1538-4357/abbf4d",
    journal = "Astrophys. J.",
    volume = "904",
    number = "1",
    pages = "73",
    year = "2020"
}

@ARTICLE{Ryu:2023b,
       author = {{Ryu}, Taeho and {Valli}, Ruggero and {Pakmor}, R{\"u}diger and {Perna}, Rosalba and {de Mink}, Selma E. and {Springel}, Volker},
        title = "{Close encounters of black hole-star binaries with stellar-mass black holes}",
      journal = {\mnras},
         year = 2023,
        month = nov,
       volume = {525},
       number = {4},
        pages = {5752-5766},
          doi = {10.1093/mnras/stad1943},
archivePrefix = {arXiv},
       eprint = {2304.01792},
 primaryClass = {astro-ph.HE},
       adsurl = {https://ui.adsabs.harvard.edu/abs/2023MNRAS.525.5752R}
}

@ARTICLE{Ryu:2023a,
       author = {{Ryu}, Taeho and {Perna}, Rosalba and {Pakmor}, Ruediger and {Ma}, Jing-Ze and {Farmer}, Rob and {de Mink}, Selma E.},
        title = "{Close encounters of tight binary stars with stellar-mass black holes}",
      journal = {\mnras},
         year = 2023,
        month = mar,
       volume = {519},
       number = {4},
        pages = {5787-5799},
          doi = {10.1093/mnras/stad079},
archivePrefix = {arXiv},
       eprint = {2211.02734},
 primaryClass = {astro-ph.HE},
       adsurl = {https://ui.adsabs.harvard.edu/abs/2023MNRAS.519.5787R}
}

@ARTICLE{Perna:2008,
       author = {{Perna}, Rosalba and {Soria}, Roberto and {Pooley}, Dave and {Stella}, Luigi},
        title = "{How rapidly do neutron stars spin at birth? Constraints from archival X-ray observations of extragalactic supernovae}",
      journal = {\mnras},
         year = 2008,
        month = mar,
       volume = {384},
       number = {4},
        pages = {1638-1648},
          doi = {10.1111/j.1365-2966.2007.12821.x},
archivePrefix = {arXiv},
       eprint = {0712.1040},
 primaryClass = {astro-ph},
       adsurl = {https://ui.adsabs.harvard.edu/abs/2008MNRAS.384.1638P}
}

@article{Belgacem:2024ohp,
    author = "Belgacem, Enis and Iacovelli, Francesco and Maggiore, Michele and Mancarella, Michele and Muttoni, Niccol{\`o}",
    title = "{The spectral density of astrophysical stochastic backgrounds}",
    eprint = "2411.04028",
    archivePrefix = "arXiv",
    primaryClass = "gr-qc",
    doi = "10.1088/1475-7516/2025/04/032",
    journal = "JCAP",
    number = "04",
    volume = "2025",
    pages = "032",
    year = "2025"
}

@article{Toscani:2025uar,
    author = "Toscani, Martina and Broggi, Luca and Sesana, Alberto and Rossi, Elena Maria",
    title = "{Updated predictions for gravitational wave emission from tidal disruption events for next-generation observatories}",
    eprint = "2505.22516",
    archivePrefix = "arXiv",
    primaryClass = "astro-ph.HE",
    doi = "10.1051/0004-6361/202555648",
    journal = "Astron. Astrophys.",
    volume = "703",
    pages = "A75",
    year = "2025"
}

@BOOK{2008gady.book.....B,
       author = {{Binney}, James and {Tremaine}, Scott},
        title = "{Galactic Dynamics: Second Edition}",
         year = 2008,
       adsurl = {https://ui.adsabs.harvard.edu/abs/2008gady.book.....B},
      publisher = {Princeton University Press},
}

@article{KAGRA:2013rdx,
    author = "Abbott, B. P. and others",
    collaboration = "KAGRA, LIGO Scientific, Virgo",
    title = "{Prospects for observing and localizing gravitational-wave transients with Advanced LIGO, Advanced Virgo and KAGRA}",
    eprint = "1304.0670",
    archivePrefix = "arXiv",
    primaryClass = "gr-qc",
    reportNumber = "LIGO-P1200087, VIR-0288A-12, JGW-P1808427",
    doi = "10.1007/s41114-020-00026-9",
    journal = "Living Rev. Rel.",
    volume = "19",
    pages = "1",
    year = "2016"
}

@misc{noise_curve_Aplusdes,
    howpublished = "\url{https://dcc.ligo.org/LIGO-T2000012-v1/public}",
    title = "Noise curves used for Simulations in the update of the Observing Scenarios Paper", 
    author = "{LIGO Scientific, Virgo, KAGRA Collaboration}",
    year = 2022,
}

@article{Virgo:2014yos,
    author = "Acernese, F. and others",
    collaboration = "Virgo",
    title = "{Advanced Virgo: a second-generation interferometric gravitational wave detector}",
    eprint = "1408.3978",
    archivePrefix = "arXiv",
    primaryClass = "gr-qc",
    doi = "10.1088/0264-9381/32/2/024001",
    journal = "Class. Quant. Grav.",
    volume = "32",
    number = "2",
    pages = "024001",
    year = "2015"
}

@article{LIGOScientific:2014pky,
    author = "Aasi, J. and others",
    collaboration = "LIGO Scientific",
    title = "{Advanced LIGO}",
    eprint = "1411.4547",
    archivePrefix = "arXiv",
    primaryClass = "gr-qc",
    doi = "10.1088/0264-9381/32/7/074001",
    journal = "Class. Quant. Grav.",
    volume = "32",
    pages = "074001",
    year = "2015"
}

@article{Punturo:2010zz,
    author = "Punturo, M. and others",
    editor = "Ricci, Fulvio",
    title = "{The Einstein Telescope: A third-generation gravitational wave observatory}",
    doi = "10.1088/0264-9381/27/19/194002",
    journal = "Class. Quant. Grav.",
    volume = "27",
    pages = "194002",
    year = "2010"
}

@article{Hild:2010id,
	title        = {{Sensitivity Studies for Third-Generation Gravitational Wave Observatories}},
	author       = {Hild, S. and others},
	year         = 2011,
	journal      = {Class. Quantum Grav.},
	volume       = 28,
	pages        = {094013},
	doi          = {10.1088/0264-9381/28/9/094013},
	eprint       = {1012.0908},
	archiveprefix = {arXiv},
	primaryclass = {gr-qc}
}

@article{ET:2025xjr,
    author = "Abac, Adrian and others",
    collaboration = "ET",
    title = "{The Science of the Einstein Telescope}",
    eprint = "2503.12263",
    archivePrefix = "arXiv",
    primaryClass = "gr-qc",
    reportNumber = "ET-0036C-25",
    doi = "10.1088/1475-7516/2026/03/081",
    journal = "JCAP",
    volume = "2026",
    number = "03",
    pages = "081",
    year = "2026"
}

@article{LISA:2017pwj,
    author = "Amaro-Seoane, Pau and others",
    collaboration = "LISA",
    title = "{Laser Interferometer Space Antenna}",
    eprint = "1702.00786",
    archivePrefix = "arXiv",
    primaryClass = "astro-ph.IM",
    month = "2",
    year = "2017",
    journal = "",
}

@article{LISA:2024hlh,
    author = "Colpi, Monica and others",
    collaboration = "LISA",
    title = "{LISA Definition Study Report}",
    eprint = "2402.07571",
    archivePrefix = "arXiv",
    primaryClass = "astro-ph.CO",
    month = "2",
    year = "2024",
    journal = "",
}

@article{Babak:2021mhe,
    author = "Babak, Stanislav and Petiteau, Antoine and Hewitson, Martin",
    title = "{LISA Sensitivity and SNR Calculations}",
    eprint = "2108.01167",
    archivePrefix = "arXiv",
    primaryClass = "astro-ph.IM",
    reportNumber = "LISA-LCST-SGS-TN-001",
    month = "8",
    year = "2021",
    journal = "",
}

@article{Iacovelli:2022mbg,
    author = "Iacovelli, Francesco and Mancarella, Michele and Foffa, Stefano and Maggiore, Michele",
    title = "{GWFAST: A Fisher Information Matrix Python Code for Third-generation Gravitational-wave Detectors}",
    eprint = "2207.06910",
    archivePrefix = "arXiv",
    primaryClass = "astro-ph.IM",
    doi = "10.3847/1538-4365/ac9129",
    journal = "Astrophys. J. Supp.",
    volume = "263",
    number = "1",
    pages = "2",
    year = "2022"
}

@article{Iacovelli:2025kwn,
    author = "Iacovelli, Francesco and Tissino, Jacopo and Harms, Jan and Berti, Emanuele",
    title = "{Gravitational-wave parameter estimation to the Moon and back: Massive binaries and the case of GW231123}",
    eprint = "2512.09978",
    archivePrefix = "arXiv",
    primaryClass = "gr-qc",
    doi = "10.1103/d2rh-btf9",
    journal = "Phys. Rev. D",
    volume = "113",
    number = "12",
    pages = "123062",
    year = "2026"
}

@article{Marsat:2020rtl,
    author = "Marsat, Sylvain and Baker, John G. and Dal Canton, Tito",
    title = "{Exploring the Bayesian parameter estimation of binary black holes with LISA}",
    eprint = "2003.00357",
    archivePrefix = "arXiv",
    primaryClass = "gr-qc",
    doi = "10.1103/PhysRevD.103.083011",
    journal = "Phys. Rev. D",
    volume = "103",
    number = "8",
    pages = "083011",
    year = "2021"
}

@article{Fumagalli:2025asw,
    author = "Fumagalli, Giulia and Gerosa, Davide and Loutrel, Nicholas",
    title = "{precession 2.1: black-hole binary spin precession on eccentric orbits}",
    eprint = "2508.21125",
    archivePrefix = "arXiv",
    primaryClass = "gr-qc",
    doi = "10.1088/1742-6596/3177/1/012117",
    journal = "J. Phys. Conf. Ser.",
    volume = "3177",
    number = "1",
    pages = "012117",
    year = "2026"
}

@ARTICLE{1964PhRv..136.1224P,
       author = {{Peters}, P.~C.},
        title = "{Gravitational Radiation and the Motion of Two Point Masses}",
      journal = {Physical Review},
         year = 1964,
        month = nov,
       volume = {136},
       number = {4B},
        pages = {1224-1232},
          doi = {10.1103/PhysRev.136.B1224},
       adsurl = {https://ui.adsabs.harvard.edu/abs/1964PhRv..136.1224P}
}

@article{Hamers:2021eir,
    author = "Hamers, Adrian S.",
    title = "{An Improved Numerical Fit to the Peak Harmonic Gravitational Wave Frequency Emitted by an Eccentric Binary}",
    eprint = "2111.08033",
    archivePrefix = "arXiv",
    primaryClass = "gr-qc",
    doi = "10.3847/2515-5172/ac3d98",
    journal = "Res. Notes AAS",
    volume = "5",
    number = "11",
    pages = "275",
    year = "2021"
}

@article{Morras:2025nlp,
    author = "Morras, Gonzalo and Pratten, Geraint and Schmidt, Patricia",
    title = "{Improved post-Newtonian waveform model for inspiralling precessing-eccentric compact binaries}",
    eprint = "2502.03929",
    archivePrefix = "arXiv",
    primaryClass = "gr-qc",
    reportNumber = "IFT-UAM/CSIC-25-12",
    doi = "10.1103/PhysRevD.111.084052",
    journal = "Phys. Rev. D",
    volume = "111",
    number = "8",
    pages = "084052",
    year = "2025"
}

@article{Prince:2002hp,
    author = "Prince, Thomas A. and Tinto, Massimo and Larson, Shane L. and Armstrong, J. W.",
    title = "{The LISA optimal sensitivity}",
    eprint = "gr-qc/0209039",
    archivePrefix = "arXiv",
    doi = "10.1103/PhysRevD.66.122002",
    journal = "Phys. Rev. D",
    volume = "66",
    pages = "122002",
    year = "2002"
}

@article{Tissino:2026skt,
    author = "Tissino, Jacopo and others",
    title = "{The geometry of lunar gravitational wave detection}",
    eprint = "2606.04918",
    archivePrefix = "arXiv",
    primaryClass = "gr-qc",
    month = "6",
    year = "2026",
    journal = "",
}

@article{Iacovelli:2022bbs,
    author = "Iacovelli, Francesco and Mancarella, Michele and Foffa, Stefano and Maggiore, Michele",
    title = "{Forecasting the Detection Capabilities of Third-generation Gravitational-wave Detectors Using GWFAST}",
    eprint = "2207.02771",
    archivePrefix = "arXiv",
    primaryClass = "gr-qc",
    doi = "10.3847/1538-4357/ac9cd4",
    journal = "Astrophys. J.",
    volume = "941",
    number = "2",
    pages = "208",
    year = "2022"
}

@article{Reitze:2019iox,
    author = "Reitze, David and others",
    title = "{Cosmic Explorer: The U.S. Contribution to Gravitational-Wave Astronomy beyond LIGO}",
    eprint = "1907.04833",
    archivePrefix = "arXiv",
    primaryClass = "astro-ph.IM",
    reportNumber = "LIGO-P1900316",
    journal = "Bull. Am. Astron. Soc.",
    volume = "51",
    number = "7",
    pages = "035",
    year = "2019"
}

@article{Evans:2021gyd,
    author = "Evans, Matthew and others",
    title = "{A Horizon Study for Cosmic Explorer: Science, Observatories, and Community}",
    eprint = "2109.09882",
    archivePrefix = "arXiv",
    primaryClass = "astro-ph.IM",
    reportNumber = "CE-P2100003-v7, Cosmic Explorer technical report CE-P2100003-v6",
    month = "9",
    year = "2021",
    journal = ""
}

@article{Evans:2023euw,
    author = "Evans, Matthew and others",
    title = "{Cosmic Explorer: A Submission to the NSF MPSAC ngGW Subcommittee}",
    eprint = "2306.13745",
    archivePrefix = "arXiv",
    primaryClass = "astro-ph.IM",
    month = "6",
    year = "2023",
    journal = ""
}

@article{Chandra:2024dhf,
    author = "Chandra, Koustav",
    title = "{gwforge: a user-friendly package to generate gravitational-wave mock data}",
    eprint = "2407.21109",
    archivePrefix = "arXiv",
    primaryClass = "gr-qc",
    doi = "10.1088/1361-6382/ad9b68",
    journal = "Class. Quant. Grav.",
    volume = "42",
    number = "2",
    pages = "025003",
    year = "2025"
}

@misc{noise_curves_gwforge,
    howpublished = "\url{https://github.com/koustavchandra/gwforge/tree/main/GWForge/ifo/noise_curves}",
    author = "Chandra, Koustav",
    year = 2025,
}

@article{LGWA:2020mma,
    author = "Harms, Jan and others",
    collaboration = "LGWA",
    title = "{Lunar Gravitational-wave Antenna}",
    eprint = "2010.13726",
    archivePrefix = "arXiv",
    primaryClass = "gr-qc",
    doi = "10.3847/1538-4357/abe5a7",
    journal = "Astrophys. J.",
    volume = "910",
    number = "1",
    pages = "1",
    year = "2021"
}

@article{Ajith:2024mie,
    author = "Ajith, Parameswaran and others",
    title = "{The Lunar Gravitational-wave Antenna: mission studies and science case}",
    eprint = "2404.09181",
    archivePrefix = "arXiv",
    primaryClass = "gr-qc",
    doi = "10.1088/1475-7516/2025/01/108",
    journal = "JCAP",
    volume = {2025},
    number = {01},
    pages = {108},
    year = "2025"
}

@ARTICLE{Ryu:2022,
       author = {{Ryu}, Taeho and {Perna}, Rosalba and {Wang}, Yi-Han},
        title = "{Close encounters of stars with stellar-mass black hole binaries}",
      journal = {\mnras},
         year = 2022,
        month = oct,
       volume = {516},
       number = {2},
        pages = {2204-2217},
          doi = {10.1093/mnras/stac2316},
archivePrefix = {arXiv},
       eprint = {2206.00603},
 primaryClass = {astro-ph.HE},
       adsurl = {https://ui.adsabs.harvard.edu/abs/2022MNRAS.516.2204R}
}

@article{Breen:2013vla,
    author = "Breen, Philip G. and Heggie, Douglas C.",
    title = "{Dynamical evolution of black hole sub-systems in idealised star clusters}",
    eprint = "1304.3401",
    archivePrefix = "arXiv",
    primaryClass = "astro-ph.GA",
    doi = "10.1093/mnras/stt628",
    journal = "Mon. Not. Roy. Astron. Soc.",
    volume = "432",
    pages = "2779",
    year = "2013"
}

@article{Fryer:2011cx,
    author = "Fryer, Chris L. and Belczynski, Krzysztof and Wiktorowicz, Grzegorz and Dominik, Michal and Kalogera, Vicky and Holz, Daniel E.",
    title = "{Compact Remnant Mass Function: Dependence on the Explosion Mechanism and Metallicity}",
    eprint = "1110.1726",
    archivePrefix = "arXiv",
    primaryClass = "astro-ph.SR",
    reportNumber = "LA-UR-11-02622",
    doi = "10.1088/0004-637X/749/1/91",
    journal = "Astrophys. J.",
    volume = "749",
    pages = "91",
    year = "2012"
}

@article{Kroupa:2002ky,
    author = "Kroupa, Pavel",
    title = "{The Initial mass function of stars: Evidence for uniformity in variable systems}",
    eprint = "astro-ph/0201098",
    archivePrefix = "arXiv",
    doi = "10.1126/science.1067524",
    journal = "Science",
    volume = "295",
    pages = "82--91",
    year = "2002"
}

@article{Hurley:2000pk,
    author = "Hurley, Jarrod R. and Pols, Onno R. and Tout, Christopher A.",
    title = "{Comprehensive analytic formulae for stellar evolution as a function of mass and metallicity}",
    eprint = "astro-ph/0001295",
    archivePrefix = "arXiv",
    doi = "10.1046/j.1365-8711.2000.03426.x",
    journal = "Mon. Not. Roy. Astron. Soc.",
    volume = "315",
    pages = "543",
    year = "2000"
}

@inproceedings{Rasio:2003sz,
    author = "Rasio, Frederic A. and Freitag, Marc and Atakan Gurkan, M.",
    title = "{Formation of massive black holes in dense star clusters}",
    booktitle = "{Carnegie Observatories Centennial Symposium. 1. Coevolution of Black Holes and Galaxies}",
    eprint = "astro-ph/0304038",
    archivePrefix = "arXiv",
    month = "4",
    year = "2003"
}

@article{Miller:2012ys,
    author = "Miller, M. Coleman and Davies, Melvyn B.",
    title = "{An upper limit to the velocity dispersion of relaxed stellar systems without massive black holes}",
    eprint = "1206.6167",
    archivePrefix = "arXiv",
    primaryClass = "astro-ph.GA",
    doi = "10.1088/0004-637X/755/1/81",
    journal = "Astrophys. J.",
    volume = "755",
    pages = "81",
    year = "2012"
}

@article{Rocha:2023xwp,
    author = "Rocha, L{\'\i}via S. and Horvath, Jorge E. and de S{\'a}, Lucas M. and Chinen, Gustavo Y. and Bar{\~a}o, Lucas G. and de Avellar, Marcio G. B.",
    title = "{Mass Distribution and Maximum Mass of Neutron Stars: Effects of Orbital Inclination Angle}",
    eprint = "2312.13244",
    archivePrefix = "arXiv",
    primaryClass = "astro-ph.HE",
    doi = "10.3390/universe10010003",
    journal = "Universe",
    volume = "10",
    number = "1",
    pages = "3",
    year = "2024"
}

@ARTICLE{2007MNRAS.375.1315K,
       author = {{Kepler}, S.~O. and {Kleinman}, S.~J. and {Nitta}, A. and {Koester}, D. and {Castanheira}, B.~G. and {Giovannini}, O. and {Costa}, A.~F.~M. and {Althaus}, L.},
        title = "{White dwarf mass distribution in the SDSS}",
      journal = {\mnras},
         year = 2007,
        month = mar,
       volume = {375},
       number = {4},
        pages = {1315-1324},
          doi = {10.1111/j.1365-2966.2006.11388.x},
archivePrefix = {arXiv},
       eprint = {astro-ph/0612277},
 primaryClass = {astro-ph},
       adsurl = {https://ui.adsabs.harvard.edu/abs/2007MNRAS.375.1315K}
}

@article{Foucart:2018rjc,
    author = "Foucart, Francois and Hinderer, Tanja and Nissanke, Samaya",
    title = "{Remnant baryon mass in neutron star-black hole mergers: Predictions for binary neutron star mimickers and rapidly spinning black holes}",
    eprint = "1807.00011",
    archivePrefix = "arXiv",
    primaryClass = "astro-ph.HE",
    doi = "10.1103/PhysRevD.98.081501",
    journal = "Phys. Rev. D",
    volume = "98",
    number = "8",
    pages = "081501",
    year = "2018"
}

@article{Pannarale:2015jia,
    author = "Pannarale, Francesco and Berti, Emanuele and Kyutoku, Koutarou and Lackey, Benjamin D. and Shibata, Masaru",
    title = "{Gravitational-wave cutoff frequencies of tidally disruptive neutron star-black hole binary mergers}",
    eprint = "1509.06209",
    archivePrefix = "arXiv",
    primaryClass = "gr-qc",
    doi = "10.1103/PhysRevD.92.081504",
    journal = "Phys. Rev. D",
    volume = "92",
    number = "8",
    pages = "081504",
    year = "2015"
}

@article{Gerosa:2013laa,
    author = "Gerosa, Davide and Kesden, Michael and Berti, Emanuele and O'Shaughnessy, Richard and Sperhake, Ulrich",
    title = "{Resonant-plane locking and spin alignment in stellar-mass black-hole binaries: a diagnostic of compact-binary formation}",
    eprint = "1302.4442",
    archivePrefix = "arXiv",
    primaryClass = "gr-qc",
    doi = "10.1103/PhysRevD.87.104028",
    journal = "Phys. Rev. D",
    volume = "87",
    pages = "104028",
    year = "2013"
}

@article{Lorimer:2001vd,
    author = "Lorimer, D. R.",
    title = "{Binary and millisecond pulsars at the new millennium}",
    eprint = "astro-ph/0104388",
    archivePrefix = "arXiv",
    doi = "10.12942/lrr-2001-5",
    journal = "Living Rev. Rel.",
    volume = "4",
    pages = "5",
    year = "2001"
}

@article{Kapil:2022blf,
    author = {Kapil, Veome and Mandel, Ilya and Berti, Emanuele and M{\"u}ller, Bernhard},
    title = "{Calibration of neutron star natal kick velocities to isolated pulsar observations}",
    eprint = "2209.09252",
    archivePrefix = "arXiv",
    primaryClass = "astro-ph.HE",
    doi = "10.1093/mnras/stad019",
    journal = "Mon. Not. Roy. Astron. Soc.",
    volume = "519",
    number = "4",
    pages = "5893--5901",
    year = "2023"
}

@article{Woosley:2002zz,
    author = "Woosley, S. E. and Heger, A. and Weaver, T. A.",
    title = "{The evolution and explosion of massive stars}",
    doi = "10.1103/RevModPhys.74.1015",
    journal = "Rev. Mod. Phys.",
    volume = "74",
    pages = "1015--1071",
    year = "2002"
}

@article{PortegiesZwart:2002iks,
    author = "Portegies Zwart, Simon F. and McMillan, Steve L. W.",
    title = "{The Runaway growth of intermediate-mass black holes in dense star clusters}",
    eprint = "astro-ph/0201055",
    archivePrefix = "arXiv",
    doi = "10.1086/341798",
    journal = "Astrophys. J.",
    volume = "576",
    pages = "899--907",
    year = "2002"
}

@ARTICLE{2014CQGra..31x4006K,
       author = {{Kruijssen}, J.~M. Diederik},
        title = "{Globular cluster formation in the context of galaxy formation and evolution}",
      journal = {Classical and Quantum Gravity},
         year = 2014,
        month = dec,
       volume = {31},
       number = {24},
          eid = {244006},
        pages = {244006},
          doi = {10.1088/0264-9381/31/24/244006},
archivePrefix = {arXiv},
       eprint = {1407.2953},
 primaryClass = {astro-ph.GA},
       adsurl = {https://ui.adsabs.harvard.edu/abs/2014CQGra..31x4006K}
}

@ARTICLE{1977MNRAS.179..433B,
       author = {{Blandford}, R.~D. and {Znajek}, R.~L.},
        title = "{Electromagnetic extraction of energy from Kerr black holes.}",
      journal = {\mnras},
         year = 1977,
        month = may,
       volume = {179},
        pages = {433-456},
          doi = {10.1093/mnras/179.3.433},
       adsurl = {https://ui.adsabs.harvard.edu/abs/1977MNRAS.179..433B}
}

@ARTICLE{2019ARA&A..57..227K,
       author = {{Krumholz}, Mark R. and {McKee}, Christopher F. and {Bland-Hawthorn}, Joss},
        title = "{Star Clusters Across Cosmic Time}",
      journal = {\araa},
         year = 2019,
        month = aug,
       volume = {57},
        pages = {227-303},
          doi = {10.1146/annurev-astro-091918-104430},
archivePrefix = {arXiv},
       eprint = {1812.01615},
 primaryClass = {astro-ph.GA},
       adsurl = {https://ui.adsabs.harvard.edu/abs/2019ARA&A..57..227K}
}

@ARTICLE{1971ApJ...164..399S,
       author = {{Spitzer}, Jr., Lyman and {Hart}, Michael H.},
        title = "{Random Gravitational Encounters and the Evolution of Spherical Systems. I. Method}",
      journal = {\apj},
         year = 1971,
        month = mar,
       volume = {164},
        pages = {399},
          doi = {10.1086/150855},
       adsurl = {https://ui.adsabs.harvard.edu/abs/1971ApJ...164..399S}
}

@article{Banerjee:2019jjs,
    author = "Banerjee, Sambaran and Belczynski, Krzysztof and Fryer, Christopher L. and Berczik, Peter and Hurley, Jarrod R. and Spurzem, Rainer and Wang, Long",
    title = "{BSE versus StarTrack: implementations of new wind, remnant-formation, and natal-kick schemes in NBODY7 and their astrophysical consequences}",
    eprint = "1902.07718",
    archivePrefix = "arXiv",
    primaryClass = "astro-ph.SR",
    doi = "10.1051/0004-6361/201935332",
    journal = "Astron. Astrophys.",
    volume = "639",
    pages = "A41",
    year = "2020"
}

@misc{RapsterGitHub,
    author = "Kritos, Konstantinos",
    year = "2022",
    howpublished = "\url{https://github.com/Kkritos/Rapster}",
}

@article{Spera:2015vkd,
    author = "Spera, Mario and Mapelli, Michela and Bressan, Alessandro",
    title = "{The mass spectrum of compact remnants from the PARSEC stellar evolution tracks}",
    eprint = "1505.05201",
    archivePrefix = "arXiv",
    primaryClass = "astro-ph.SR",
    doi = "10.1093/mnras/stv1161",
    journal = "Mon. Not. Roy. Astron. Soc.",
    volume = "451",
    number = "4",
    pages = "4086--4103",
    year = "2015"
}

@ARTICLE{2021MNRAS.505.1053W,
       author = {{Wang}, Yi-Han and {Leigh}, Nathan W.~C. and {Liu}, Bin and {Perna}, Rosalba},
        title = "{SpaceHub: A high-performance gravity integration toolkit for few-body problems in astrophysics}",
      journal = {\mnras},
         year = 2021,
        month = jul,
       volume = {505},
       number = {1},
        pages = {1053-1070},
          doi = {10.1093/mnras/stab1189},
archivePrefix = {arXiv},
       eprint = {2104.06413},
 primaryClass = {astro-ph.SR},
       adsurl = {https://ui.adsabs.harvard.edu/abs/2021MNRAS.505.1053W}
}

@article{Stein:2026qmg,
    author = "Stein, Robert and others",
    title = "{TDE 2025abcr: A Tidal Disruption Event in the Outskirts of a Massive Galaxy}",
    eprint = "2602.10180",
    archivePrefix = "arXiv",
    primaryClass = "astro-ph.HE",
    doi = "10.3847/2041-8213/ae77f3",
    journal = "Astrophys. J. Lett.",
    volume = "1006",
    number = "2",
    pages = "L57",
    year = "2026"
}

@article{Li:2025mae,
    author = "Li, Dongyue and others",
    title = "{A fast powerful X-ray transient from possible tidal disruption of a white dwarf}",
    eprint = "2509.25877",
    archivePrefix = "arXiv",
    primaryClass = "astro-ph.HE",
    doi = "10.1016/j.scib.2025.12.050",
    journal = "Sci. Bull.",
    volume = "71",
    pages = "538--546",
    year = "2026"
}

@article{Soria:2017ght,
    author = "Soria, Roberto and Musaeva, Aina and Wu, Kinwah and Zampieri, Luca and Federle, Sara and Urquhart, Ryan and van der Helm, Edwin and Farrell, Sean",
    title = "{Outbursts of the intermediate-mass black hole HLX-1: a wind instability scenario}",
    eprint = "1704.05468",
    archivePrefix = "arXiv",
    primaryClass = "astro-ph.HE",
    doi = "10.1093/mnras/stx888",
    journal = "Mon. Not. Roy. Astron. Soc.",
    volume = "469",
    number = "1",
    pages = "886--905",
    year = "2017"
}

@article{Yao:2025dbw,
    author = "Yao, Yuhan and others",
    title = "{A Massive Black Hole 0.8 kpc from the Host Nucleus Revealed by the Offset Tidal Disruption Event AT2024tvd}",
    eprint = "2502.17661",
    archivePrefix = "arXiv",
    primaryClass = "astro-ph.GA",
    doi = "10.3847/2041-8213/add7de",
    journal = "Astrophys. J. Lett.",
    volume = "985",
    number = "2",
    pages = "L48",
    year = "2025"
}

@article{Lin:2018dev,
    author = "Lin, Dacheng and others",
    title = "{A luminous X-ray outburst from an intermediate-mass black hole in an off-centre star cluster}",
    eprint = "1806.05692",
    archivePrefix = "arXiv",
    primaryClass = "astro-ph.HE",
    doi = "10.1038/s41550-018-0493-1",
    journal = "Nature Astron.",
    volume = "2",
    number = "8",
    pages = "656--661",
    year = "2018"
}

@article{Jin:2025izu,
    author = "Jin, C. -C. and others",
    title = "{An Intermediate-mass Black Hole Lurking in A Galactic Halo Caught Alive during Outburst}",
    eprint = "2501.09580",
    archivePrefix = "arXiv",
    primaryClass = "astro-ph.HE",
    month = "1",
    year = "2025",
    journal = "",
}

@article{Chang:2025ucz,
    author = "Chang, Yi-Chi and Soria, Roberto and Kong, Albert K. H. and Graham, Alister W. and Grishin, Kirill A. and Chilingarian, Igor V.",
    title = "{Multiwavelength Study of a Hyperluminous X-Ray Source near NGC 6099: A Strong IMBH Candidate}",
    eprint = "2503.00904",
    archivePrefix = "arXiv",
    primaryClass = "astro-ph.HE",
    doi = "10.3847/1538-4357/adbbee",
    journal = "Astrophys. J.",
    volume = "983",
    number = "2",
    pages = "109",
    year = "2025"
}

@article{Grotova:2025tdm,
    author = "Grotova, Iuliia and others",
    title = "{The population of tidal disruption events discovered with eROSITA}",
    eprint = "2504.08424",
    archivePrefix = "arXiv",
    primaryClass = "astro-ph.HE",
    doi = "10.1051/0004-6361/202553669",
    journal = "Astron. Astrophys.",
    volume = "697",
    pages = "A159",
    year = "2025"
}

@article{Ryu:2024utf,
    author = "Ryu, Taeho and Perna, Rosalba and Cantiello, Matteo",
    title = "{Tidal Disruption Encores}",
    eprint = "2402.15590",
    archivePrefix = "arXiv",
    primaryClass = "astro-ph.HE",
    doi = "10.3847/2041-8213/ad3946",
    journal = "Astrophys. J. Lett.",
    volume = "965",
    number = "2",
    pages = "L25",
    year = "2024"
}

@article{Gupta:2019nwj,
    author = "Gupta, Anuradha and Gerosa, Davide and Arun, K. G. and Berti, Emanuele and Farr, Will M. and Sathyaprakash, B. S.",
    title = "{Black holes in the low mass gap: Implications for gravitational wave observations}",
    eprint = "1909.05804",
    archivePrefix = "arXiv",
    primaryClass = "gr-qc",
    reportNumber = "LIGO-P1900271",
    doi = "10.1103/PhysRevD.101.103036",
    journal = "Phys. Rev. D",
    volume = "101",
    number = "10",
    pages = "103036",
    year = "2020"
}

@article{Perna:2021fbq,
    author = "Perna, Rosalba and Tagawa, Hiromichi and Haiman, Zoltan and Bartos, Imre",
    title = "{Accretion-Induced Collapse of Neutron Stars in the Disks of Active Galactic Nuclei}",
    eprint = "2103.10963",
    archivePrefix = "arXiv",
    primaryClass = "astro-ph.HE",
    doi = "10.3847/1538-4357/abfdb4",
    journal = "Astrophys. J.",
    volume = "915",
    number = "1",
    pages = "10",
    year = "2021"
}

@article{McPike:2026ugj,
    author = "McPike, Emily and others",
    title = "{McFACTS IV: Electromagnetic Counterparts to AGN Disk Embedded Binary Black Hole Mergers}",
    eprint = "2602.04135",
    archivePrefix = "arXiv",
    primaryClass = "astro-ph.HE",
    month = "2",
    year = "2026",
    journal = "",
}

@article{Toscani:2021bzr,
    author = "Toscani, Martina and Lodato, Giuseppe and Price, Daniel J. and Liptai, David",
    title = "{Gravitational waves from tidal disruption events: an open and comprehensive catalog}",
    eprint = "2111.05145",
    archivePrefix = "arXiv",
    primaryClass = "astro-ph.HE",
    doi = "10.1093/mnras/stab3384",
    journal = "Mon. Not. Roy. Astron. Soc.",
    volume = "510",
    number = "1",
    pages = "992--1001",
    year = "2022"
}

@article{Ivanova:2010ia,
    author = "Ivanova, N. and Chaichenets, S. and Fregeau, J. and Heinke, C. O. and Lombardi, Jr., J. C. and Woods, T.",
    title = "{Formation of black-hole X-ray binaries in globular clusters}",
    eprint = "1001.1767",
    archivePrefix = "arXiv",
    primaryClass = "astro-ph.HE",
    doi = "10.1088/0004-637X/717/2/948",
    journal = "Astrophys. J.",
    volume = "717",
    pages = "948--957",
    year = "2010"
}

@article{Kremer:2023sof,
    author = "Kremer, Kyle and Mockler, Brenna and Piro, Anthony L. and Lombardi, James C.",
    title = "{Wind-reprocessed transients from stellar-mass black hole Tidal Disruption Events}",
    eprint = "2305.08905",
    archivePrefix = "arXiv",
    primaryClass = "astro-ph.HE",
    doi = "10.1093/mnras/stad2239",
    journal = "Mon. Not. Roy. Astron. Soc.",
    volume = "524",
    number = "4",
    pages = "6358--6373",
    year = "2023"
}

@article{Volonteri:2005fj,
    author = "Volonteri, Marta and Rees, Martin J.",
    title = "{Rapid growth of high redshift black holes}",
    eprint = "astro-ph/0506040",
    archivePrefix = "arXiv",
    doi = "10.1086/466521",
    journal = "Astrophys. J.",
    volume = "633",
    pages = "624--629",
    year = "2005"
}

@article{Madau:2016jbv,
    author = "Madau, Piero and Fragos, Tassos",
    title = "{Radiation Backgrounds at Cosmic Dawn: X-Rays from Compact Binaries}",
    eprint = "1606.07887",
    archivePrefix = "arXiv",
    primaryClass = "astro-ph.GA",
    doi = "10.3847/1538-4357/aa6af9",
    journal = "Astrophys. J.",
    volume = "840",
    number = "1",
    pages = "39",
    year = "2017"
}

@article{Yang:2019cbr,
    author = "Yang, Yang and others",
    title = "{Hierarchical Black Hole Mergers in Active Galactic Nuclei}",
    eprint = "1906.09281",
    archivePrefix = "arXiv",
    primaryClass = "astro-ph.HE",
    doi = "10.1103/PhysRevLett.123.181101",
    journal = "Phys. Rev. Lett.",
    volume = "123",
    number = "18",
    pages = "181101",
    year = "2019"
}

@article{Tagawa:2020dxe,
    author = "Tagawa, Hiromichi and Haiman, Zoltan and Bartos, Imre and Kocsis, Bence",
    title = "{Spin Evolution of Stellar-mass Black Hole Binaries in Active Galactic Nuclei}",
    eprint = "2004.11914",
    archivePrefix = "arXiv",
    primaryClass = "astro-ph.HE",
    doi = "10.3847/1538-4357/aba2cc",
    journal = "Astrophys. J.",
    volume = "899",
    number = "1",
    pages = "26",
    year = "2020"
}

@article{Kiroglu:2024xpc,
    author = "K{\i}ro{\u{g}}lu, Fulya and Kremer, Kyle and Biscoveanu, Sylvia and Prieto, Elena Gonz{\'a}lez and Rasio, Frederic A.",
    title = "{Black Hole Accretion and Spin-up through Stellar Collisions in Dense Star Clusters}",
    eprint = "2410.01879",
    archivePrefix = "arXiv",
    primaryClass = "astro-ph.HE",
    doi = "10.3847/1538-4357/ada26b",
    journal = "Astrophys. J.",
    volume = "979",
    number = "2",
    pages = "237",
    year = "2025"
}

@misc{rockfish,
    howpublished = {\url{https://www.arch.jhu.edu/}}
}

@article{Samsing:2018isx,
    author = "Samsing, Johan and D'Orazio, Daniel J.",
    title = "{Black Hole Mergers From Globular Clusters Observable by LISA I: Eccentric Sources Originating From Relativistic $N$-body Dynamics}",
    eprint = "1804.06519",
    archivePrefix = "arXiv",
    primaryClass = "astro-ph.HE",
    doi = "10.1093/mnras/sty2334",
    journal = "Mon. Not. Roy. Astron. Soc.",
    volume = "481",
    number = "4",
    pages = "5445--5450",
    year = "2018"
}

@article{Berry:2010gt,
    author = "Berry, Christopher P. L. and Gair, Jonathan R.",
    title = "{Gravitational wave energy spectrum of a parabolic encounter}",
    eprint = "1010.3865",
    archivePrefix = "arXiv",
    primaryClass = "gr-qc",
    doi = "10.1103/PhysRevD.82.107501",
    journal = "Phys. Rev. D",
    volume = "82",
    pages = "107501",
    year = "2010"
}

@article{LIGOScientific:2020zkf,
    author = "Abbott, R. and others",
    collaboration = "LIGO Scientific, Virgo",
    title = "{GW190814: Gravitational Waves from the Coalescence of a 23 Solar Mass Black Hole with a 2.6 Solar Mass Compact Object}",
    eprint = "2006.12611",
    archivePrefix = "arXiv",
    primaryClass = "astro-ph.HE",
    reportNumber = "LIGO-P190814",
    doi = "10.3847/2041-8213/ab960f",
    journal = "Astrophys. J. Lett.",
    volume = "896",
    number = "2",
    pages = "L44",
    year = "2020"
}

@article{Lattimer:2012nd,
    author = "Lattimer, James M.",
    title = "{The nuclear equation of state and neutron star masses}",
    eprint = "1305.3510",
    archivePrefix = "arXiv",
    primaryClass = "nucl-th",
    doi = "10.1146/annurev-nucl-102711-095018",
    journal = "Ann. Rev. Nucl. Part. Sci.",
    volume = "62",
    pages = "485--515",
    year = "2012"
}

@article{Farr:2010tu,
    author = "Farr, Will M. and Sravan, Niharika and Cantrell, Andrew and Kreidberg, Laura and Bailyn, Charles D. and Mandel, Ilya and Kalogera, Vicky",
    title = "{The Mass Distribution of Stellar-Mass Black Holes}",
    eprint = "1011.1459",
    archivePrefix = "arXiv",
    primaryClass = "astro-ph.GA",
    doi = "10.1088/0004-637X/741/2/103",
    journal = "Astrophys. J.",
    volume = "741",
    pages = "103",
    year = "2011"
}

@article{Olejak:2022zee,
    author = "Olejak, Aleksandra and Fryer, Chris L. and Belczynski, Krzysztof and Baibhav, Vishal",
    title = "{The role of supernova convection for the lower mass gap in the isolated binary formation of gravitational wave sources}",
    eprint = "2204.09061",
    archivePrefix = "arXiv",
    primaryClass = "astro-ph.HE",
    doi = "10.1093/mnras/stac2359",
    journal = "Mon. Not. Roy. Astron. Soc.",
    volume = "516",
    number = "2",
    pages = "2252--2271",
    year = "2022"
}

@article{Rosswog:2009fh,
    author = "Rosswog, S. and Kasen, D. and Guillochon, J. and Ramirez-Ruiz, E.",
    title = "{Collisions of white dwarfs as a new progenitor channel for type Ia supernovae}",
    eprint = "0907.3196",
    archivePrefix = "arXiv",
    primaryClass = "astro-ph.HE",
    doi = "10.1088/0004-637X/705/2/L128",
    journal = "Astrophys. J. Lett.",
    volume = "705",
    pages = "L128--L132",
    year = "2009"
}

@article{Ye:2023fpb,
    author = "Ye, Claire S. and Fragione, Giacomo and Perna, Rosalba",
    title = "{On the Tidal Capture of White Dwarfs by Intermediate-mass Black Holes in Dense Stellar Environments}",
    eprint = "2303.07375",
    archivePrefix = "arXiv",
    primaryClass = "astro-ph.HE",
    doi = "10.3847/1538-4357/ace1eb",
    journal = "Astrophys. J.",
    volume = "953",
    number = "2",
    pages = "141",
    year = "2023"
}

@ARTICLE{2020MNRAS.495.1061F,
       author = {{Fragione}, Giacomo and {Metzger}, Brian D. and {Perna}, Rosalba and {Leigh}, Nathan W.~C. and {Kocsis}, Bence},
        title = "{Electromagnetic transients and gravitational waves from white dwarf disruptions by stellar black holes in triple systems}",
      journal = {\mnras},
         year = 2020,
        month = jun,
       volume = {495},
       number = {1},
        pages = {1061-1072},
          doi = {10.1093/mnras/staa1192},
archivePrefix = {arXiv},
       eprint = {1908.00987},
 primaryClass = {astro-ph.HE},
       adsurl = {https://ui.adsabs.harvard.edu/abs/2020MNRAS.495.1061F}
}

@ARTICLE{2020A&ARv..28....4N,
       author = {{Neumayer}, Nadine and {Seth}, Anil and {B{\"o}ker}, Torsten},
        title = "{Nuclear star clusters}",
      journal = {\aapr},
         year = 2020,
        month = jul,
       volume = {28},
       number = {1},
          eid = {4},
        pages = {4},
          doi = {10.1007/s00159-020-00125-0},
archivePrefix = {arXiv},
       eprint = {2001.03626},
 primaryClass = {astro-ph.GA},
       adsurl = {https://ui.adsabs.harvard.edu/abs/2020A&ARv..28....4N}
}

@article{McKernan:2012rf,
    author = "McKernan, B. and Ford, K. E. S. and Lyra, W. and Perets, H. B.",
    title = "{Intermediate mass black holes in AGN disks: I. Production {\&} Growth}",
    eprint = "1206.2309",
    archivePrefix = "arXiv",
    primaryClass = "astro-ph.GA",
    doi = "10.1111/j.1365-2966.2012.21486.x",
    journal = "Mon. Not. Roy. Astron. Soc.",
    volume = "425",
    pages = "460",
    year = "2012"
}

@article{Bartos:2026xlt,
    author = "Bartos, I. and Haiman, Z.",
    title = "{High-Spin BBH Subpopulation from AGN Accretion}",
    eprint = "2605.09351",
    archivePrefix = "arXiv",
    primaryClass = "astro-ph.HE",
    month = "5",
    year = "2026",
    journal = "",
}

@article{Kritos:2024kpn,
    author = "Kritos, Konstantinos and Reali, Luca and Gerosa, Davide and Berti, Emanuele",
    title = "{Minimum gas mass accreted by spinning intermediate-mass black holes in stellar clusters}",
    eprint = "2409.15439",
    archivePrefix = "arXiv",
    primaryClass = "astro-ph.HE",
    doi = "10.1103/PhysRevD.110.123017",
    journal = "Phys. Rev. D",
    volume = "110",
    number = "12",
    pages = "123017",
    year = "2024"
}

@article{Mandel:2021smh,
    author = "Mandel, Ilya and Broekgaarden, Floor S.",
    title = "{Rates of compact object coalescences}",
    eprint = "2107.14239",
    archivePrefix = "arXiv",
    primaryClass = "astro-ph.HE",
    doi = "10.1007/s41114-021-00034-3",
    journal = "Living Rev. Rel.",
    volume = "25",
    number = "1",
    pages = "1",
    year = "2022"
}

@article{Neights:2025keq,
    author = "Neights, Eliza and others",
    title = "{GRB~250702B: discovery of a gamma-ray burst from a black hole falling into a star}",
    eprint = "2509.22792",
    archivePrefix = "arXiv",
    primaryClass = "astro-ph.HE",
    doi = "10.1093/mnras/staf2019",
    journal = "Mon. Not. Roy. Astron. Soc.",
    volume = "545",
    number = "2",
    pages = "staf2019",
    year = "2025"
}

@article{Ho:2021fyb,
    author = "Ho, Anna Y. Q. and others",
    title = "{A Search for Extragalactic Fast Blue Optical Transients in ZTF and the Rate of AT2018cow-like Transients}",
    eprint = "2105.08811",
    archivePrefix = "arXiv",
    primaryClass = "astro-ph.HE",
    doi = "10.3847/1538-4357/acc533",
    journal = "Astrophys. J.",
    volume = "949",
    number = "2",
    pages = "120",
    year = "2023"
}

@article{Rodriguez:2018rmd,
    author = "Rodriguez, Carl L. and Loeb, Abraham",
    title = "{Redshift Evolution of the Black Hole Merger Rate from Globular Clusters}",
    eprint = "1809.01152",
    archivePrefix = "arXiv",
    primaryClass = "astro-ph.HE",
    doi = "10.3847/2041-8213/aae377",
    journal = "Astrophys. J. Lett.",
    volume = "866",
    number = "1",
    pages = "L5",
    year = "2018"
}

@article{Hobbs:2005yx,
    author = "Hobbs, George and Lorimer, D. R. and Lyne, A. G. and Kramer, M.",
    title = "{A Statistical study of 233 pulsar proper motions}",
    eprint = "astro-ph/0504584",
    archivePrefix = "arXiv",
    doi = "10.1111/j.1365-2966.2005.09087.x",
    journal = "Mon. Not. Roy. Astron. Soc.",
    volume = "360",
    pages = "974--992",
    year = "2005"
}

@article{Mandel:2007hi,
    author = "Mandel, Ilya and Brown, Duncan A. and Gair, Jonathan R. and Miller, M. Coleman",
    title = "{Rates and Characteristics of Intermediate-Mass-Ratio Inspirals Detectable by Advanced LIGO}",
    eprint = "0705.0285",
    archivePrefix = "arXiv",
    primaryClass = "astro-ph",
    doi = "10.1086/588246",
    journal = "Astrophys. J.",
    volume = "681",
    pages = "1431--1447",
    year = "2008"
}

@ARTICLE{2018MNRAS.481.3278R,
       author = {{Ricarte}, Angelo and {Natarajan}, Priyamvada},
        title = "{The observational signatures of supermassive black hole seeds}",
      journal = {\mnras},
         year = 2018,
        month = dec,
       volume = {481},
       number = {3},
        pages = {3278-3292},
          doi = {10.1093/mnras/sty2448},
archivePrefix = {arXiv},
       eprint = {1809.01177},
 primaryClass = {astro-ph.GA},
       adsurl = {https://ui.adsabs.harvard.edu/abs/2018MNRAS.481.3278R}
}

@article{LIGOScientific:2026ctl,
    author = {Abac, A. G. and others},
    collaboration = "LIGO Scientific, Virgo, KAGRA",
    title = "{GWTC-5.0: Population Properties of Merging Compact Binaries}",
    eprint = "2605.27226",
    archivePrefix = "arXiv",
    primaryClass = "astro-ph.HE",
    reportNumber = "LIGO-P2600045",
    month = "5",
    year = "2026",
    journal = "",
}

@ARTICLE{2019MNRAS.482.4528E,
       author = {{El-Badry}, Kareem and {Quataert}, Eliot and {Weisz}, Daniel R. and {Choksi}, Nick and {Boylan-Kolchin}, Michael},
        title = "{The formation and hierarchical assembly of globular cluster populations}",
      journal = {\mnras},
         year = 2019,
        month = feb,
       volume = {482},
       number = {4},
        pages = {4528-4552},
          doi = {10.1093/mnras/sty3007},
archivePrefix = {arXiv},
       eprint = {1805.03652},
 primaryClass = {astro-ph.GA},
       adsurl = {https://ui.adsabs.harvard.edu/abs/2019MNRAS.482.4528E}
}

@article{Mapelli:2021gyv,
    author = "Mapelli, Michela and Bouffanais, Yann and Santoliquido, Filippo and Sedda, Manuel Arca and Artale, M. Celeste",
    title = "{The cosmic evolution of binary black holes in young, globular, and nuclear star clusters: rates, masses, spins, and mixing fractions}",
    eprint = "2109.06222",
    archivePrefix = "arXiv",
    primaryClass = "astro-ph.HE",
    doi = "10.1093/mnras/stac422",
    journal = "Mon. Not. Roy. Astron. Soc.",
    volume = "511",
    number = "4",
    pages = "5797--5816",
    year = "2022"
}

@ARTICLE{1974ApJS...27...21O,
       author = {{Oke}, J.~B.},
        title = "{Absolute Spectral Energy Distributions for White Dwarfs}",
      journal = {\apjs},
         year = 1974,
        month = feb,
       volume = {27},
        pages = {21},
          doi = {10.1086/190287},
       adsurl = {https://ui.adsabs.harvard.edu/abs/1974ApJS...27...21O}
}

@article{Berti:2008af,
    author = "Berti, Emanuele and Volonteri, Marta",
    title = "{Cosmological black hole spin evolution by mergers and accretion}",
    eprint = "0802.0025",
    archivePrefix = "arXiv",
    primaryClass = "astro-ph",
    doi = "10.1086/590379",
    journal = "Astrophys. J.",
    volume = "684",
    pages = "822--828",
    year = "2008"
}

@ARTICLE{1970Natur.226...64B,
       author = {{Bardeen}, James M.},
        title = "{Kerr Metric Black Holes}",
      journal = {\nat},
         year = 1970,
        month = apr,
       volume = {226},
       number = {5240},
        pages = {64-65},
          doi = {10.1038/226064a0},
       adsurl = {https://ui.adsabs.harvard.edu/abs/1970Natur.226...64B}
}

@ARTICLE{1994ApJ...424..823C,
       author = {{Cook}, Gregory B. and {Shapiro}, Stuart L. and {Teukolsky}, Saul A.},
        title = "{Rapidly Rotating Neutron Stars in General Relativity: Realistic Equations of State}",
      journal = {\apj},
         year = 1994,
        month = apr,
       volume = {424},
        pages = {823},
          doi = {10.1086/173934},
       adsurl = {https://ui.adsabs.harvard.edu/abs/1994ApJ...424..823C}
}

@article{Breu:2016ufb,
    author = "Breu, Cosima and Rezzolla, Luciano",
    title = "{Maximum mass, moment of inertia and compactness of relativistic stars}",
    eprint = "1601.06083",
    archivePrefix = "arXiv",
    primaryClass = "gr-qc",
    doi = "10.1093/mnras/stw575",
    journal = "Mon. Not. Roy. Astron. Soc.",
    volume = "459",
    number = "1",
    pages = "646--656",
    year = "2016"
}

@ARTICLE{1974ApJ...191..507T,
       author = {{Thorne}, Kip S.},
        title = "{Disk-Accretion onto a Black Hole. II. Evolution of the Hole}",
      journal = {\apj},
         year = 1974,
        month = jul,
       volume = {191},
        pages = {507-520},
          doi = {10.1086/152991},
       adsurl = {https://ui.adsabs.harvard.edu/abs/1974ApJ...191..507T}
}

@ARTICLE{2010A&A...520A..16B,
       author = {{Bejger}, M. and {Zdunik}, J.~L. and {Haensel}, P.},
        title = "{Approximate analytic expressions for circular orbits around rapidly rotating compact stars}",
      journal = {\aap},
         year = 2010,
        month = sep,
       volume = {520},
          eid = {A16},
        pages = {A16},
          doi = {10.1051/0004-6361/201015513},
archivePrefix = {arXiv},
       eprint = {1008.0384},
 primaryClass = {astro-ph.SR},
       adsurl = {https://ui.adsabs.harvard.edu/abs/2010A&A...520A..16B}
}

@article{Akmal:1998cf,
    author = "Akmal, A. and Pandharipande, V. R. and Ravenhall, D. G.",
    title = "{The Equation of state of nucleon matter and neutron star structure}",
    eprint = "nucl-th/9804027",
    archivePrefix = "arXiv",
    doi = "10.1103/PhysRevC.58.1804",
    journal = "Phys. Rev. C",
    volume = "58",
    pages = "1804--1828",
    year = "1998"
}

@article{Lattimer:2004nj,
    author = "Lattimer, James M. and Schutz, Bernard F.",
    title = "{Constraining the equation of state with moment of inertia measurements}",
    eprint = "astro-ph/0411470",
    archivePrefix = "arXiv",
    doi = "10.1086/431543",
    journal = "Astrophys. J.",
    volume = "629",
    pages = "979--984",
    year = "2005"
}

@article{Pappas:2013naa,
    author = "Pappas, George and Apostolatos, Theocharis A.",
    title = "{Effectively universal behavior of rotating neutron stars in general relativity makes them even simpler than their Newtonian counterparts}",
    eprint = "1311.5508",
    archivePrefix = "arXiv",
    primaryClass = "gr-qc",
    doi = "10.1103/PhysRevLett.112.121101",
    journal = "Phys. Rev. Lett.",
    volume = "112",
    pages = "121101",
    year = "2014"
}

@article{Laarakkers:1997hb,
    author = "Laarakkers, William G. and Poisson, Eric",
    title = "{Quadrupole moments of rotating neutron stars}",
    eprint = "gr-qc/9709033",
    archivePrefix = "arXiv",
    doi = "10.1086/306732",
    journal = "Astrophys. J.",
    volume = "512",
    pages = "282--287",
    year = "1999"
}

@article{Yagi:2013awa,
    author = "Yagi, Kent and Yunes, Nicolas",
    title = "{I-Love-Q Relations in Neutron Stars and their Applications to Astrophysics, Gravitational Waves and Fundamental Physics}",
    eprint = "1303.1528",
    archivePrefix = "arXiv",
    primaryClass = "gr-qc",
    doi = "10.1103/PhysRevD.88.023009",
    journal = "Phys. Rev. D",
    volume = "88",
    number = "2",
    pages = "023009",
    year = "2013"
}

@article{Luk:2018xmt,
    author = "Luk, Shun-Sun and Lin, Lap-Ming",
    title = "{Universal relations for innermost stable circular orbits around rapidly rotating neutron stars}",
    eprint = "1805.10813",
    archivePrefix = "arXiv",
    primaryClass = "astro-ph.HE",
    doi = "10.3847/1538-4357/aac8d6",
    journal = "Astrophys. J.",
    volume = "861",
    number = "2",
    pages = "141",
    year = "2018"
}

@ARTICLE{1972ApJ...178..347B,
       author = {{Bardeen}, James M. and {Press}, William H. and {Teukolsky}, Saul A.},
        title = "{Rotating Black Holes: Locally Nonrotating Frames, Energy Extraction, and Scalar Synchrotron Radiation}",
      journal = {\apj},
         year = 1972,
        month = dec,
       volume = {178},
        pages = {347-370},
          doi = {10.1086/151796},
       adsurl = {https://ui.adsabs.harvard.edu/abs/1972ApJ...178..347B}
}

@inproceedings{Buonanno:2007yg,
    author = "Buonanno, Alessandra",
    title = "{Gravitational waves}",
    booktitle = "{Les Houches Summer School - Session 86: Particle Physics and Cosmology: The Fabric of Spacetime}",
    eprint = "0709.4682",
    archivePrefix = "arXiv",
    primaryClass = "gr-qc",
    month = "9",
    year = "2007"
}

@article{Kremer:2022xgm,
    author = "Kremer, Kyle and Lombardi, James C. and Lu, Wenbin and Piro, Anthony L. and Rasio, Frederic A.",
    title = "{Hydrodynamics of Collisions and Close Encounters between Stellar Black Holes and Main-sequence Stars}",
    eprint = "2201.12368",
    archivePrefix = "arXiv",
    primaryClass = "astro-ph.HE",
    doi = "10.3847/1538-4357/ac714f",
    journal = "Astrophys. J.",
    volume = "933",
    number = "2",
    pages = "203",
    year = "2022"
}

@ARTICLE{2010ARA&A..48..431P,
       author = {{Portegies Zwart}, Simon F. and {McMillan}, Stephen L.~W. and {Gieles}, Mark},
        title = "{Young Massive Star Clusters}",
      journal = {\araa},
         year = 2010,
        month = sep,
       volume = {48},
        pages = {431-493},
          doi = {10.1146/annurev-astro-081309-130834},
archivePrefix = {arXiv},
       eprint = {1002.1961},
 primaryClass = {astro-ph.GA},
       adsurl = {https://ui.adsabs.harvard.edu/abs/2010ARA&A..48..431P}
}

@article{Wang:2021poh,
    author = "Wang, Yi-Han and Perna, Rosalba and Armitage, Philip J.",
    title = "{Partial tidal disruption events by stellar mass black holes: gravitational instability of stream and impact from remnant core}",
    eprint = "2103.09238",
    archivePrefix = "arXiv",
    primaryClass = "astro-ph.HE",
    doi = "10.1093/mnras/stab802",
    journal = "Mon. Not. Roy. Astron. Soc.",
    volume = "503",
    number = "4",
    pages = "6005--6015",
    year = "2021"
}

@article{Maguire:2020lad,
    author = "Maguire, Kate and Eracleous, Michael and Jonker, Peter G. and MacLeod, Morgan and Rosswog, Stephan",
    title = "{Tidal Disruptions of White Dwarfs: Theoretical Models and Observational Prospects}",
    eprint = "2004.00146",
    archivePrefix = "arXiv",
    primaryClass = "astro-ph.HE",
    doi = "10.1007/s11214-020-00661-2",
    journal = "Space Sci. Rev.",
    volume = "216",
    number = "3",
    pages = "39",
    year = "2020"
}

@article{Sesana:2008zc,
    author = "Sesana, A. and Vecchio, A. and Eracleous, M. and Sigurdsson, S.",
    title = "{Observing white dwarfs orbiting massive black holes in the gravitational wave and electro-magnetic window}",
    eprint = "0806.0624",
    archivePrefix = "arXiv",
    primaryClass = "astro-ph",
    doi = "10.1111/j.1365-2966.2008.13904.x",
    journal = "Mon. Not. Roy. Astron. Soc.",
    volume = "391",
    pages = "718--726",
    year = "2008"
}

@article{Perets:2016pwr,
    author = "Perets, Hagai B. and Li, Zhuo and Lombardi, James C. and Milcarek, Stephen R.",
    title = "{Micro - tidal disruption events by stellar compact objects and the production of ultra-long GRBs}",
    eprint = "1602.07698",
    archivePrefix = "arXiv",
    primaryClass = "astro-ph.HE",
    doi = "10.3847/0004-637X/823/2/113",
    journal = "Astrophys. J.",
    volume = "823",
    number = "2",
    pages = "113",
    year = "2016"
}

@ARTICLE{2024MNRAS.530.3043P,
       author = {{Pomeroy}, Richard T. and {Norris}, Mark A.},
        title = "{A search for intermediate-mass black holes in compact stellar systems through optical emissions from tidal disruption events}",
      journal = {\mnras},
         year = 2024,
        month = may,
       volume = {530},
       number = {3},
        pages = {3043-3050},
          doi = {10.1093/mnras/stae960},
archivePrefix = {arXiv},
       eprint = {2404.09144},
 primaryClass = {astro-ph.GA},
       adsurl = {https://ui.adsabs.harvard.edu/abs/2024MNRAS.530.3043P}
}

@article{Gezari:2021bmb,
    author = "Gezari, Suvi",
    title = "{Tidal Disruption Events}",
    eprint = "2104.14580",
    archivePrefix = "arXiv",
    primaryClass = "astro-ph.HE",
    doi = "10.1146/annurev-astro-111720-030029",
    journal = "Ann. Rev. Astron. Astrophys.",
    volume = "59",
    pages = "21--58",
    year = "2021"
}

@article{Kremer:2020cne,
    author = "Kremer, Kyle and Lu, Wenbin and Piro, Anthony L. and Chatterjee, Sourav and Rasio, Frederic A. and Ye, Claire S.",
    title = "{Fast Optical Transients from Stellar-Mass Black Hole Tidal Disruption Events in Young Star Clusters}",
    eprint = "2012.02796",
    archivePrefix = "arXiv",
    primaryClass = "astro-ph.HE",
    doi = "10.3847/1538-4357/abeb14",
    journal = "Astrophys. J.",
    volume = "911",
    number = "2",
    pages = "104",
    year = "2021"
}

@article{Kremer:2019zql,
    author = "Kremer, Kyle and Lu, Wenbin and Rodriguez, Carl L. and Lachat, Mitchell and Rasio, Frederic",
    title = "{Tidal Disruptions of Stars by Black Hole Remnants in Dense Star Clusters}",
    eprint = "1904.06353",
    archivePrefix = "arXiv",
    primaryClass = "astro-ph.HE",
    doi = "10.3847/1538-4357/ab2e0c",
    month = "4",
    year = "2019",
    journal = "",
}

@ARTICLE{2026A&A...707A.217R,
       author = {{Rastello}, Sara and {Iorio}, Giuliano and {Gieles}, Mark and {Wang}, Long},
        title = "{Micro-tidal disruption events in young star clusters}",
      journal = {\aap},
         year = 2026,
        month = mar,
       volume = {707},
          eid = {A217},
        pages = {A217},
          doi = {10.1051/0004-6361/202556781},
archivePrefix = {arXiv},
       eprint = {2509.07067},
 primaryClass = {astro-ph.HE},
       adsurl = {https://ui.adsabs.harvard.edu/abs/2026A&A...707A.217R}
}

@ARTICLE{2019ApJ...877...56L,
       author = {{Lopez}, Jr., Martin and {Batta}, Aldo and {Ramirez-Ruiz}, Enrico and {Martinez}, Irvin and {Samsing}, Johan},
        title = "{Tidal Disruptions of Stars by Binary Black Holes: Modifying the Spin Magnitudes and Directions of LIGO Sources in Dense Stellar Environments}",
      journal = {\apj},
         year = 2019,
        month = may,
       volume = {877},
       number = {1},
          eid = {56},
        pages = {56},
          doi = {10.3847/1538-4357/ab1842},
archivePrefix = {arXiv},
       eprint = {1812.01118},
 primaryClass = {astro-ph.HE},
       adsurl = {https://ui.adsabs.harvard.edu/abs/2019ApJ...877...56L}
}

@article{Kritos:2022ggc,
    author = "Kritos, Konstantinos and Strokov, Vladimir and Baibhav, Vishal and Berti, Emanuele",
    title = "{Dynamical formation of black hole binaries in dense star clusters: Rapid cluster evolution code}",
    eprint = "2210.10055",
    archivePrefix = "arXiv",
    primaryClass = "astro-ph.HE",
    doi = "10.1103/PhysRevD.110.043023",
    journal = "Phys. Rev. D",
    volume = "110",
    number = "4",
    pages = "043023",
    year = "2024"
}

@BOOK{1969mech.book.....L,
       author = {{Landau}, Lev Davidovich and {Lifshitz}, E.~M.},
        title = "{Mechanics}",
         year = 1969,
       adsurl = {https://ui.adsabs.harvard.edu/abs/1969mech.book.....L},
      publisher = "Pergamon Press"
}

@ARTICLE{2019PASP..131a8002B,
       author = {{Bellm}, Eric C. and {Kulkarni}, Shrinivas R. and {Graham}, Matthew J. and others},
        title = "{The Zwicky Transient Facility: System Overview, Performance, and First Results}",
      journal = {\pasp},
         year = 2019,
        month = jan,
       volume = {131},
       number = {995},
        pages = {018002},
          doi = {10.1088/1538-3873/aaecbe},
archivePrefix = {arXiv},
       eprint = {1902.01932},
 primaryClass = {astro-ph.IM},
       adsurl = {https://ui.adsabs.harvard.edu/abs/2019PASP..131a8002B}
}

@article{Graham:2019qsw,
    author = "Graham, Matthew J. and others",
    title = "{The Zwicky Transient Facility: Science Objectives}",
    eprint = "1902.01945",
    archivePrefix = "arXiv",
    primaryClass = "astro-ph.IM",
    doi = "10.1088/1538-3873/ab006c",
    journal = "Publ. Astron. Soc. Pac.",
    volume = "131",
    number = "1001",
    pages = "078001",
    year = "2019"
}

@article{Aso:2013eba,
    author = "Aso, Yoichi and Michimura, Yuta and Somiya, Kentaro and Ando, Masaki and Miyakawa, Osamu and Sekiguchi, Takanori and Tatsumi, Daisuke and Yamamoto, Hiroaki",
    collaboration = "KAGRA",
    title = "{Interferometer design of the KAGRA gravitational wave detector}",
    eprint = "1306.6747",
    archivePrefix = "arXiv",
    primaryClass = "gr-qc",
    doi = "10.1103/PhysRevD.88.043007",
    journal = "Phys. Rev. D",
    volume = "88",
    number = "4",
    pages = "043007",
    year = "2013"
}

\end{document}